\documentclass[english,preprintnumbers]{revtex4}
\usepackage[T1]{fontenc}
\usepackage[latin9]{inputenc}
\usepackage{xcolor}
\usepackage{pdfcolmk}
\usepackage{amsmath}
\usepackage{amssymb}
\usepackage{esint}
\PassOptionsToPackage{normalem}{ulem}
\usepackage{ulem}

\makeatletter

\providecommand{\tabularnewline}{\\}
\providecolor{lyxadded}{rgb}{0.99,0.0078,1}
\providecolor{lyxdeleted}{rgb}{1,0,0}

\DeclareRobustCommand{\lyxsout}[1]{\ifx\\#1\else\sout{#1}\fi}

\@ifundefined{textcolor}{}
{%
 \definecolor{BLACK}{gray}{0}
 \definecolor{WHITE}{gray}{1}
 \definecolor{RED}{rgb}{1,0,0}
 \definecolor{GREEN}{rgb}{0,1,0}
 \definecolor{BLUE}{rgb}{0,0,1}
 \definecolor{CYAN}{cmyk}{1,0,0,0}
 \definecolor{MAGENTA}{cmyk}{0,1,0,0}
 \definecolor{YELLOW}{cmyk}{0,0,1,0}
}

\usepackage{graphics}\usepackage{epstopdf}\usepackage{epsfig}\@ifundefined{definecolor}{\usepackage{color}}{}
\usepackage{amsfonts}\usepackage{bm}\usepackage{amscd}

\input{tcilatex}

\usepackage{babel}

\usepackage{babel}

\usepackage{hyperref}

\makeatother

\usepackage{babel}
\begin{document}
\title{Ultrafast spintronics with geometric effects in non-adiabatic wave-packet
dynamics}
\author{Matisse Wei-Yuan Tu}
\email{kerustemiro@gmail.com}
\affiliation{Fritz Haber Center for Molecular Dynamics, Institute of Chemistry,
The Hebrew University of Jerusalem, Jerusalem 91904 Israel}
\affiliation{Center for Theoretical and Computation Physics, Department of Physics,
National Sun Yat-sen University, Kaohsiung 80424, Taiwan}
\author{Li-Sheng Lin}
\affiliation{Graduate School of Advanced Technology, National Taiwan University,
Taipei 106319, Taiwan}
\author{Chung-Yu Wang}
\affiliation{Fritz Haber Center for Molecular Dynamics, Institute of Chemistry,
The Hebrew University of Jerusalem, Jerusalem 91904 Israel}
\affiliation{Physics Division, National Center for Theoretical Sciences, Taipei 10617, Taiwan}
\author{Jyh-Pin Chou}
\affiliation{Department of Physics, National Changhua University of Education,
Changhua 50007, Taiwan}
\affiliation{Graduate School of Advanced Technology, National Taiwan University,
Taipei 106319, Taiwan}
\affiliation{Physics Division, National Center for Theoretical Sciences, Taipei 10617, Taiwan}
\author{Sin-Yi Wei }
\affiliation{Department of Electrophysics, National Yang Ming Chiao Tung University,
Hsinchu 30010, Taiwan}
\author{Chien-Ming Tu}
\affiliation{Department of Electrophysics, National Yang Ming Chiao Tung University,
Hsinchu 30010, Taiwan}
\author{Chia-Nung Kuo }
\affiliation{Department of Physics, National Cheng Kung University, Tainan 70101,
Taiwan}
\author{Chin-Shan Lue}
\affiliation{Department of Physics, National Cheng Kung University, Tainan 70101,
Taiwan}
\affiliation{Taiwan Consortium of Emergent Crystalline Materials (TCECM), National
Science and Technology Council, Taipei 10601, Taiwan}
\author{Chih-Wei Luo}
\affiliation{Department of Electrophysics and Institute of Physics, National Yang
Ming Chiao Tung University, Hsinchu 30010, Taiwan}
\affiliation{National Synchrotron Radiation Research Center, Hsinchu 30076, Taiwan}
\affiliation{Taiwan Consortium of Emergent Crystalline Materials (TCECM), National
Science and Technology council, Taipei 10601, Taiwan}
\begin{abstract}
Motivated by the intriguing possibilities of steering ultrafast non-adiabatic
processes through the geometric properties of bands in quantum materials
by laser pulses, we extend a wave-packet transport theory, previously
well-established in the adiabatic regime that intuitively captured
geometric properties of bands, to the transient and non-adiabatic
regime. This extension facilitates us to investigate macroscopic ways
of manifesting microscopic band-geometric effects that highlight the
special capability of non-adiabatic drivings not available to adiabatic
drivings. These include imprinting band-geometric properties to the
current rate after switching off the laser pulses and the induction
of intrinsic macroscopic spin polarisation with an orientation not
accessible by adiabatic processes. In particular, the microscopic
geometrically-rooted intrinsic spin coherence is shown to underlie
the spin-mediated parts of the macroscopic photocurrents. Through
explicit calculations of an example with Rashba spin-orbit coupling,
the spin-mediated part is shown to be discernible from the non-spin-mediated
part in terms of the anisotropy of the photocurrents. Working principles
behind the above theoretical results allegedly applicable beyond the
Rashba example are distilled to inspect experimental data collected
for SnSe, exhibiting considerable anisotropic effects. Consistency
between theory and experiment is observed, paving the way of further
exploration into the above intended direction.
\end{abstract}
\maketitle

\section{Introduction}

Geometric and topological properties of band structures are fundamental
physical assets vital for developing technologies for quantum materials.
These properties can be manifested through the electronic currents,
the primary observable induced in both electric transport and optoelectronic
experiments. Three temporal forms of the currents are usually investigated,
namely, steady-state, time-periodic, and transient. A previously established
theoretical framework based on the picture of wave packets (see its
introduction later) has been witnessed to be useful for intuitively
understanding the band-geometric and -topological effects through
steady-state as well as time-periodic currents. Particularly, with
ultrashort laser pulses, the induced transient photocurrents are expected
to reveal the underlying band-structure effects in a way that is not
obviously visible in the steady-state and time-periodic currents.
We are thus motivated here to develop extensions of this theoretically
intuitive framework into the transient regime to address the band-geometric
effects in laser-pulse-induced real-time photocurrents. It is well-known
that nonzero Berry curvature is the defining ingredient of geometric
band-structure effects that have been revealed in both steady-state
and time-periodic regimes \cite{Karplus541154,Thouless82405,Nagaosa101539,Murakami031348,Xiao07236809,Sodemann15216806,Gorbachev14448,Sui151027,Komatsu18eaaq0194,Xu18900,Kang19324}.
To highlight the transient regime, here we will demonstrate interesting
band-geometric effects by applications to systems with zero Berry
curvatures. 

To theoretically address the band-geometric and -topological effects
on the currents, there are primarily two frameworks available in the
literatures. The first is the Green-function framework \cite{Rammer86323,Mahan00book,Haug08book}
which had been developed for treating general properties of materials
and not specifically for the above purpose. The second framework,
born out of the above purpose, is the wave-packet-based transport
theory \cite{Chang951348,Chang967010,Xiao101959}. In the first framework,when
setting to the linear response regime, the resulting Kubo formula
has been turned into TKNN formula for illustrating the role of topological
effects in near-equilibrium transport and opened the venue for the
studies of band topology in quantum materials \cite{Thouless82405,KOHMOTO1985343}.
Following the rigorous quantum mechanical rules, the Green-function
framework had been well-established over the years and its applications
related to the intrinsic band-geometric and -topological properties
are mainly concentrated on tackling the extrinsic scattering effects,
owing to its characteristic ability of systematically performing perturbative
calculations using diagrams (see examples \cite{Parker19045121,Michishita22125114}).
Despite being fully rigorous on the quantum mechanical ground, the
way that band-geometric effects on transport are manifested through
the heavy machinery of the Green function formulation is not as transparent
as the second framework based on the picture of electron wave packets,
here called the wave-packet transport theory (WPT). 

The WPT is quantum mechanically rigorous on the level of a single
electron as a wave packet moving among the bands. It is phenomenological
on the level of electron gases as ensembles of electron wave packets.
The main advantage of the WPT is its transparency of revealing band-geometric
effects with clear physical intuition. It achieves this physical insight
by identifying the anomalous velocity with the Berry curvatures of
the bands that in turn manifest in the Hall currents \cite{Karplus541154,Chang951348,Chang967010,Xiao101959}.
It has found numerous applications in exploring the band-geometric
effects in quantum materials \cite{Xiao05137204,Olson07035114,Chang08193202,Son13104412,Jho17205113,Gorbar18045203,Misaki18075122}.
Nevertheless, the application of this framework to the study of transient
photocurrents induced by short laser pulses is not a straightforward
and easy task. The emergence of the Berry curvature in the anomalous
velocity is a result of the adiabatic limit in which the applied bias
to transport the electrons is relatively small in comparison to the
band gaps in question. This limit enables adiabatic approximation
as well as perturbation treatment of the external field. In narrowly
gapped materials, featured by gaps of tens of meV at the anti-crossing
\cite{Gorbachev14448,Shimazaki151032,Deng20895}, it becomes possible
to break the adiabatic condition with moderate bias. Even under a
constant-in-time electric field, extending the WPT to the non-adiabatic
regime already met challenges due to multiple-band effects \cite{Tu20045423}.
Still in the steady-state regime, the non-adiabatic extension of the
WPT readily yields interesting results previously unseen with the
adiabatic and perturbative treatments, such as non-adiabatic renormalization
of valley Hall current and crystal anisotropy manifestation of non-Abelian
multiple-band characters \cite{Tu20045004,Li21045012}. We are thus
naturally motivated to further extend this non-adiabatic WPT from
steady-states to the transient regime. 

A natural playground of the transient and non-adiabatic regime concerning
the band-geometric effects is the ultrafast spintronics with non-magnetic
materials that mainly exploit spin-orbit coupling (SOC). There are
a number of interesting materials where the SOC induced spin-split
gaps range from tens to hundreds of meV \cite{Hu18235404,Sino216608,Gomes15085406,Tao21113001}
implying the possibilities of activating the non-adiabatic dynamics
by attainable electric field amplitudes not more than 100meV/$\text{\AA}$
(see later discussions for details of estimation). Because of the
non-adiabatic nature, the aspects of the geometric properties to be
focused on here can be quite different from others. The geometric
properties of the bands are mathematically-defined derivatives of
the band wavefunctions with respect to Bloch momentum \cite{Xiao101959,Torma03240001}.
Under small perturbating electric field, the field amplitudes can
be factorised from the finite-order response coefficients in which
derivatives of the band wavefunctions naturally appear \cite{Deyo091917v1,Moore10026805,Sodemann15216806,Tsirkin22039}.
These finite-order response coefficients, e.g., the linear conductivity
as the most famous one, tell of the native properties of the materials
in terms of geometric quantities as the focus of other established
studies quoted above. Here in the regime of non-adiabatic drivings
exceeding the validity of order-by-order analysis, the properties
brought up by the drivings themselves, not just natively of the materials,
become crucial for transiently steering the ultrafast electronic/spintronic
processes due to the non-perturbation involvement of the laser fields.
Many experiments related to SOC-supported band-geometric effects have
demonstrated exquisite capacities of manipulating ultrafast lasers,
hinting feasibility of tuning the driving amplitudes to activate the
non-adiabatic regime \cite{Kastl156617,Braun1613259,Tu17195407,Reimann18396,Ma23170,Borsch23668}.
Based on SOC-supported topological surface states, intriguing transient
effects have been experimentally witnessed, including near unity fidelity
\cite{Kastl156617}, bond-specific ultrafast charge transfer \cite{Braun1613259}
and polarity-tunable terahertz emission \cite{Tu17195407}, just to
name a few. To have a grounded starting point for exploring this relatively
less charted area of non-adiabatic steering of transient dynamics
using band-geometric effects in ultrafast spintronics, instead of
the topological surface states with complicated geometrical properties
\cite{Fu09266801}, we should go back to the archetypal Rashba SOC
as one foundational setup of spintronics \cite{Bychkov198478,Datta90665,Sinova04126603,Manchon15871,Bandyopadhyay23015001}.

For being self-content, the previously established WPT was briefly
reviewed in Sec. \ref{WPT-basics-adia}. A non-adiabatic transient
extension of this WPT (NADT-WPT) is introduced in Sec. \ref{WPT-basics-nadia}.
Details of derivations are left in the supplementary material (SM).
In Sec. \ref{WPT-basics-Uniqnadia}, we discuss in general possible
ways that NADT-WPT can manifest band-geometric effects, especially
through macroscopic observables, e.g., currents, under non-adiabatic
drivings where the adiabatic WPT is not applicable. We demonstrate
these effects by examples with zero Berry curvatures in Sec. \ref{sec-examples}.
The first example is pedagogical and realised by the Su-Schrieffer-Heeger
(SSH) system in Sec. \ref{sec-examples-SSH}. The second example in
Sec. \ref{sec-examples-Rashba} is a prototype system with anisotropic
Rashba SOC for the foundational interests of spintronics. With short
laser pulses, the time domain is divided into a in-pulse (where the
laser field exists) and a post-pulse region (in which the laser driving
has been switched off). Our main findings are summarised here. (1):
In-pulse non-adiabatic drivings allow one to manipulate post-pulse
dynamics by band-geometric-enabled history-dependence effects that
are not available by in-pulse adiabatic drivings. (2): The geometric
effects of the SOC-governed bands can be macroscopically revealed
via the intrinsic (instead of extrinsic) current-induced spin polarisation
triggered by non-adiabatic drivings only. With anisotropic Rashba-SOC,
such manifestation in terms of macroscopic spin polarisation is dependent
on the anisotropy. (3): This anisotropic dependence furnishes a way
to differentiate spin-mediated from bond-mediated processes behind
the photocurrents. In Sec. \ref{conclude-sec} for conclusions, we
draw theoretical implications for various types of experiments. We
also perform ultrafast terahertz (THz) emission experiments on SnSe
showing considerable anisotropic SOC effects. Consistency between
NADT-WPT results deduced independently of the Rashba SOC and experimental
data is seen. Limitations and possibilities of improvements of the
present theoretical development are addressed.

\section{Formulation of a wave-packet transport theory}

\label{WPT-basics}The WPT takes as input the band structure of the
electronic material in question. The output of the theory is the electronic
current $\boldsymbol{j}\left(t\right)$ under a prescribed external
electric field $\boldsymbol{E}\left(t\right)$ that is in principle
dependent on time $t$. For the input, the Hamiltonian for the pristine
crystalline material is prescribed as $\mathcal{H}\left(\boldsymbol{k}\right)$
where $\boldsymbol{k}$ denotes the Bloch momentum in the Brillouin
zone (BZ). By writing $\mathcal{H}\left(\boldsymbol{k}\right)\left\vert u_{n}\left(\boldsymbol{k}\right)\right\rangle =\varepsilon_{n}\left(\boldsymbol{k}\right)\left\vert u_{n}\left(\boldsymbol{k}\right)\right\rangle $
with $n$ being the band index, the band structure is then specified
by $\left\{ \text{\ensuremath{\varepsilon_{n}\left(\boldsymbol{k}\right)}}\right\} _{n}$
the band energies and $\left\{ \left\vert u_{n}\left(\boldsymbol{k}\right)\right\rangle \right\} _{n}$
the periodic parts of the Bloch wavefunctions. The band wavefunctions
$\left\{ \left\vert u_{n}\left(\boldsymbol{k}\right)\right\rangle \right\} _{n}$
collected over all the bands at any given $\boldsymbol{k}$ spans
a Hilbert space called the band space. The band geometry is charaterised
by the properties of local variations of the band wavefunctions $\left\{ \left\vert u_{n}\left(\boldsymbol{k}\right)\right\rangle \right\} _{n}$
over the Bloch momentum $\boldsymbol{k}$. The Berry connection $\boldsymbol{\mathcal{R}}\left(\boldsymbol{k}\right)$,
which is a matrix indexed by the bands, has its elements defined by
\begin{equation}
\left[\mathcal{\mathcal{R}}_{\boldsymbol{k}}\right]_{n,m}=i\left\langle u_{n}\left(\boldsymbol{k}\right)\left\vert \frac{\partial}{\partial\boldsymbol{k}}\right\vert u_{m}\left(\boldsymbol{k}\right)\right\rangle .\label{BerryCon-def-Mk}
\end{equation}
It manifests the variations of the band wavefunctions over $\boldsymbol{k}$
and serves as a starting point to discuss band geometry.

To relate the input band structure to the output current, the WPT
is constructed in two steps \cite{Ashcroft76book,Xiao101959}. In
the first step, one considers the dynamics of a single electron wave
packet carrying a charge $-e$. Its state is characterised by its
centre-of-mass (CM) coordinate in the phase space, namely, $\left(\boldsymbol{x}_{t},\boldsymbol{k}_{t}\right)$,
with $\boldsymbol{x}_{t}$ the position and $\boldsymbol{k}_{t}$
the momentum, which follows

\begin{equation}
\dot{\boldsymbol{x}}_{t}=\left\langle \left[\hbar^{-1}\mathcal{D}_{\boldsymbol{k}_{t}},\mathcal{H}\left(\boldsymbol{k}_{t}\right)\right]\right\rangle ,\label{xdot-nra}
\end{equation}
\begin{equation}
\hbar\dot{\boldsymbol{k}}_{t}=\left(-e\right)\boldsymbol{E}\left(t\right),\label{driving-E1}
\end{equation}
where $\left[\mathcal{D}_{\boldsymbol{k}_{t}}\right]_{n,m}=\delta_{n,m}\partial/\partial\boldsymbol{k}_{t}-i\left[\mathcal{\mathcal{R}}_{\boldsymbol{k}_{t}}\right]_{n,m}$
is the covariant derivative (see derivations in SM-Sec. \ref{SI-wpt-deriv-single}).
In the second step, the electronic current $\boldsymbol{j}\left(t\right)$
through the electron gas as an ensemble of wave packets is constructed
from collecting the velocities of all the wave packets. This is formally
defined by

\begin{equation}
\boldsymbol{j}\left(t\right)=-e\int\text{d}^{D}\boldsymbol{k}\text{Tr}\left(\boldsymbol{V}\left(\boldsymbol{k}_{t}\right)\varrho_{\boldsymbol{k}}\left(t\right)\right),\label{j-def0}
\end{equation}
where $D$ is the spatial dimension of the BZ,
\begin{equation}
\boldsymbol{V}\left(\boldsymbol{k}_{t}\right)\equiv\frac{\partial\mathcal{H}\left(\boldsymbol{k}_{t}\right)}{\partial\hbar\boldsymbol{k}_{t}},\label{vopt-def}
\end{equation}
is the velocity operator acting on the band space and $\varrho_{\boldsymbol{k}}\left(t\right)$
is the one-body reduced density matrix (called the density matrix
shortly afterwards) for the state of the ensemble. At time $t\le t_{0}$
we assume the electron gas remains in equilibrium, namely, $\varrho_{\boldsymbol{k}}\left(t\le t_{0}\right)=\varrho_{\boldsymbol{k}}^{eq}$,
where 
\begin{equation}
\varrho_{\boldsymbol{k}}^{eq}=\sum_{n}f_{FD}^{\mu,T}\left(\varepsilon_{n}\left(\boldsymbol{k}\right)\right)\left\vert u_{n}\left(\boldsymbol{k}\right)\right\rangle \left\langle u_{n}\left(\boldsymbol{k}\right)\right\vert ,\label{eq_1RDM-def}
\end{equation}
in which $f_{FD}^{\mu,T}\left(\varepsilon\right)=\left\{ \exp\left[\left(\varepsilon-\mu\right)/k_{B}T\right]+1\right\} ^{-1}$
is the Fermi-Dirac distribution with $\mu$ the chemical potential,
$T$ the temperature and $k_{B}$ the Boltzmann constant. The driving
begins for $t\ge t_{0}$. In the integrand of Eq.~(\ref{j-def0}),
the relation Eq. (\ref{driving-E1}) gives \begin{subequations}\label{k-shift}
\begin{equation}
\boldsymbol{k}_{t}=\boldsymbol{k}+\boldsymbol{\mathcal{K}}_{t},\label{nox-sol-k}
\end{equation}
with $\boldsymbol{k}_{t=t_{0}}=\boldsymbol{k}$ and 
\begin{equation}
\boldsymbol{\mathcal{K}}_{t}=-\frac{e}{\hbar}\int_{t_{0}}^{t}\text{d}t^{\prime\prime}\boldsymbol{E}\left(t^{\prime\prime}\right),\label{nox-sol-K}
\end{equation}
\end{subequations} is the net wave-vector displacement caused by
the electric field accumulated from $t_{0}$ to $t$. 

Starting from the initial state of the ensemble as Eq. (\ref{eq_1RDM-def}),
it is instructive to consider the intrinsic limit (no scatterings)
by defining 
\begin{equation}
\left\vert \phi_{n,\boldsymbol{k}}\left(t,t_{0}\right)\right\rangle :=U_{\boldsymbol{k}}\left(t,t_{0}\right)\left\vert u_{n}\left(\boldsymbol{k}\right)\right\rangle ,\label{intrinsic-stat_nk}
\end{equation}
 as the wave-packet state at time $t$ initiated from $\left\vert u_{n}\left(\boldsymbol{k}\right)\right\rangle $
at time $t_{0}$, where 
\begin{equation}
U_{\boldsymbol{k}}\left(t,t^{\prime}\right)=\hat{T}\exp\left\{ -\frac{i}{\hbar}\int_{t^{\prime}}^{t}\text{d}t^{\prime\prime}\mathcal{H}\left(\boldsymbol{k}_{t^{\prime\prime}}\right)\right\} ,\label{unievol-1}
\end{equation}
is the time-ordered evolution operator with $t\ge t^{\prime}$. Consequently,
the ensemble of the wave packets evolves as $\varrho_{\boldsymbol{k}}\left(t\right)\rightarrow\varrho_{\boldsymbol{k}}^{id}\left(t\right)$,
where
\begin{equation}
\varrho_{\boldsymbol{k}}^{id}\left(t\right)=\sum_{n}f_{FD}^{\mu,T}\left(\varepsilon_{n}\left(\boldsymbol{k}\right)\right)\left\vert \phi_{n,\boldsymbol{k}}\left(t,t_{0}\right)\right\rangle \left\langle \phi_{n,\boldsymbol{k}}\left(t,t_{0}\right)\right\vert .\label{ideal-dmtx}
\end{equation}
The purpose of the WPT is to relate microscopic $\boldsymbol{k}$-resolved
geometric properties of the bands (see, e.g. Eq. (\ref{BerryCon-def-Mk}))
to macroscopic $\boldsymbol{k}$-integrated observable (see e.g.,
the current Eq. (\ref{j-def0})).

\subsection{Adiabatic transport of wave packets }

\label{WPT-basics-adia}

\subsubsection{Formulation of adiabatic wave-packet dynamics and electronic currents}

Under the adiabatic condition, we re-denote $\left\vert \phi_{n,\boldsymbol{k}}\left(t,t_{0}\right)\right\rangle $
by $\left\vert \phi_{n,\boldsymbol{k}}^{ad}\left(t\right)\right\rangle $.
By driving $\boldsymbol{k}_{t}$ sufficiently slow with small enough
electric field, a wave packet starting as $\left\vert u_{n}\left(\boldsymbol{k}\right)\right\rangle $
at time $t_{0}$ will evolve to $\left\vert u_{n}\left(\boldsymbol{k}_{t}\right)\right\rangle $
at time $t$ with negligible occupations on $\left\vert u_{m\ne n}\left(\boldsymbol{k}_{t}\right)\right\rangle $
but with the inter-band coherence between bands $n$ and $m\ne n$
accounted to the first order in the driving field. Explicitly, this
so-called adiabatic driving leads to \cite{Boehm03book} 
\begin{equation}
\left\vert \phi_{n,\boldsymbol{k}}^{ad}\left(t\right)\right\rangle =e^{id_{n}\left(t\right)}\left[e^{i\gamma_{n}\left(\boldsymbol{k}_{t}\right)}\left\vert u_{n}\left(\boldsymbol{k}_{t}\right)\right\rangle +\sum_{m\ne n}r_{m,n}\left(\boldsymbol{k}_{t},\boldsymbol{E}\left(t\right)\right)e^{i\gamma_{m}\left(\boldsymbol{k}_{t}\right)}\left\vert u_{m}\left(\boldsymbol{k}_{t}\right)\right\rangle \right],\label{n-ad-uni}
\end{equation}
where
\begin{equation}
r_{m,n}\left(\boldsymbol{k}_{t},\boldsymbol{E}\left(t\right)\right)=\frac{\left[\mathcal{\bar{\mathcal{R}}}_{\boldsymbol{k}_{t}}\right]_{m,n}\cdot\left(-e\right)\boldsymbol{E}\left(t\right)}{\left(\varepsilon_{m}-\varepsilon_{n}\right)\left(\boldsymbol{k}_{t}\right)},\label{nonad-meas-def}
\end{equation}
$d_{n}\left(t\right)$ is the dynamical phase, $\gamma_{n}\left(\boldsymbol{k}_{t}\right)=\int_{\boldsymbol{k}_{t_{0}}}^{\boldsymbol{k}_{t}}\text{d}\boldsymbol{k}^{\prime}\cdot\left[\mathcal{\mathcal{R}}_{\boldsymbol{k}^{\prime}}\right]_{n,n}$
is the geometric phase and $\left[\mathcal{\bar{\mathcal{R}}}_{\boldsymbol{k}_{t}}\right]_{m,n}=e^{-i\gamma_{m}\left(\boldsymbol{k}_{t}\right)}\left[\mathcal{\mathcal{R}}_{\boldsymbol{k}_{t}}\right]_{m,n}e^{i\gamma_{n}\left(\boldsymbol{k}_{t}\right)}.$
By substituting Eq. (\ref{n-ad-uni}) to Eq. (\ref{xdot-nra}), the
most well-known result in this first step of the WPT is 
\begin{equation}
\dot{\boldsymbol{x}}_{t}=\frac{\partial\varepsilon\left(\boldsymbol{k}_{t}\right)}{\partial\boldsymbol{k}_{t}}-\left(-\frac{e}{\hbar}\right)\boldsymbol{E}\left(t\right)\times\boldsymbol{\Omega}\left(\boldsymbol{k}_{t}\right),\label{nox-EOMx-ad1}
\end{equation}
in which $\varepsilon\left(\boldsymbol{k}\right)=\varepsilon_{n}\left(\boldsymbol{k}\right)$,
and $\boldsymbol{\Omega}\left(\boldsymbol{k}\right)$ given by 
\begin{equation}
\boldsymbol{\Omega}\left(\boldsymbol{k}\right)=\boldsymbol{\nabla}_{\boldsymbol{k}}\times\left[\mathcal{\mathcal{R}}_{\boldsymbol{k}}\right]_{n,n},\label{AbelianBrcv-defMk}
\end{equation}
is known as the Berry curvature. For brevity under the adiabatic condition,
the band index $n$ is usually dropped off. Restoring the band index,
we have $\dot{\boldsymbol{x}}_{t}\rightarrow\dot{\boldsymbol{x}}_{t}^{n}=\boldsymbol{v}_{n}^{b}\left(\boldsymbol{k}_{t}\right)+\boldsymbol{v}_{n}^{r}\left(\boldsymbol{k}_{t},\boldsymbol{E}\left(t\right)\right)$.
Here the first term $\boldsymbol{v}_{n}^{b}\left(\boldsymbol{k}_{t}\right)=\partial\varepsilon_{n}\left(\boldsymbol{k}_{t}\right)/\partial\boldsymbol{k}_{t}$
due to the occupation on band $n$ is the known as the normal velocity
or group velocity. The second term, which arises from the first-order
effects in inter-band coherence, given by $\boldsymbol{v}_{n}^{r}\left(\boldsymbol{k}_{t},\boldsymbol{E}\left(t\right)\right)=\left(e/\hbar\right)\boldsymbol{E}\left(t\right)\times\boldsymbol{\Omega}_{n}\left(\boldsymbol{k}_{t}\right)$
in Eq. (\ref{nox-EOMx-ad1}), is known as the anomalous velocity that
carries the band-geometric information as the Berry curvature. In
the intrinsic limit with $\left\vert \phi_{n,\boldsymbol{k}}\left(t,t_{0}\right)\right\rangle $
in Eq. (\ref{ideal-dmtx}) replaced by $\left\vert \phi_{n,\boldsymbol{k}}^{ad}\left(t\right)\right\rangle $
due to the adiabatic condition, the current Eq. (\ref{j-def0}) is
simply evaluated by $\boldsymbol{j}\left(t\right)=-e\int\text{d}^{D}\boldsymbol{k}\sum_{n}f_{FD}^{\mu,T}\left(\varepsilon_{n}\left(\boldsymbol{k}\right)\right)\dot{\boldsymbol{x}}_{t}^{n}$
.

When one is away from the intrinsic limit, it is almost impossible
to obtain the density matrix exactly without any approximations. The
WPT in practice approximates $\varrho_{\boldsymbol{k}}\left(t\right)$
by phenomenologically incorporating effects extrinsic to the pristine
band structure via the scattering rates $\left\{ \tau_{\boldsymbol{k}}^{-1}\right\} _{\boldsymbol{k}}$
under the relaxation-time approximation (RTA) \cite{Ashcroft76book,Xiao101959}.
Combining the RTA for the ensemble and the adiabatic condition for
single wave packet, one arrives 

\begin{equation}
\boldsymbol{j}\left(t\right)=-e\int\text{d}^{D}\boldsymbol{k}_{t}f\left(\boldsymbol{k}_{t},t\right)\dot{\boldsymbol{x}}_{t},\label{Jad-tdep0}
\end{equation}
from Eq. (\ref{j-def0}) (see more details in SM-Sec. \ref{nlqh}).
In Eq. (\ref{Jad-tdep0}), the time-dependent distribution function
$f\left(\boldsymbol{k}_{t},t\right)$ is given by \begin{subequations}\label{non-eq-ad-tt}
\begin{equation}
f\left(\boldsymbol{k}_{t},t\right)=f^{0}\left(\boldsymbol{k}_{t}\right)+\delta f\left(\boldsymbol{k}_{t},t\right),\label{non-eq-t0}
\end{equation}
in which $f^{0}\left(\boldsymbol{k}_{t}\right)=f_{FD}^{\mu,T}\left(\varepsilon_{n}\left(\boldsymbol{k}_{t}\right)\right)$
accounts for the equilibrium part of the distribution and deviation
from equilibrium is 
\begin{align}
\delta f\left(\boldsymbol{k}_{t},t\right)= & -\int_{t_{0}}^{t}\text{d}t^{\prime}P\left(\boldsymbol{k}_{t},\boldsymbol{k}_{t^{\prime}}\right)\frac{\partial f^{0}}{\partial\boldsymbol{k}_{t^{\prime}}}\cdot\dot{\boldsymbol{k}}_{t^{\prime}}.\label{non-eq-t1}
\end{align}
\end{subequations} Here,
\begin{equation}
P\left(\boldsymbol{k}_{t},\boldsymbol{k}_{t^{\prime}}\right)=e^{-\int_{t^{\prime}}^{t}\text{d}t^{\prime\prime}\tau_{\boldsymbol{k}_{t^{\prime\prime}}}^{-1}},\label{non-scatt-prob1}
\end{equation}
is the probability that a wave packet carrying momentum $\boldsymbol{k}_{t^{\prime}}$
at time $t^{\prime}$ will arrive at time $t$ to the state $\boldsymbol{k}_{t}$
without being scattered for the whole time interval $\left[t^{\prime},t\right]$\cite{Ashcroft76book}.
Equation (\ref{Jad-tdep0}) has been widely applied for studying band-geometric
effects on transport \cite{Chang951348,Chang967010,Deyo091917v1,Moore10026805,Sodemann15216806,Dantas21L201105,Kovalev202451}. 

In principle, one does not need to restrict how $\boldsymbol{E}\left(t\right)$
depends on time for deploying Eq.(\ref{Jad-tdep0}) to compute electronic
current $\boldsymbol{j}\left(t\right)$ as long as the adiabatic condition
is satisfied. Substituting Eq. (\ref{nox-EOMx-ad1}) for $\dot{\boldsymbol{x}}_{t}$
into Eq. (\ref{Jad-tdep0}) decomposes the current to $\boldsymbol{j}\left(t\right)=\boldsymbol{j}^{H}\left(t\right)+\boldsymbol{j}^{L}\left(t\right)$
where $\boldsymbol{j}^{H}\left(t\right)=-e\int\text{d}^{D}\boldsymbol{k}_{t}f\left(\boldsymbol{k}_{t},t\right)\boldsymbol{v}^{r}\left(\boldsymbol{k}_{t},\boldsymbol{E}\left(t\right)\right)$
is known as the Hall current coming from the anomalous velocity and
$\boldsymbol{j}^{L}\left(t\right)=-e\int\text{d}^{D}\boldsymbol{k}_{t}\delta f\left(\boldsymbol{k}_{t},t\right)\boldsymbol{v}^{b}\left(\boldsymbol{k}_{t}\right)$
is the longitudinal current due to the group velocity. The most well-known
result in this final step of the WPT occurs with $D=2$ via the linear-response
Hall conductivity $\sigma_{xy}$ defined by $j_{x}^{H}\left(t\right)=\sigma_{xy}E_{y}\left(t\right)$.
At zero temperature, it is given by $\sigma_{xy}=-\left(e^{2}/\hbar\right)\int\text{d}^{2}\boldsymbol{k}\Omega^{z}\left(\boldsymbol{k}\right)$,
where the geometric properties of the bands rooted to the derivatives
of band wavefunctions, appear in the form of the Berry curvature \cite{Thouless82405,Chang951348,Xiao101959}.
Response coefficients second orders in $\boldsymbol{E}\left(t\right)$
appear as higher derivatives in band wavefunctions, such as Berry
curvature dipole, and serve as the basis for the helicity-dependent
rectification of photocurrent \cite{Moore10026805} and the nonlinear
Hall effect \cite{Deyo091917v1,Sodemann15216806} obtained under periodic
drivings. The all-order effects from the time-periodic driving fields
$\boldsymbol{E}\left(t\right)$ encoded in $\delta f\left(\boldsymbol{k}_{t},t\right)$
have been shown to result in non-perturbation intra-band processes
significant for the HHG signals in Weyl semimetals \cite{Dantas21L201105,Kovalev202451}.
However, as long as the external field $\boldsymbol{E}\left(t\right)$
is set to induce non-negligible occupations on bands not indexed by
$n$, then the transport process will not remain in the adiabatic
regime, which then renders Eq.(\ref{Jad-tdep0}) inapplicable. In
previous works, the above WPT (here called AD-WPT to be distinguished
from NADT-WPT later) has been extended to non-adiabatic regimes but
only for the steady states at $t\rightarrow\infty$ with $\boldsymbol{E}\left(t\rightarrow\infty\right)=\boldsymbol{E}$
being a time-independent constant \cite{Tu20045423,Tu20045004,Li21045012}.
Before we continue such non-adiabatic extension of the AD-WPT to the
transient regime with arbitrary time dependence of $\boldsymbol{E}\left(t\right)$,
we briefly inspect the potential relevance of this regime to real
materials.

\subsubsection{Breaking of the adiabatic condition with moderate electric fields
in the transients}

\begin{figure}[h] \includegraphics[width=10cm, height=7.5cm]{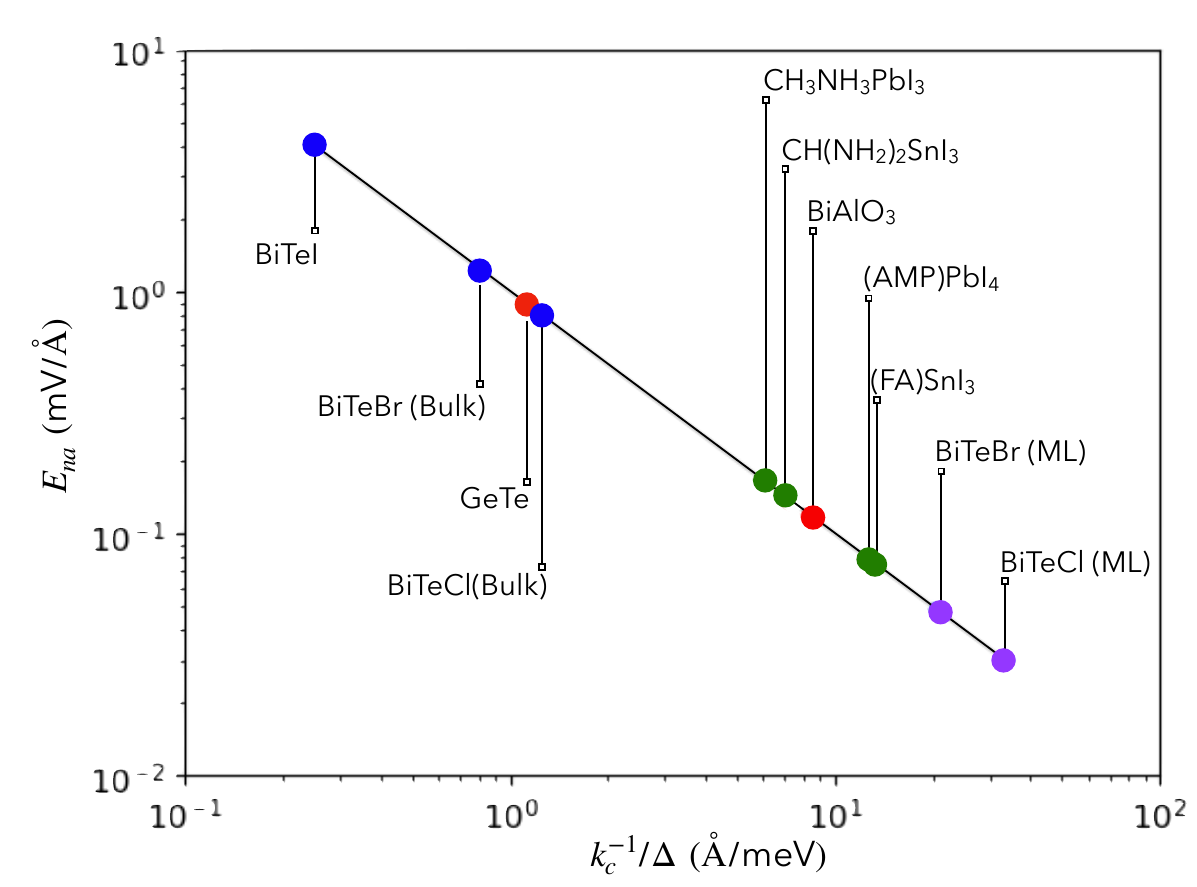} 
\caption{An illustrative list of materials with $E_{na}<10\text{mV}/\text{\AA}$, in which $E_{na}$ (shown as the straight line against the material-inherent parameter $k_{c}^{-1}/\Delta$ in log scale) is defined to be the required external field strength that reaches the value $r=100\%$ for the non-adiabatic measure (see the texts).  The enlisted Rashba/Dresselhaus materials include polar components (blue spots: $\mathrm{BiTeI}$,  $\mathrm{BiTeBr}$,  $\mathrm{BiTeCl}$ \cite{Eremeev12246802}), ferroelectrics (red spots: $\mathrm{GeTe}$ \cite{DiSante13509}, $\mathrm{BiAlO_{3}}$ \cite{daSilveira16245159}),  perovskites (green spots: $\mathrm{CH}_3\mathrm{NH}_3\mathrm{PbI}3$, $\mathrm{CH(NH_{2})_{2}SnI_{3}}$ \cite{Kepenekian1511557},  $\mathrm{(AMP)PbI_{4}}$ \cite{Wang20183}, $\mathrm{(FA)SnI_{3}}$ \cite{Stroppa145900},  and 2D Janus materials (purple spots: $\mathrm{BiTeBr}$,  $\mathrm{BiTeCl}$ \cite{Bafekry2115216}). The ML for purple spots stands for monolayer.  Here $\mathrm{CH(NH_{2})_{2}SnI_{3}}$ and $\mathrm{(FA)SnI_{3}}$ are Dresselhaus-dominated while others are Rashba-dominated.
The estimation is taken for their conduction-band spin splittings, whose geometric-connection effects in spin dynamics are potentially accessible by optical excitations.  
} 
\label{nonadmat} 
\end{figure}

The adequacy of the adiabatic condition underlying the use of AD-WPT
can be assessed by the non-adiabatic measure $r\equiv\left\vert r_{m,n}\right\vert $
given by Eq. (\ref{nonad-meas-def}) that appears in Eq. (\ref{n-ad-uni})
describing adiabatic wave-packet dynamics. Clearly, $r$ should be
small, e.g., $r<2\%$, for the adiabatic condition to be held. Noting
that $r\sim k_{c}^{-1}\left(eE_{ext}\right)/\Delta$, where $k_{c}^{-1}\sim\left\vert \left[\mathcal{\bar{\mathcal{R}}}_{\boldsymbol{k}}\right]_{m,n}\right\vert $
and $\Delta$ are respectively the magnitudes of the inter-band geometric
connection strength and the gap, the adiabatic condition can be intentionally
broken by tuning the external electric-field amplitude $E_{ext}$
in accordance with $k_{c}^{-1}/\Delta$ as an inherent material parameter.
We therefore call driving scenarios $\boldsymbol{E}\left(t\right)$
that keep the smallness of $r$ as the adiabatic drivings and otherwise
non-adiabatic drivings. A definite breaking of the adiabatic condition
given by $r=100\%$ corresponds to $E_{ext}\rightarrow E_{na}\equiv e^{-1}\Delta/k_{c}^{-1}$.
Note that an overwhelmingly large $E_{ext}$ can spoil the material
basis for hosting the targeted geometric properties of bands. This
is concomitant to two reasons for implementing non-adiabatic dynamics
particularly in the transient regime. First of all, exposing materials
to strong fields for long time can inflict radiation damages. It is
therefore more practical to study the non-perturbation effects of
$E_{ext}$ in the short-time transient regime \cite{McIver2038,Ito23696}.
Secondly, the sought-for intrinsic geometric effects of pure electronic
origin are also anticipated to be more manifested transiently before
relaxation and the dynamics of driven nuclei on longer time scales
come into play. With the above consideration, to enquire the feasibility
of reaching $r=100\%$ in real materials minding that $E_{na}\propto\Delta$,
we prescribe a limit $\Delta\le200$meV to survey the literature of
first-principle calculations to obtain $\Delta$ and $k_{c}^{-1}$.
While $\Delta$ can be easily found, $k_{c}^{-1}$ is usually unavailable.
We then surveyed only those that provide Rashba-Dresselhaus parameters
so as to estimate $E_{na}$ using the Rashba-Dresselhaus SOC Hamiltonian
to fix $k_{c}^{-1}$ \cite{FootnoteRSOCkc}. In Fig. \ref{nonadmat},
a number of materials with their $E_{na}$ values are listed along
the $E_{na}$-versus-$k_{c}^{-1}/\Delta$ relation in log scale with
$E_{na}<$10mV/$\text{\AA}$. The diversity of the enlisted materials'
categories (see captions of Fig. \ref{nonadmat}), in addition to
endorse the needs of extending the AD-WPT to NADT-WPT, also suggests
us to start with simple band-structure models characterised by well-defined
geometric properties to avoid complications arising from other properties,
e.g., polar/ferroelectric, etc. 

\subsection{Non-adiabatic transport of wave packets}

\label{WPT-basics-nadia}Knowing the feasibility of breaking the adiabatic
condition in real materials potentially of interests for band-geometric
properties, we now introduce the formulation of NADT-WPT.

\subsubsection{Formulation of ensemble dynamics of non-adiabatically driven wave
packets }

Keeping RTA without imposing the adiabatic condition on single-wave-packet
dynamics results in (see details in SM-Sec. \ref{SI-wpt-deriv-ensemble})
\begin{subequations}\label{incoh-rho-1}

\begin{align}
\varrho_{\boldsymbol{k}}\left(t\right)=\bar{\varrho}_{\boldsymbol{k}}\left(t\right)+\delta\varrho_{\boldsymbol{k}}\left(t\right),\label{incoh-rho-1sum}
\end{align}
where 
\begin{align}
\bar{\varrho}_{\boldsymbol{k}}\left(t\right)=\sum_{i}g_{i}^{0}\left(\boldsymbol{k}_{t}\right)\left\vert \mathfrak{u}_{i}\left(\boldsymbol{k}_{t}\right)\right\rangle \left\langle \mathfrak{u}_{i}\left(\boldsymbol{k}_{t}\right)\right\vert ,\label{incoh-rho-1bar}
\end{align}
and 
\begin{align}
\delta\varrho_{\boldsymbol{k}}\left(t\right)=-\int_{t_{0}}^{t}\text{d}t^{\prime}P\left(\boldsymbol{k}_{t},\boldsymbol{k}_{t^{\prime}}\right)\frac{\partial}{\partial t^{\prime}}\bar{\varrho}_{\boldsymbol{k}}^{U}\left(t,t^{\prime}\right),\label{incoh-rho-1del}
\end{align}
in which 
\begin{align}
\bar{\varrho}_{\boldsymbol{k}}^{U}\left(t,t^{\prime}\right)=U_{\boldsymbol{k}}\left(t,t^{\prime}\right)\bar{\varrho}_{\boldsymbol{k}}\left(t^{\prime}\right)U_{\boldsymbol{k}}\left(t^{\prime},t\right),\label{incoh-rho-1evm}
\end{align}
\end{subequations} with $U_{\boldsymbol{k}}\left(t^{\prime},t\right)=U_{\boldsymbol{k}}\left(t,t^{\prime}\right)^{\dagger}$
defined for $t\ge t^{\prime}$ (see $U_{\boldsymbol{k}}\left(t,t^{\prime}\right)$
defined in Eq. (\ref{unievol-1})). Here $\left\vert \mathfrak{u}_{i}\left(\boldsymbol{k}_{t}\right)\right\rangle $,
called the $i$th hybridised band, is obtained by solving $\bar{\mathcal{H}}\left(\boldsymbol{k}_{t},\boldsymbol{E}\left(t\right)\right)\left\vert \mathfrak{u}_{i}\left(\boldsymbol{k}_{t}\right)\right\rangle =\mathcal{E}_{i}\left(\boldsymbol{k}_{t}\right)\left\vert \mathfrak{u}_{i}\left(\boldsymbol{k}_{t}\right)\right\rangle $,
in which $\bar{\mathcal{H}}_{n,m}\left(\boldsymbol{k}_{t},\boldsymbol{E}\left(t\right)\right)=\delta_{n,m}\varepsilon_{n}\left(\boldsymbol{k}_{t}\right)-(1-\delta_{n,m})\left[\mathcal{\bar{\mathcal{R}}}_{\boldsymbol{k}_{t}}\right]_{n,m}\cdot\left(-e\right)\boldsymbol{E}\left(t\right)$
is known as the moving-frame Hamiltonian \cite{Tu20045423,Tu20045004}.
The energy of the $i$th hybridised band is $\mathcal{E}_{i}\left(\boldsymbol{k}_{t}\right)$
and $g_{i}^{0}\left(\boldsymbol{k}_{t}\right)=f_{FD}^{\mu,T}\left(\mathcal{E}_{i}\left(\boldsymbol{k}_{t}\right)\right)$.
Equation (\ref{incoh-rho-1}) formally describes general non-adiabatic
evolution of an ensemble of wave packets incorporating both effects
due to the intrinsic band structure (through $\left\vert \mathfrak{u}_{i}\left(\boldsymbol{k}_{t}\right)\right\rangle $
and $U_{\boldsymbol{k}}\left(t,t^{\prime}\right)$) and extrinsic
scatterings (through $P\left(\boldsymbol{k}_{t},\boldsymbol{k}_{t^{\prime}}\right)$
defined in Eq. (\ref{non-scatt-prob1})). By substituting Eq. (\ref{incoh-rho-1})
to Eq. (\ref{j-def0}), we find $\boldsymbol{j}\left(t\right)=\bar{\boldsymbol{j}}\left(t\right)+\delta\boldsymbol{j}\left(t\right)$
with $\bar{\boldsymbol{j}}\left(t\right)=-e\int\text{d}^{D}\boldsymbol{k}_{t}\text{Tr}\left[\boldsymbol{V}\left(\boldsymbol{k}_{t}\right)\bar{\varrho}_{\boldsymbol{k}}\left(t\right)\right]$
and $\delta\boldsymbol{j}\left(t\right)=-e\int\text{d}^{D}\boldsymbol{k}_{t}\text{Tr}\left[\boldsymbol{V}\left(\boldsymbol{k}_{t}\right)\delta\varrho_{\boldsymbol{k}}\left(t\right)\right]$
reproducing the above mentioned Hall current and the longitudinal
current respectively in the adiabatic limit (see details in SM-Secs.
\ref{SI-wpt-deriv-ensemble} and \ref{nlqh}, including reproduction
of Eq. (\ref{Jad-tdep0}) and the nonlinear Hall effect). 

\subsubsection{Differences between AD-WPT and NADT-WPT in interpretations of geometric
effects }

In a general WPT we picture that an electron wave packet possesses
both the CM and band-pseudospin degrees of freedom (DOF). The geometric
effects in wave-packet dynamics arise from the coupling between the
band-pseudospin DOF and the CM DOF. The CM DOF is specified by $\left(\boldsymbol{x}_{t},\boldsymbol{k}_{t}\right)$
with $\dot{\boldsymbol{k}}_{t}$ directly driven by the external field,
Eq. (\ref{driving-E1}). This coupling matrix element between the
CM and the band-pseudospin DOF is explicitly given by,
\begin{equation}
V_{n,m}^{\text{CM-BPSpin}}\left(\dot{\boldsymbol{k}}_{t},\boldsymbol{k}_{t}\right)=-\hbar\dot{\boldsymbol{k}}_{t}\cdot\left[\mathcal{\bar{\mathcal{R}}}_{\boldsymbol{k}_{t}}\right]_{n,m},\label{CM-spin-coup1}
\end{equation}
for $n\ne m$ that enables the rotation of the band-pseudospin as
transitions between bands $n$ and $m$ via the CM motion $\dot{\boldsymbol{k}}_{t}$.
This directly follows from that the band-pseudospin DOF specified
by the amplitude $\boldsymbol{\eta}\left(t\right)$ is subject to
$i\hbar\dot{\boldsymbol{\eta}}\left(t\right)=\bar{\mathcal{H}}\left(\boldsymbol{k}_{t},\boldsymbol{E}\left(t\right)\right)\boldsymbol{\eta}\left(t\right)$
(see SM-Sec.\ref{SI-wpt-deriv-single}) in which $\bar{\mathcal{H}}_{n,m}\left(\boldsymbol{k}_{t},\boldsymbol{E}\left(t\right)\right)=V_{n,m}^{\text{CM-BPSpin}}\left(\dot{\boldsymbol{k}}_{t},\boldsymbol{k}_{t}\right)$
for $n\ne m$. The consequence of interplays between the drivings
(through $\boldsymbol{k}_{t}$ of the CM DOF) and the geometric properties
of bands (through the band pseudospin) is revealed in $\dot{\boldsymbol{x}}_{t}$
(see Eq. (\ref{xdot-nra})), the wave-packet velocity. In NADT-WPT,
this CM-band-pseudospin coupling can non-perturbatively drive considerable
rotations of the band-pseudospin DOF. In AD-WPT, the smallness of
the non-adiabatic measure (see Eq. (\ref{nonad-meas-def})) restricts
the band-pseudospin state $\left\vert \phi_{n,\boldsymbol{k}}\left(t,t_{0}\right)\right\rangle $
to $\left\vert \phi_{n,\boldsymbol{k}}^{ad}\left(t\right)\right\rangle $
(see Eq. (\ref{n-ad-uni})). The existence of couplings among multiple
bands (the fundamental reason of taking the view of band pseudospin)
just modifies one band's velocity $\dot{\boldsymbol{x}}_{t}^{n}$
from having only its own group velocity $\boldsymbol{v}_{n}^{b}\left(\boldsymbol{k}_{t}\right)=\partial\varepsilon_{n}\left(\boldsymbol{k}_{t}\right)/\partial\boldsymbol{k}_{t}$
to also have a Berry-curvature-governed anomalous velocity $\boldsymbol{v}_{n}^{r}\left(\boldsymbol{k}_{t},\boldsymbol{E}\left(t\right)\right)=-\left(-e/\hbar\right)\boldsymbol{E}\left(t\right)\times\boldsymbol{\Omega}_{n}\left(\boldsymbol{k}_{t}\right)$
as the dominant geometric effects (see Sec. \ref{WPT-basics-adia}).
In contrast with the availability of NADT-WPT, the multiplicity of
bands, coupled by collaboration between non-adiabatic drivings and
geometric connections, is not to be reduced to modification of one
band's properties. The interpretation of geometric effects relies
on a more complete picture of the band-pseudospin dynamics whose particular
transient non-adiabatic consequences are discussed below.

\subsection{Geometric effects encoded in transient non-adiabatic dynamics of
macroscopic observables}

\label{WPT-basics-Uniqnadia}The main interests of studying transient
currents lie in the ultrafast responses to short laser pulses. A pulsing
driving field $\boldsymbol{E}\left(t\right)$ is essentially nonzero
only for a finite time window $t_{0}\le t\le t_{\text{off}}$ and
for time $t>t_{\text{off}}$ the electric field is switched off, 
\begin{equation}
\boldsymbol{E}\left(t>t_{\text{off}}\right)=0.\label{pulse-Eft}
\end{equation}
It is therefore natural to separate the in-pulse ($t_{0}\le t\le t_{\text{off}}$)
and the post-pulse ($t>t_{\text{off}}$) time regions. Derived from
Eq. (\ref{incoh-rho-1}), the geometric properties of the bands with
non-perturbation effects from the transient non-adiabatic drivings
can be manifested for macroscopic observables through (I): the post-pulse
current rate with geometric effects embedded in the quantum acceleration
shaped by transient in-pulse non-adiabatic drivings, (II): intrinsic
current-induced spin polarisation and (III): spin-mediated part of
the photocurrents with intrinsic spin coherence. Below we first explain
these abstract points respectively, leaving concrete examples to Sec.
\ref{sec-examples}.

\subsubsection{Reading off geometric effects after switching off the laser pulses}

\label{WPT-basics-QnAccHistory}

The transient characters of the time-dependent current can be pronounced
in its rate, namely, $\dot{\boldsymbol{j}}\left(t\right):=\text{d}\boldsymbol{j}\left(t\right)/\text{d}t$.
In the intrinsic limit $\varrho_{\boldsymbol{k}}\left(t\right)\rightarrow\varrho_{\boldsymbol{k}}^{id}\left(t\right)$
(see Eq. (\ref{ideal-dmtx})), the post-pulse current rate reads $\dot{\boldsymbol{j}}\left(t>t_{\text{off}}\right)\rightarrow\dot{\boldsymbol{j}}^{geo}\left(t\right)$
where \begin{subequations}\label{jdt-qcc}
\begin{equation}
\dot{\boldsymbol{j}}^{geo}\left(t\right)=-e\int\text{d}^{D}\boldsymbol{k}\text{Tr}\left(\vec{\mathbb{A}}\left(\boldsymbol{k}_{t}\right)\varrho_{\boldsymbol{k}}\left(t\right)\right),\label{gcomp-jdt-free}
\end{equation}
with 
\begin{equation}
\vec{\mathbb{A}}\left(\boldsymbol{k}_{t}\right):=-\frac{i}{\hbar}\left[\boldsymbol{V}\left(\boldsymbol{k}_{t}\right),\mathcal{H}\left(\boldsymbol{k}_{t}\right)\right].\label{q-acc-opt-def}
\end{equation}
\end{subequations}See derivation details in SM-Sec. \ref{SI-wpt-deriv-ppcrntrt}.
Here $\vec{\mathbb{A}}\left(\boldsymbol{k}_{t}\right)$ is called
the quantum acceleration since in the Heisenberg picture the commutator
of an observable (here the velocity) with the Hamiltonian gives its
time derivative (henceforth the acceleration). The geometric content
of $\dot{\boldsymbol{j}}^{geo}\left(t\right)$ (as indicated by the
superscript $geo$) is more revealed by writing the quantum acceleration
expectation value of a wave packet of with CM momentum $\boldsymbol{k}_{t}$
in the moving basis $\left\{ \left\vert u_{n}\left(\boldsymbol{k}_{t}\right)\right\rangle \right\} _{n}$,
namely, 
\begin{equation}
\text{Tr}\left(\vec{\mathbb{A}}\left(\boldsymbol{k}_{t}\right)\varrho_{\boldsymbol{k}}\left(t\right)\right)=\frac{1}{\hbar^{2}}\sum_{n}\varepsilon_{n}\left(\boldsymbol{k}_{t}\right)\sum_{m\ne n}\left(\varepsilon_{m}-\varepsilon_{n}\right)\left(\boldsymbol{k}_{t}\right)\left[\mathcal{\mathcal{R}}_{\boldsymbol{k}_{t}}\right]_{n,m}\left\langle u_{m}\left(\boldsymbol{k}_{t}\right)\right\vert \varrho_{\boldsymbol{k}}\left(t\right)\left\vert u_{n}\left(\boldsymbol{k}_{t}\right)\right\rangle .\label{jdt-qcc-bndrep}
\end{equation}
The quantum acceleration at $\boldsymbol{k}_{t}$ thus is nonzero
only when the band pseudospin has developed nonzero inter-band coherence
$\left.\left\langle u_{m}\left(\boldsymbol{k}_{t}\right)\right\vert \varrho_{\boldsymbol{k}}\left(t\right)\left\vert u_{n}\left(\boldsymbol{k}_{t}\right)\right\rangle \right\vert _{m\ne n}\ne0$
for which nonzero geometric connection $\left.\left[\mathcal{\mathcal{R}}_{\boldsymbol{k}_{t}}\right]_{n,m}\right\vert _{m\ne n}\ne0$
is indispensable. Since $\dot{\boldsymbol{j}}\left(t>t_{\text{off}}\right)=\dot{\boldsymbol{j}}^{geo}\left(t\right)$,
the main message conveyed by Eqs. (\ref{jdt-qcc}) and (\ref{jdt-qcc-bndrep})
is that the current rate $\dot{\boldsymbol{j}}\left(t\right)$ in
the post-pulse time region $t>t_{\text{off}}$ is a macroscopic manifestation
of the microscopic geometric effect embedded in the quantum acceleration.
Furthermore, such manifestation through $\dot{\boldsymbol{j}}\left(t>t_{\text{off}}\right)$
is only possible for non-adiabatic drivings during $t_{0}\le t\le t_{\text{off}}$
but not adiabatic drivings. With extrinsic scatterings, the same conclusion
about the exclusivity to non-adiabatic drivings is maintained (see
SM-Sec. \ref{SI-wpt-deriv-ppcrntrt-extrinsic}). A nonzero current
after switching off the laser pulse implies that (i): interplays between
drivings and band geometries are imprinted to the post-pulse dynamics
of $\dot{\boldsymbol{j}}\left(t>t_{\text{off}}\right)$ and (ii):
$\dot{\boldsymbol{j}}\left(t>t_{\text{off}}\right)$ carries with
it the history of the past non-adiabatic drivings (see explanations
in SM-Secs. \ref{SI-wpt-deriv-ppcrntrt-postpulse}). 

The above (i) and (ii) have immediate implications for transient steering
of electron dynamics using laser pulses in coordination with the geometric
properties of the bands. It is impossible to affect post-pulse dynamics
by in-pulse adiabatic drivings since the current ceases right after
the pulse. In contrast, non-adiabatic drivings during the in-pulse
time region determines post-pulse current rate through the geometric
properties of the bands embedded in the quantum acceleration as Eq.
(\ref{jdt-qcc-bndrep}) (see demonstrations in Sec. \ref{sec-examples}).
Note that although non-adiabatic geometric effects have been studied
based on Landau-Zener tunnelling scenarios \cite{Kitamura2063,Takayoshi21075},
the main focus there is the asymptotic limit, not the transient regime
focused here. The essential Landau-Zener tunneling probability formula
can be reached via the Markov approximation, namely, neglecting the
history dependence \cite{Glasbrenner23104001}. Here we want to explore
the geometric effects underlying the history dependence induced by
non-adiabatic drivings. 

\subsubsection{Intrinsic current-induced spin polarisation (CISP) }

\label{WPT-basics-spntxt} We now turn our attention to geometric
properties of bands with spin splitting induced by SOC, or briefly,
SOC-geometric effects. We take the widely applied convention to single
out the spin DOF with a two-band minimal description of SOC effects
\cite{Johansson16195440,Saberi-Pouya17075411,Tao21113001}, namely,
\begin{equation}
\mathcal{H}\left(\boldsymbol{k}\right)=\mathcal{T}_{0}\left(\boldsymbol{k}\right)+\mathcal{H}_{so}\left(\boldsymbol{k}\right),\label{SOC-H-gDef}
\end{equation}
where $\mathcal{T}_{0}\left(\boldsymbol{k}\right)$ is the kinetic
energy and 
\begin{equation}
\mathcal{H}_{so}\left(\boldsymbol{k}\right)=\boldsymbol{\Lambda}_{so}\left(\boldsymbol{k}\right)\cdot\boldsymbol{\sigma},\label{ani-SO-H}
\end{equation}
is the SOC in which the spin-orbit field $\boldsymbol{\Lambda}_{so}\left(\boldsymbol{k}\right)$
is a real-valued 3D vector specified by the type of SOC under consideration
and $\boldsymbol{\sigma}$ is the three-component vector of Pauli
matrices. It is customary to define the spin texture, namely, 
\begin{equation}
\left\langle \boldsymbol{\sigma}\right\rangle _{n}^{0}\left(\boldsymbol{k}\right):=\left\langle u_{n}\left(\boldsymbol{k}\right)\right\vert \boldsymbol{\sigma}\left\vert u_{n}\left(\boldsymbol{k}\right)\right\rangle ,\label{spintxt-native-def}
\end{equation}
which is related to inter-spin geometric connections $\left[\mathcal{\mathcal{R}}_{\boldsymbol{k}}\right]_{n,\bar{n}}$
where $\bar{n}=\mp$ denotes the spin opposite of $n=\pm$ by 
\begin{equation}
\frac{\partial\left\langle \boldsymbol{\sigma}\right\rangle _{n}^{0}\left(\boldsymbol{k}\right)}{\partial k_{\beta}}=2\text{Im}\left[\left\langle u_{n}\left(\boldsymbol{k}\right)\right\vert \boldsymbol{\sigma}\left\vert u_{\bar{n}}\left(\boldsymbol{k}\right)\right\rangle \left[\mathcal{\mathcal{R}}_{k_{\beta}}\right]_{n,\bar{n}}\right],\label{sptxt-geoct-r1}
\end{equation}
where $k_{\beta}$ is the $\beta$th spatial component of $\boldsymbol{k}$.

Henceforth, in the absence of any external drivings the spin texture
$\left\langle \boldsymbol{\sigma}\right\rangle _{n}^{0}\left(\boldsymbol{k}\right)$
can visualise the geometric properties of bands. In the presence of
transient non-adiabatic drivings, this picture of spin texture can
be extended to the time domain by defining 
\begin{equation}
\left\langle \boldsymbol{\sigma}\right\rangle \left(\boldsymbol{k},t\right):=\text{Tr}\left(\boldsymbol{\sigma}\varrho_{\boldsymbol{k}}\left(t\right)\right).\label{spintxt-tmdep-def}
\end{equation}
We call this non-equilibrium spin texture. For $t\le t_{0}$, Eq.
(\ref{spintxt-tmdep-def}) describes the equilibrium spin texture,
namely, $\left\langle \boldsymbol{\sigma}\right\rangle \left(\boldsymbol{k},t\le t_{0}\right)=\text{Tr}\left(\boldsymbol{\sigma}\varrho_{\boldsymbol{k}}^{eq}\right)=\sum_{n}f_{FD}^{\mu,T}\left(\varepsilon_{n}\left(\boldsymbol{k}\right)\right)\left\langle \boldsymbol{\sigma}\right\rangle _{n}^{0}\left(\boldsymbol{k}\right)$.
In the intrinsic limit, Eq. (\ref{ideal-dmtx}), the non-equilibrium
spin texture becomes $\left\langle \boldsymbol{\sigma}\right\rangle \left(\boldsymbol{k},t\right)=\sum_{n}f_{FD}^{\mu,T}\left(\varepsilon_{n}\left(\boldsymbol{k}\right)\right)\left\langle \boldsymbol{\sigma}\right\rangle _{n}\left(\boldsymbol{k},t\right)$
where 
\begin{equation}
\left\langle \boldsymbol{\sigma}\right\rangle _{n}\left(\boldsymbol{k},t\right):=\left\langle \phi_{n,\boldsymbol{k}}\left(t,t_{0}\right)\right\vert \boldsymbol{\sigma}\left\vert \phi_{n,\boldsymbol{k}}\left(t,t_{0}\right)\right\rangle ,\label{spintxt-drivn-def}
\end{equation}
(see definition for $\left\vert \phi_{n,\boldsymbol{k}}\left(t,t_{0}\right)\right\rangle $
in Eq. (\ref{intrinsic-stat_nk})). We then see that $\left\langle \boldsymbol{\sigma}\right\rangle _{n}\left(\boldsymbol{k},t\right)$,
as an intrinsic quantity, can directly be compared to $\left\langle \boldsymbol{\sigma}\right\rangle _{n}^{0}\left(\boldsymbol{k}\right)$
because $\left\langle \boldsymbol{\sigma}\right\rangle _{n}\left(\boldsymbol{k},t\le t_{0}\right)=\left\langle \boldsymbol{\sigma}\right\rangle _{n}^{0}\left(\boldsymbol{k}\right)$.
So $\left\langle \boldsymbol{\sigma}\right\rangle _{n}\left(\boldsymbol{k},t\right)$
is also a spin texture that arises in time due to the external driving.
For disambiguation, $\left\langle \boldsymbol{\sigma}\right\rangle _{n}^{0}\left(\boldsymbol{k}\right)$
and $\left\langle \boldsymbol{\sigma}\right\rangle _{n}\left(\boldsymbol{k},t\right)$
are called native and driven spin textures respectively. As the CM
of the wave packet is driven from $\boldsymbol{k}$ to $\boldsymbol{k}_{t}$,
the non-vanishing variation of the native spin texture $\partial\left\langle \boldsymbol{\sigma}\right\rangle _{n}^{0}\left(\boldsymbol{k}_{t}\right)/\partial\boldsymbol{k}_{t}\ne0\Rightarrow\left[\mathcal{\mathcal{R}}_{\boldsymbol{k}_{t}}\right]_{n,\bar{n}}\ne0$
(see Eq. (\ref{sptxt-geoct-r1})) along the CM trajectory in BZ provides
the needed CM-spin couplings to induce intrinsic spin coherence $\left\langle \boldsymbol{\sigma}\right\rangle _{n}\left(\boldsymbol{k},t>t_{0}\right)\ne\left\langle \boldsymbol{\sigma}\right\rangle _{n}^{0}\left(\boldsymbol{k}_{t}\right)$
as the microscopic manifestation of the SOC-geometric effects. Such
effects are thus not expected for persistent spin texture \cite{Tao21113001}
but could be ubiquitous for Rashba-Dresselhaus spin textures (see
explanations in SM-Sec. \ref{SI-spin-basics-txtcoh}). 

To macroscopically manifest the above microscopic SOC-geometric effects,
the non-equilibrium spin texture $\left\langle \boldsymbol{\sigma}\right\rangle \left(\boldsymbol{k},t\right)$
is directly linked to the spin polarisation $\boldsymbol{S}\left(t\right)$
as a macroscopic observable, namely, 
\begin{equation}
\boldsymbol{S}\left(t\right):=\frac{\hbar}{2}\int\text{d}^{D}\boldsymbol{k}\left\langle \boldsymbol{\sigma}\right\rangle \left(\boldsymbol{k},t\right).\label{St-neq-0}
\end{equation}
Therefore, any intrinsic geometric effects that are possibly contained
in the non-equilibrium spin texture also underlies the spin polarisation
by the definition of Eq. (\ref{St-neq-0}). For non-magnetic materials
endowed with SOC, the time-reversal symmetry (TRS) is respected. Henceforth,
before the electric field driving is turned on, the spin polarisation
is always zero, namely,
\begin{equation}
\boldsymbol{S}_{\text{eq}}=\boldsymbol{S}\left(t\le t_{0}\right)=0.\label{zeroSP-TRSsptxt-1}
\end{equation}
The possibility of applying an electric field to induce a spin polarisation
via the SOC is known as the current-induced spin polarisation (CISP).
Various circumstances of the CISP have been intensively studied but
the focus is largely on the steady states in the adiabatic regime
and not much interest from the geometric perspective is shown \cite{Edelstein90233,Aronov91537,Johansson16195440,Li2016749,Tao21085438}.
To explore peculiarities of transient non-adiabatic aspects (as our
main theme here) of the CISP, we first remind that the adiabatically
induced CISP linear in the electric field does not retain properties
of the intrinsic spin coherence in the resulting macroscopic spin
polarisation (see details in SM-Sec. \ref{SI-spin-basics-adiaCISP}).
This leaves the possibilities of manifesting the intrinsic spin coherence
as SOC-geometric effects to non-adiabatically induced transient CISP.

\subsubsection{Spin-mediated part of the photocurrents with intrinsic spin coherence.}

\label{sec-examples-SOC-isc} The general separation of kinetic energy
and the SOC in Eq. (\ref{SOC-H-gDef}) in conjunction with Eq. (\ref{vopt-def})
leads to a natural decomposition of the velocity matrix as \begin{subequations}\label{Vdecompose}
\begin{equation}
\boldsymbol{V}\left(\boldsymbol{k}_{t}\right)=\boldsymbol{V}^{K}\left(\boldsymbol{k}_{t}\right)+\boldsymbol{V}^{so}\left(\boldsymbol{k}_{t}\right),\label{Vdecompose-tot}
\end{equation}
where
\begin{equation}
\boldsymbol{V}^{K}\left(\boldsymbol{k}_{t}\right):=\frac{\partial\mathcal{T}_{0}\left(\boldsymbol{k}_{t}\right)}{\partial\hbar\boldsymbol{k}_{t}},\label{Vdecompose-K}
\end{equation}
comes from the kinetic energy and
\begin{equation}
\boldsymbol{V}^{so}\left(\boldsymbol{k}_{t}\right):=\frac{\partial\mathcal{H}_{so}\left(\boldsymbol{k}_{t}\right)}{\partial\hbar\boldsymbol{k}_{t}}.\label{Vdecompose-so}
\end{equation}
\end{subequations} is due to the SOC. Substituting Eq. (\ref{Vdecompose})
to Eq. (\ref{j-def0}) further decomposes the electronic current to
\begin{subequations}\label{JtoKnL}
\begin{equation}
\boldsymbol{j}\left(t\right)=\boldsymbol{j}_{K}\left(t\right)+\boldsymbol{j}_{so}\left(t\right),\label{JtoKnL-def0}
\end{equation}
where
\begin{equation}
\boldsymbol{j}_{K}\left(t\right)=\left(-e\right)\int\text{d}^{D}\boldsymbol{k}\text{Tr}\left\{ \boldsymbol{V}^{K}\left(\boldsymbol{k}_{t}\right)\varrho_{\boldsymbol{k}}\left(t\right)\right\} ,\label{JtoKnL-defK}
\end{equation}
and $\boldsymbol{j}_{so}\left(t\right)=\left(-e\right)\int\text{d}^{D}\boldsymbol{k}\text{Tr}\left\{ \boldsymbol{V}^{so}\left(\boldsymbol{k}_{t}\right)\varrho_{\boldsymbol{k}}\left(t\right)\right\} $
which by Eq. (\ref{ani-SO-H}) can be alternatively written as
\begin{equation}
\boldsymbol{j}_{so}\left(t\right)=\left(-e\right)\sum_{\beta\in\left\{ x,y,z\right\} }\int\text{d}^{D}\boldsymbol{k}\frac{\partial\Lambda_{so,\beta}\left(\boldsymbol{k}_{t}\right)}{\partial\hbar\boldsymbol{k}_{t}}\left\langle \mathcal{\sigma}_{\beta}\right\rangle \left(\boldsymbol{k},t\right),\label{JtoKnL-defL}
\end{equation}
\end{subequations} where $\left\langle \mathcal{\sigma}_{\beta}\right\rangle \left(\boldsymbol{k},t\right)$
and $\Lambda_{so,\beta}\left(\boldsymbol{k}_{t}\right)$ are the $\beta$th
components of $\left\langle \boldsymbol{\sigma}\right\rangle \left(\boldsymbol{k},t\right)$
(defined in Eq. (\ref{spintxt-tmdep-def})) and $\boldsymbol{\Lambda}_{so}\left(\boldsymbol{k}_{t}\right)$
respectively. Since . We call $\boldsymbol{j}_{so}\left(t\right)$
the spin-mediated part of the photocurrent for obvious reasons. The
other part of the current, $\boldsymbol{j}_{K}\left(t\right)$, that
exists without any consideration of spin is due to bonding among neighbouring
unit cells. Here we call it the bond-mediated part. Since the intrinsic
part of the non-equilibrium spin texture $\left\langle \boldsymbol{\sigma}\right\rangle \left(\boldsymbol{k},t\right)$
contains the SOC-geometric effects, the spin-mediated current $\boldsymbol{j}_{so}\left(t\right)$
as a macroscopic observable can potentially manifest such effects. 

\section{Examples}

\label{sec-examples} Here in this section, we illustrate the above
abstract points (I-III) of Sec. \ref{WPT-basics-Uniqnadia} with concrete
examples of band structures. The examples addressed here are all with
zero Berry curvatures so as to focus on the manifestation of geometric
properties through the transient non-adiabatic dynamics. 

\subsection{Transient non-adiabatic manifestation of geometric effects in SSH
system}

\label{sec-examples-SSH} 

Hereby with SSH system we demonstrate the point (I) outlined in Sec.
\ref{WPT-basics-QnAccHistory}. The SSH system consists of bonded
sublattices arranged periodically in one dimension with $D=1$ so
$\boldsymbol{k}\rightarrow k$, $\boldsymbol{\mathcal{K}}_{t}\rightarrow\mathcal{K}_{t}$,
and $\dot{\boldsymbol{j}}\left(t\right)\rightarrow\dot{j}\left(t\right)$.
By the simplicity of 1D BZ, the Berry curvature is zero \cite{footnoteZeroBCat1D}.
The SSH Hamiltonian $\mathcal{H}_{SSH}\left(k\right)$ can be seen
as a case of Eq. (\ref{SOC-H-gDef}) specified by $\mathcal{T}_{0}\left(\boldsymbol{k}\right)=0$
and $\boldsymbol{\Lambda}_{so}\left(\boldsymbol{k}\right)$ replaced
by $\boldsymbol{\Lambda}_{SSH}\left(k\right)=\left(b_{1}+b_{2}\cos\left(ka\right),b_{2}\sin\left(ka\right),0\right)$
in which $b_{1}$ and $b_{2}$ are bonding strengths and $a$ is the
lattice constant \cite{SU19791698}. The relations among spin-geometric
effects, intrinsic spin coherence, native and driven spin textures,
enunciated in Sec. \ref{WPT-basics-spntxt} supplemented by SM-Secs.
\ref{SI-spin-basics-txtcoh} and \ref{SI-spin-basics-ppdys} equally
apply here via the substitution of spin by sublattice pseudospin. 

\begin{figure}[h] \includegraphics[width=9cm, height=7.5cm]{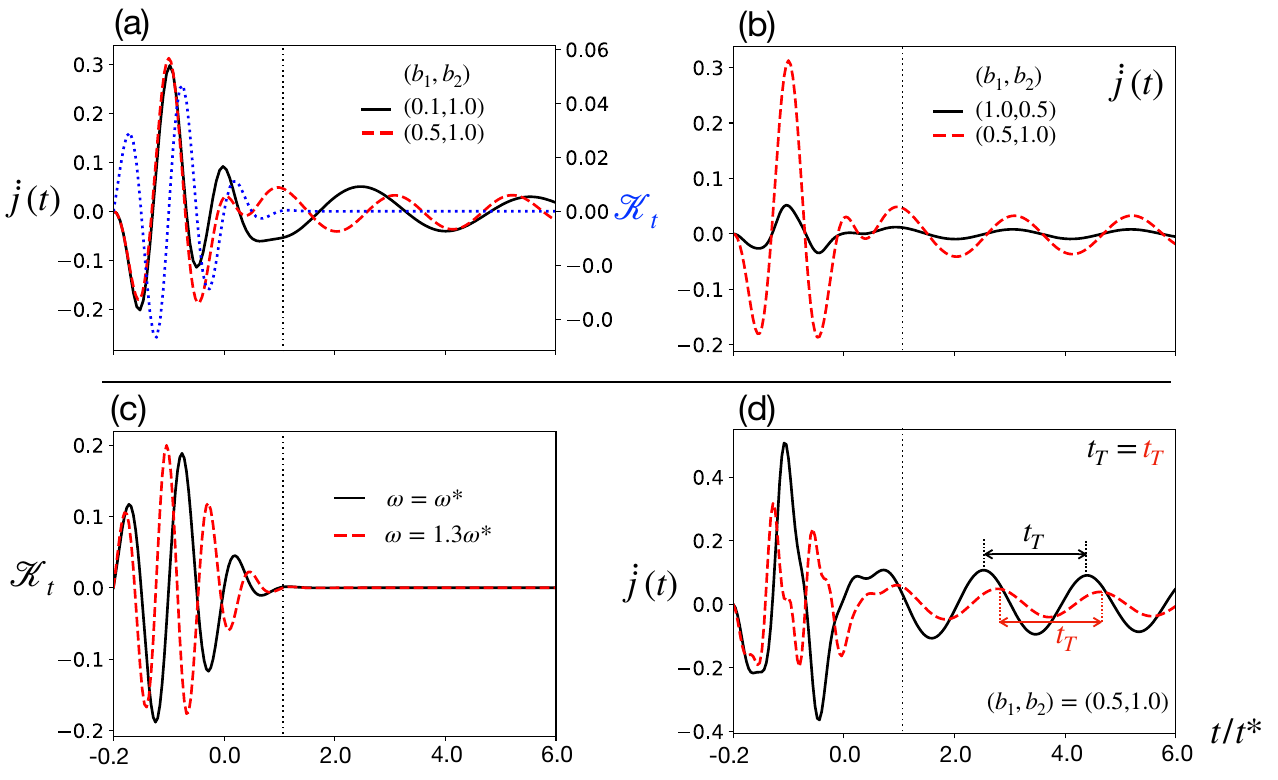} 
\caption{Imprints of in-pulse  interplays between drivings and band-geometry on post-pulse current rates  ((a), (b)) and driving-history dependencies  ((c), (d)).   The vertical dotted line in each plot marks the time the pulse ceases $t_{\text{off}}$. The blue dotted line in (a),  whose value goes by the right vertical axis, is the waveform $\mathcal{K}_{t}$ of the pulse  (see more details in \cite{FootnoteSSHNumSet}) with the amplitude $\mathcal{K}^{*}=0.05\pi/a$ and central frequency $\omega=\omega^{*}$.  (a): Different settings of  bonds $(b_1,b_2)$ (see legends) under the same topology $b_1<b_2$ show distinct post-pulse oscillation phases and amplitudes. (b): Distinct topologies (see legends for both $b_1<b_2$ and $b_1>b_2$) with the same dispersion show the same post-pulse oscillation periods but with sizably different amplitudes.  In (c),  two different pulses differing only in central frequencies (see legends) otherwise identical (with $\mathcal{K}^{*}=0.2\pi/a$) are portrayed.  Correspondingly in (d),  two different  current rates (with the same band structure specified by $(b_1,b_2)$ in the legend) are plotted.  During in-pulse time region, different driving frequencies naturally result in different transient oscillation frequencies of the current rates. No difference in oscillation frequencies show up  after the pulse is off (see marks of the equal periods).   However, the difference in driving histories shows up  as different phases and amplitudes of post-pulse oscillations. 
} 
\label{SSH-jdot} 
\end{figure}

Here the native sublattice-pseudospin texture determined by $\boldsymbol{\Lambda}_{SSH}\left(k\right)$
dictates that $\left\langle \boldsymbol{\sigma}\right\rangle _{n}^{0}\left(k\right)\cdot\hat{\boldsymbol{z}}=0$.
The quantum acceleration defined by Eq. (\ref{q-acc-opt-def}) gives
$\mathbb{A}\left(k_{t}\right)=-2a\hbar^{-2}\left(b_{2}^{2}+b_{1}b_{2}\cos k_{t}a\right)\sigma_{z}$.
So Eq. (\ref{jdt-qcc}) and $\mathbb{A}\left(k_{t}\right)\propto\sigma_{z}$
for the SSH system together signify that the non-vanishing of the
post-pulse macroscopic current rate $\dot{j}\left(t>t_{\text{off}}\right)=\dot{j}^{geo}\left(t\right)$
testifies microscopic intrinsic pseudospin coherences $\left\langle \sigma_{z}\right\rangle _{n}\left(k,t\right)\ne\left\langle \boldsymbol{\sigma}\right\rangle _{n}^{0}\left(k_{t}\right)\cdot\hat{\boldsymbol{z}}=0$.
We demonstrate these effects explicitly in Fig. \ref{SSH-jdot} with
laser pulses $E\left(t\right)$ subject to Eq. (\ref{pulse-Eft})
(see \cite{FootnoteSSHNumSet} for pulse setting and units). As expected
by the sub-point (i) of (I), different geometric properties (here
set up by choices of $b_{1}$ and $b_{2}$) of the bands can leave
their imprints to the post-pulse current rates. This is exemplified
in Fig. \ref{SSH-jdot}(a) where the two curves of $\dot{j}\left(t>t_{\text{off}}\right)$
differ in all three attributes of an oscillating current rates, namely,
amplitudes, frequencies and phases. By setting the two compared band
structures to share the same gaps but different geometric properties,
encoded by $b_{1}<b_{2}$ and $b_{1}>b_{2}$ also as different topologies,
in Fig. \ref{SSH-jdot}(b), the two curves of $\dot{j}\left(t>t_{\text{off}}\right)$
appear to have the same frequency but considerably different amplitudes.
To illustrates the sub-point (ii) of (I), driving-history dependence,
two pulses of the same duration but different central frequencies
are depicted in Fig. \ref{SSH-jdot}(c) and the two corresponding
current rates are shown in Fig. \ref{SSH-jdot}(d). The in-pulse transient
oscillations displayed by the two curves of $\dot{j}\left(t<t_{\text{off}}\right)$
driven by these two pulses obviously differ in oscillation frequencies.
Nevertheless, for the post-pulse oscillations $t>t_{\text{off}}$,
they differ more apparently in the oscillation phases/amplitudes but
with equal periods, attesting driving-history dependence as geometric
effects explained formally in more details in SM-Sec. \ref{SI-spin-basics-ppdys}.

\subsection{SOC-geometric effects in ultrafast opto-spintronics with Rashba SOC }

\label{sec-examples-Rashba} 

Hereby we demonstrate the points (II) and (III) abstracted in Secs.
\ref{WPT-basics-spntxt} and \ref{sec-examples-SOC-isc} with Rashba
SOC. The Rashba spin-orbit field is defined in $D=2$. This endows
the system with the possibility of in-plane anisotropy \cite{Simon10235438,Popovi15035135,Hu18235404,Sino216608}
whose interesting physical effects have been witnessed, such as the
fermi surface fragmentation \cite{Johansson16195440} and modulation
of optical conductivities under continuous irradiation \cite{Saberi-Pouya17075411}.
We therefore include the anisotropy in our following investigations. 

As has been established in Refs. \cite{Popovi15035135,Saberi-Pouya17075411},
the Rashba Hamiltonian with effective-mass anisotropy has its kinetic
energy part given by
\begin{equation}
\mathcal{T}_{0}\left(\boldsymbol{k}\right)=\frac{\hbar^{2}k_{x}^{2}}{2m_{x}}+\frac{\hbar^{2}k_{y}^{2}}{2m_{y}},\label{ani-kinetic-1}
\end{equation}
in which $m_{x/y}$ is the effective mass along the $x/y$-direction
and its SOC part specified by the spin-orbit field given by
\begin{equation}
\boldsymbol{\Lambda}_{so}\left(\boldsymbol{k}\right)=\alpha_{R}\left[\mathcal{X}_{K}\boldsymbol{k}\right]\times\hat{\boldsymbol{z}},\label{sorb-fld-1}
\end{equation}
in which $\alpha_{R}$ is the Rashba coefficient and $\mathcal{X}_{K}=\left(\begin{array}{cc}
r_{X} & 0\\
0 & r_{Y}
\end{array}\right)$ anisotropically rescales the components $k_{x}$ and $k_{y}$ of
the Bloch momentum $\boldsymbol{k}=k_{x}\hat{\boldsymbol{x}}+k_{y}\hat{\boldsymbol{y}}$
to $\mathcal{X}_{K}\boldsymbol{k}=r_{X}k_{x}\hat{\boldsymbol{x}}+r_{Y}k_{y}\hat{\boldsymbol{y}}$
with $r_{X}=\sqrt{m_{y}/m_{x}}$ and $r_{Y}=\sqrt{m_{x}/m_{y}}$.
To simplify the description of the anisotropy, we define
\begin{equation}
r_{A}:=m_{y}/m_{x}.\label{mass-ani-1}
\end{equation}
By setting $r_{A}=1$, Eq. (\ref{sorb-fld-1}) reduces to the familiar
isotropic Rashba SOC with $\boldsymbol{\Lambda}_{so}\left(\boldsymbol{k}\right)=\alpha_{R}\boldsymbol{k}\times\hat{\boldsymbol{z}}$.
When the anisotropy is present, namely, $r_{A}\ne1$, the system then
possesses $C_{2v}$ symmetry \cite{Johansson16195440} in which the
$x$-$z$ and $y$-$z$ planes are the two mirror planes. For brevity
in the present $D=2$ analysis, we call the $x$- and $y$- axes the
mirror lines. For investigating the effects of the anisotropy, we
fix the orientation of $\boldsymbol{E}\left(t\right)$ to be independent
of time, namely, $\boldsymbol{E}\left(t\right)=\hat{\boldsymbol{e}}E\left(t\right)$
with $\hat{\boldsymbol{e}}=\hat{\boldsymbol{x}}\cos\varphi_{e}+\hat{\boldsymbol{y}}\sin\varphi_{e}$,
specified by an angle $0\le\varphi_{e}\le\pi$. The anisotropy can
then be revealed through the dependence on $\varphi_{e}$. Arbitrary
vectorial physical quantity, e.g., the macroscopic spin polarisation
$\boldsymbol{S}$, can then be decomposed as $\boldsymbol{S}=\hat{\boldsymbol{e}}S^{\parallel}+\left(\hat{\boldsymbol{z}}\times\hat{\boldsymbol{e}}\right)S^{\perp}+\hat{\boldsymbol{z}}S^{z}$
where the symbols $\perp$ and $\parallel$ stand for in-plane transverse
and longitudinal components to track anisotropic effects. Regardless
whether or not $r_{A}=1$, the Rashba spin-orbit field (\ref{sorb-fld-1})
straightforwardly gives zero Berry curvature according to Eq. (\ref{AbelianBrcv-defMk}).
So the SOC-geometric effects here are not caused by the Berry curvature. 

\subsubsection{SOC-geometric effects behind transient non-adiabatic CISP}

The Rashba SOC given by Eq. (\ref{sorb-fld-1}) shows the native spin
texture $\left\langle \boldsymbol{\sigma}\right\rangle _{\pm}^{0}\left(\boldsymbol{k}\right)=\pm\mathcal{X}_{K}\boldsymbol{k}\times\hat{\boldsymbol{z}}/\left\vert \mathcal{X}_{K}\boldsymbol{k}\right\vert $
with $\left\langle \sigma_{z}\right\rangle _{\pm}^{0}\left(\boldsymbol{k}\right)=0$.
A nonzero out-of-plane driven spin texture $\left\langle \sigma_{z}\right\rangle _{\pm}\left(\boldsymbol{k},t\right)\ne0$
therefore indicates the existence of intrinsic spin coherence raised
by the external field. This also raises the possibility of getting
$S^{z}\left(t\right)\ne0$ whose existence then represents a macroscopic
manifestation of the microscopic intrinsic spin coherence rooted to
the Rashba-SOC geometric effects. Through a detailed investigation
using Eq. (\ref{incoh-rho-1}) including both the intrinsic limit
and with extrinsic scatterings (see SM-Secs. \ref{sec-examples-Rashba-nadCISP}
and \ref{SI-spin-basics-adiaCISP}), we found that although both adiabatic
and non-adiabatic drivings can induce microscopic spin coherence $\left\langle \sigma_{z}\right\rangle _{\pm}\left(\boldsymbol{k},t\right)\ne0$,
a non-vanishing macroscopic out-of-plane CISP is a result of non-adiabatic
drivings that break the anisotropy-supported mirror symmetry by $\varphi_{e}\ne0,\pi/2$
(see a demonstration in Fig. \ref{CISP-Szpp}(a)). We summerise these
in the table: 

\begin{tabular}{|c|c|c|c|c|c|}
\hline 
$\left\langle \sigma_{z}\right\rangle _{\pm}\left(\boldsymbol{k},t\right)\ne0$ & adiabatic & non-adiabatic & $S^{z}\left(t\right)\ne0$ & adiabatic & non-adiabatic\tabularnewline
\hline 
\hline 
$r_{A}=1$ & yes & yes & $r_{A}=1$ & no & no\tabularnewline
\hline 
$r_{A}\ne1$ & yes & yes & $r_{A}\ne1$ & no & yes\tabularnewline
\hline 
\end{tabular} Indeed, the association of anisotropic characteristics, i.e., dependence
on the orientation angle $\varphi_{e}$, with the geometric properties
of bands has already been hinted by the coupling between the wave-packet
CM and its band pseudospin in Eq. (\ref{CM-spin-coup1}). There $\hbar\dot{\boldsymbol{k}}_{t}\parallel\hat{\boldsymbol{e}}$
participates in this scalar product by its orientation relative to
the geometric connection. This hint has been brought to light explicitly
by the non-adiabatic transient out-of-plane CISP effects realised
with anisotropic Rashba SOC.

\subsubsection{Anisotropic characteristics differentiated between spin-mediated
and bond-mediated parts of the photocurrent}

\begin{figure}[h] \includegraphics[width=9cm, height=7.5cm]{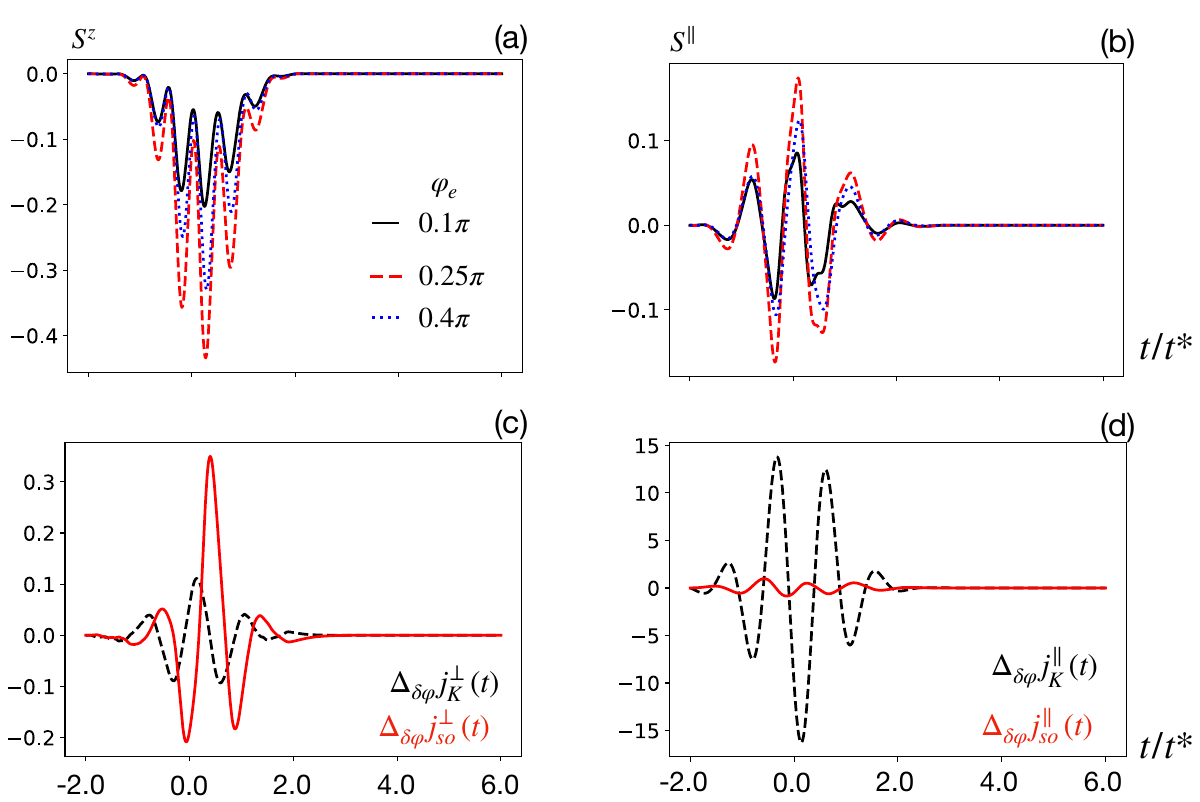} 
\caption{ Spin ((a),(b)) and photocurrent ((c),(d)) dynamics revealing anisotropic effect, demonstrated with $r_{A}=3$ and the laser pulse is set as $E\left(t\right)=-\dot{A}\left(t\right)$ with $A\left(t\right)=(E_{0}/\omega)\mathcal{K}_{t}/\mathcal{K}^{*}$ in which $\mathcal{K}_{t}$ is given by the form used in \cite{FootnoteSSHNumSet}. The Rashba SOC defines a reciprocal length scale, $k_{\text{so}}^{*}=m_{r}\alpha_{R}/\hbar^{2}$ and an energy scale, $\varepsilon_{\text{so}}^{*}=m_{r}\alpha_{R}^{2}/\left(2\hbar^{2}\right)$, so time and electric field are measured in unit of $t^{*}\equiv2\pi\hbar/\varepsilon_{\text{so}}^{*}$ and $E_{0}^{*}=k_{\text{so}}^{*}\varepsilon_{\text{so}}^{*}/e$ respectively. For simplicity, here we use $\tau_{P}=t^{*}$, $\omega=2\pi/t^{*}$ and $E_{0}=E_{0}^{*}$ in all the plots concerning the Rashba system.
(a)/(b): Rise of $S^{z/\parallel}(t)$ (in unit of $(\hbar/2)(k^{*}_{\text{so}})^{2}$  as spin density) due to varied amount of mirror-symmetry breaking as different $\varphi_{e}$'s (see legends).  These spin polarisation components are maximised at maximum mirror-symmetry breaking, namely $\varphi_{e}=0.25\pi$.  (c)/(d): Anisotropic asymmetry (with $\delta\varphi=0.1\pi$) in transverse/longitudinal components of the bond-mediated (black dashed) and spin-mediated (red solid) parts of the photocurrent (in unit of $\left(-e/\hbar\right)k_{\text{so}}^{*}\varepsilon_{\text{so}}^{*}$ as current density for 2D). The transverse and the longitudinal components are dominated by the spin-mediated and bond-mediated parts respectively.  Other parameters used are $\mu=-0.5\varepsilon^{*}_{\text{so}}$, $k_{B}T=0.01\varepsilon^{*}_{\text{so}}$ and $\tau=0.2t^{*}$.
} 
\label{CISP-Szpp} 
\end{figure}

The manifestation of SOC-geometric effects in the out-of-plane spin
polarisation through non-adiabatically breaking the mirror symmetry
($\hat{\boldsymbol{e}}\nparallel\hat{\boldsymbol{x}}/\hat{\boldsymbol{y}}$)
has further implications for the spin-mediated processes in the photocurrents
as discussed in Sec. \ref{sec-examples-SOC-isc}. We make the dependence
on $\hat{\boldsymbol{e}}$ (specified by $\varphi_{e}$) explicit
by writing $\left.\boldsymbol{j}\left(t\right)\right\vert _{\varphi_{e}}=\boldsymbol{j}\left(t,\varphi_{e}\right)$.
It becomes convenient to define the anisotropic asymmetry, namely,

\begin{align}
\Delta_{\delta\varphi}j_{K/so}^{\alpha}\left(t\right):=j_{K/so}^{\alpha}\left(t,\delta\varphi\right)-j_{K/so}^{\alpha}\left(t,\pi/2-\delta\varphi\right).\label{ani-asym-def}
\end{align}
It quantifies the efficacy of the anisotropy in the $\alpha$-component
of the bond/spin-mediated current (see Eq. (\ref{JtoKnL})) under
the circumstance of breaking the mirror symmetry by deviating of $\hat{\boldsymbol{e}}$
from both of the mirror lines by an angle $\delta\varphi$. Analysing
bond/spin-mediated parts of the photocurrent respectively with Eq.
(\ref{ani-asym-def}) (see details in SM-Sec. \ref{sec-examples-Rashba-mrr})
shows that \begin{subequations}\label{so-K-domir}
\begin{equation}
\left\vert \Delta_{\delta\varphi}j_{so}^{\parallel}\left(t\right)\right\vert \ll\left\vert \Delta_{\delta\varphi}j_{K}^{\parallel}\left(t\right)\approx\Delta_{\delta\varphi}j^{\parallel}\left(t\right)\right\vert ,\label{so-K-domir-LG}
\end{equation}
\begin{equation}
\left\vert \Delta_{\delta\varphi}j^{\perp}\left(t\right)\approx\Delta_{\delta\varphi}j_{so}^{\perp}\left(t\right)\right\vert \gg\left\vert \Delta_{\delta\varphi}j_{K}^{\perp}\left(t\right)\right\vert .\label{so-K-domir-TV}
\end{equation}
\end{subequations} We make two remarks on Eq. (\ref{so-K-domir-TV}).
First, for small $\delta\varphi$, $\Delta_{\delta\varphi}j_{so}^{\perp}\left(t\right)$
is determined by $S^{\parallel}$ (see Eq. (\ref{aniasym-jso-perp})).
Second, under non-adiabatic drivings, $S^{\parallel}\left(t,\varphi_{e}\right)$
rises simultaneously with $S^{z}\left(t,\varphi_{e}\right)$ (see
analytical indications in SM-Sec. \ref{sec-examples-Rashba-nadCISP}
and numerical demonstrations in Fig. \ref{CISP-Szpp}(b) that accompanies
Fig. \ref{CISP-Szpp}(a)). We already know $S^{z}\left(t,\varphi_{e}\right)$
signifies SOC-geometric effect from previous discussions. Henceforth,
the above two remarks plus the dominance of $\Delta_{\delta\varphi}j_{so}^{\perp}\left(t\right)$
in $\Delta_{\delta\varphi}j^{\perp}\left(t\right)$ indicated by Eq.
(\ref{so-K-domir-TV}) is also a signature of SOC-geometric effects.
The result Eq. (\ref{so-K-domir}) tells that spin-mediated part $\boldsymbol{j}_{so}\left(t\right)$
can be differentiated from the bond-mediated part $\boldsymbol{j}_{K}\left(t\right)$
of the photocurrent $\boldsymbol{j}\left(t\right)=\boldsymbol{j}_{K}\left(t\right)+\boldsymbol{j}_{so}\left(t\right)$
by their respective dominances in the anisotropic asymmetry of the
$\perp$ and $\parallel$ components, see Fig. \ref{CISP-Szpp} (c)
(for Eq. (\ref{so-K-domir-TV})) and (d) (for Eq. (\ref{so-K-domir-LG})).
This requires to discern $\perp$ from $\parallel$ components of
the full photocurrent in terms of $\varphi_{e}$-dependence characters.
Such discernment can be arrived by exploiting the anisotropy-furnished
mirror symmetry, namely,
\begin{equation}
j^{\perp}\left(t,M\hat{\boldsymbol{e}}\right)\ne j^{\perp}\left(t,\hat{\boldsymbol{e}}\right),j^{\parallel}\left(t,M\hat{\boldsymbol{e}}\right)=j^{\parallel}\left(t,\hat{\boldsymbol{e}}\right),\label{mirror-perp-pp}
\end{equation}
in which $M\hat{\boldsymbol{e}}$ reflects $\hat{\boldsymbol{e}}$
by either of the two orthogonal mirror lines (see details in SM-Sec.
\ref{sec-examples-Rashba-mrr}). 

\section{Conclusions }

\label{conclude-sec}

\subsection{Summary}

\label{concl-theosum}With the development of a NADT-WPT as an extension
of well-established AD-WPT, we have explored three distinct ways of
manifesting band-geometric effects in macroscopic observables for
transient non-adiabatic drivings, namely, the current rates, the spin
polarisation, and the spin-mediated part of the photocurrents. We
have shown geometric properties of bands leave their imprints on the
patterns of current rate dynamics after turning off the laser pulses.
This is through the history-dependence effects enabled by non-adiabatic
in-pulse drivings via the mechanism of quantum acceleration. We have
demonstrated the above mechanisms using the example of the SSH system
and anticipate similar phenomena to be found in other systems as well
\cite{FootnoteSSHsimDirac}. The macroscopic spin polarisation is
revealed to contain band-geometric effects that are exclusive to non-adiabatic
drivings due to its embedding of intrinsic spin coherences. This contrasts
with the adiabatic CISP that is proportional to the relaxation time
characterising extrinsic scatterings. With Rashba SOC defined with
in-plane native spin texture, we demonstrated how the out-of-plane
CISP signifies such SOC-band-geometric effects albeit restricted by
the SOC anisotropy. The spin-mediated part of the photocurrent signifies
SOC-band-geometric effects due to the same reason of its association
with the intrinsic spin coherence. The dependence of CISP as well
as of the photocurrents on the anisotropy of the SOC furnishes a way
to differentiate spin-mediated from the bond-mediated parts in the
full photocurrent by the former's distinct dominance in the anisotropic
asymmetry of the transverse component of the photocurrent. The above
spin-related effects, rooted to intrinsic spin coherence, have also
been examined including extrinsic scatterings via NADT-WPT. The salient
features of their manifestations are not altered by the extrinsic
scatterings, owing to the transient dominance of intrinsic geometric
properties induced by non-adiabatic drivings.

\subsection{Implications for experiments}

\label{concl-EXPIM}With the application of ultrafast laser pulses
to quantum materials, three types of detections are commonly seen
in experiments, namely, the time- and angle-resolved photoemission
spectroscopy (Tr-ARPES), time-integrated electronic currents at contacts,
and time-resolved THz emissions \cite{Ma23170}. We have investigated
both microscopic momentum-resolved electron dynamics and macroscopic
observables like the current. The former has been experimentally targeted
in general through the Tr-ARPES down to subcycle time resolutions
\cite{Ma23170}. Retrieving band structure information from the electronic
currents excited by laser pulses has also been experimentally practiced
\cite{Ma23170}. For time-resolved THz emissions, within the framework
of electromagnetism, the emitted electric fields $\boldsymbol{E}_{em}\left(t\right)$
have been shown to be proportional to $\text{d}\boldsymbol{j}\left(t\right)/\text{d}t$
\cite{Shan04book}. This provides a link between experimental signals
and the present theory which computes $\boldsymbol{j}\left(t\right)$
and $\text{d}\boldsymbol{j}\left(t\right)/\text{d}t$. So broadly
speaking, the present theoretical endeavour is related to all three
types of experiments. More specifically, concerning the particular
geometric effects and their manifestations through macroscopic observables,
there are subtleties and caveats for each type of the experiments
(see discussions in SM-Sec. \ref{SI-EXP-broadcmmts}). As experiments
are performed for specific materials, before we apply NADT-WPT to
band structures tailor-made for the targeted materials, the followings
should be noted. Purely on a theoretical ground, a more complicated
band structure in which the bands are not only distinguished by their
spin DOF like the two-band systems used here, the spin-orbit entanglement
is generically anticipated. How such entanglement coupled with other
possible material-dependent intrinsic properties, e.g., orbital characters,
affects the band-geometric effects with non-adiabatic transient drivings
are themselves sufficiently complicated to be treated as separate
research topics accessible by NADT-WPT. These purely theoretical intrinsic
issues are readily beyond the scope of this initial-stage investigation.

Nevertheless, despite the complexities arising from the band compositions
for band-geometric effects, the present results comprehending Secs.
\ref{WPT-basics-spntxt}, \ref{sec-examples-SOC-isc} and \ref{sec-examples-Rashba}
readily furnish implications relevant for experimental studies of
certain aspects of SOC effects (discussed below) within the broader
context of spintronics. Notably, the decomposition $\boldsymbol{j}\left(t\right)=\boldsymbol{j}_{so}\left(t\right)+\boldsymbol{j}_{K}\left(t\right)$,
Eq. (\ref{JtoKnL}), is held not only for two-band descriptions. It
is inherited to any Hamiltonian describing an electron moving in periodic
potentials with an SOC part and the other part that exists readily
without SOC. Recall from Sec. \ref{sec-examples-SOC-isc} that the
spin-mediated part $\boldsymbol{j}_{so}\left(t\right)$ differentiates
itself from the bond-mediated part $\boldsymbol{j}_{K}\left(t\right)$
of the photocurrent $\boldsymbol{j}\left(t\right)=\boldsymbol{j}_{so}\left(t\right)+\boldsymbol{j}_{K}\left(t\right)$
readily in terms of their microscopic underpinning processes, without
referring to specific type of SOC. With the Rashba SOC in Sec. \ref{sec-examples-Rashba},
we showed that $j^{\perp}\left(t,\varphi_{e}\right)$ and $j^{\parallel}\left(t,\varphi_{e}\right)$
are distinctively dominated by $j_{so}^{\perp}\left(t,\varphi_{e}\right)$
and $j_{K}^{\parallel}\left(t,\varphi_{e}\right)$ in terms of the
anisotropic asymmetry Eq. (\ref{so-K-domir}). Apart from the result
of Eq. (\ref{so-K-domir}) that is specific to the Rashba SOC example
of Sec. \ref{sec-examples-Rashba}, the above NADT-WPT results generally
testify three points that are independent of the SOC type. (i): The
spin and the usual electronic bond mediate fundamentally different
processes behind the photocurrents. (ii) The responses along $\parallel$
and $\perp$ directions are induced by quite different mechanisms
(differentiated by Eq. (\ref{mirror-perp-pp}) as an example) so that
(iii): their respective ways of encoding spin-mediated and bond-mediated
processes, e.g., through anisotropic characteristics, are also different,
exemplified by Eq. (\ref{so-K-domir}). These theoretical implications
indicate the latent potency of structural anisotropy in experimentally
revealing the underlying spin-related processes. Hereby we perform
THz-emission experiments on SnSe whose anisotropic effects have already
received much attention in thermoelectric properties \cite{Zhao14373}
but not yet fully explored for the spintronic processes with implications
from NADT-WPT.

The 3D bulk single-crystal SnSe has an orthorhombic structure which
naturally possesses a termination plane (called the sample plane)
with two orthogonal non-equivalent crystal axes, similar to the two
orthogonal mirror lines addressed in Sec. \ref{sec-examples-Rashba}.
To relate to the above points (i-iiii), we collected THz emissions
with both $\parallel$ and $\perp$ components generated by linearly
polarised (LP) pulses with various polarisation angles (to map out
its anisotropic characters) together with those induced by circularly
polarised (CP) pulses. For details of experiments, see SM-Secs. \ref{SI-EXP-exp}.
Deductions made purely from the CP data and that purely from the LP
data aided by NADT-WPT lead to the same conclusion that the processes
on the sample plane dominate the detected signals (see SM-Sec. \ref{SI-EXP-planeforSOC}
aided by SM-Sec. \ref{SI-EXP-fit}). An analysis of these data without
a priori assumption of the SOC type shows consistency between the
experimental data and the points (i-iii) above (details in SM-Sec.
\ref{SI-EXP-qualitative}). This consistency result on one hand supports
the meaningful existences of the spin-mediated and the bond-mediated
parts of the photocurrents. On the other hand, it also helps to illustrate
the potential feasibility of exploiting the anisotropy for spintronic
interests through the additional the use of LP pulses, complementary
to the familiar association of SOC effects to CP-differentiated responses
in THz-emission experiments.

\subsection{Limitations and possibilities of improvements}

\label{concl-lim}This work shows one step toward the big question
of how band-geometric properties play their roles in transient and
non-adiabatic steering of micro/macroscopic processes. We proposed
a wave-packet-based framework to analyse the driving-dominated revelation
of band-geometric effects. At this stage, we then have only explicitly
calculated results from two-band systems. More refined band-structure
and relaxation times customised for specific materials can always
be incorporated as inputs to NADT-WPT for future studies of these
materials. We also have only compared two big categories of drivings,
namely, adiabatic and non-adiabatic drivings. This is because how
band-geometric properties are carried with adiabatic transport is
readily well known.The logical first step is then to identify manifestations
of geometric effects that are exclusively available for non-adiabatic
drivings. As experimental techniques of tailoring ultrafast laser
pulses are readily advanced, comparison among non-adiabatic transient
drivings of different characters await. 
\begin{acknowledgments}
We thank Prof. Feng-Chuan Chuang for raising our attention to the
on-going research interests about anisotropic Rashba SOC effects.
We also thank useful discussion with Prof. Ting-Kuo Lee. MWYT acknowledges
financial support from European Research Council (ERC) under the European
Unions Horizon 2020 research and innovation programme (Grant Agreement
No. ERC-2017-AdG-788890). JPC acknowledges financial support from
the National Science and Technology Council, Taiwan (Grant No. 112-2112-M-018-005-MY3).
CWL thanks the support by National Science and Technology Council,
Taiwan (Grant No.\textquoteright s 109-2124-M-009-003-MY3 and 112-2119-
M-A49-012-MBK).
\end{acknowledgments}

\newpage
 \appendix




\begin{center}
  \textbf{\large Supplementary Materials for Ultrafast spintronics with geometric effects
in non-adiabatic wave-packet dynamics}
\end{center}

For being self-content, in Sec. \ref{SI-wpt-deriv}, we give a derivation
of how the previously established WPT (or AD-WPT) is extended to the
transient and non-adiabatic regime (NADT-WPT) as the new development
here, supplementing primarily the main text contents of Sec. \ref{WPT-basics-adia}
and \ref{WPT-basics-nadia}. In Sec. \ref{SI-spin-basics} of this
SM, we link the well-established concepts of spin texture to wave-packet
dynamics for enunciation of spin-related geometric properties of bands.
These links underlie the primary discussions abstracted in Sec. \ref{WPT-basics-Uniqnadia}
and demonstrated with concrete examples in Sec. \ref{sec-examples}
of the main text. We have briefed in Sec. \ref{sec-examples-Rashba}
of the main text the primary results of the particular example of
anisotropic Rashba SOC. Here in Sec. \ref{SI-RSOC-dys} of this SM,
we supplement these results by systematic investigations using the
NADT-WPT applied to the Rashba systems under consideration. We describe
implications to experiments with ultrafast lasers of several types
in general in Sec. \ref{SI-EXP-broadcmmts}. More specifically we
detail the THz-emission experiments on SnSe with the subsequent analysis
of the data in Sec. \ref{SI-EXP-SnSe}. The SM-Sec. \ref{SI-EXP}
underlies the discussions in Sec. \ref{conclude-sec} of the main
text.

\section{A wave-packet transport theory}

\label{SI-wpt-deriv}

As mentioned in the main text, the WPT is constructed in two steps
\cite{Ashcroft76book,Xiao101959}. In the first step, a single electron
wave packet is considered (detailed in Sec. \ref{SI-wpt-deriv-single})
while for the second step, macroscopic current in an electron gas
is derived from an ensemble of wave packets (detailed in Sec. \ref{SI-wpt-deriv-ensemble}).
While reproduction of steady-state AD-WPT from the non-adiabatic but
static extension of AD-WPT has been detailed in \cite{Tu20045423},
here the reproduction of time-dependent AD-WPT results (including
the well-known nonlinear Hall effect \cite{Sodemann15216806}) from
NADT-WPT is described in Sec. \ref{nlqh}. In Sec. \ref{SI-wpt-deriv-ppcrntrt},
the real-time non-adiabatic current formula from NADT-WPT is analysed
for the post-pulse time region.

\subsection{Dynamics of a single wave packet with multiple bands}

\label{SI-wpt-deriv-single}

The equations of motion followed by a single wave packet with multiple
bands, namely, Eqs. (\ref{xdot-nra}) and (\ref{driving-E1}) of the
main text, have been derived using variational-principle on wave-packet
Lagrangian in \cite{Culcer05085110,Shindou05399}. They have been
re-derived without using the Lagrangian-based variational formalism
in Ref. \cite{Tu20045423}. In this subsection, we give a short summary
of the much more detailed derivation given in \cite{Tu20045423} to
obtain Eqs. (\ref{xdot-nra}) and (\ref{driving-E1}) and take the
adiabatic limit to arrive Eq. (\ref{nox-EOMx-ad1}) given in the main
text.

The starting point is an electron with charge $-e$ moving in a periodic
potential described by the Hamiltonian $H\left(\boldsymbol{r},\boldsymbol{p},\boldsymbol{\sigma}\right)$
where $\boldsymbol{r}$, $\boldsymbol{p}$ and $\boldsymbol{\sigma}$
are the coordinate, momentum and Pauli operators for the spin respectively.
Under a spatially uniform electric field $\boldsymbol{E}\left(t\right)=-\dot{\boldsymbol{A}}\left(t\right)$
generated by a vector potential $\boldsymbol{A}\left(t\right)$ with
the minimum coupling principle, $\boldsymbol{p}\rightarrow\boldsymbol{p}-\left(-e\right)\boldsymbol{A}\left(t\right)$,
one obtains a time-dependent Schr\"{o}dinger equation $H\left(t\right)\left\vert \Phi\left(t\right)\right\rangle =i\hbar\left\vert \dot{\Phi}\left(t\right)\right\rangle $
with the time-dependent Hamiltonian $H\left(t\right):=H\left(\boldsymbol{r},\boldsymbol{p}-\left(-e\right)\boldsymbol{A}\left(t\right),\boldsymbol{\sigma}\right)$.
Under the Born--von Karman boundary condition, we can take the ansatz:
$\left\langle \boldsymbol{x},s\left\vert \Phi\left(t\right)\right.\right\rangle =\left(1/\sqrt{\mathcal{V}}\right)e^{i\boldsymbol{k}\cdot\boldsymbol{x}}\left\langle \boldsymbol{x},s\left\vert \chi_{t}\right.\right\rangle $
in which $\boldsymbol{r}\left\vert \boldsymbol{x},s\right\rangle =\boldsymbol{x}\left\vert \boldsymbol{x},s\right\rangle $,
$\boldsymbol{\sigma}\cdot\hat{\boldsymbol{n}}\left\vert \boldsymbol{x},s=\pm\right\rangle =\pm\left\vert \boldsymbol{x},\pm\right\rangle $
for an arbitrarily prescribed real 3D unit vector $\hat{\boldsymbol{n}}$.
Here $\left\vert \chi_{t}\right\rangle $, normalised to unity within
a primitive cell, is a cell-periodic function, namely, $\left\langle \boldsymbol{x},s\left\vert \chi_{t}\right.\right\rangle =\left\langle \boldsymbol{x}+\boldsymbol{R},s\left\vert \chi_{t}\right.\right\rangle $
with $\boldsymbol{R}$ being an arbitrary lattice vector and $1/\sqrt{\mathcal{V}}$
is a normalisation constant. In the space of periodic functions, we
then have $i\hbar\left\vert \dot{\chi}_{t}\right\rangle =\mathcal{H}\left(\boldsymbol{k}_{t}\right)\left\vert \chi_{t}\right\rangle $
where $\mathcal{H}\left(\boldsymbol{k}_{t}\right)=H\left(\boldsymbol{r},\boldsymbol{p}+\hbar\boldsymbol{k}_{t},\boldsymbol{\sigma}\right)$
in which $\hbar\boldsymbol{k}_{t}=\hbar\boldsymbol{k}-\left(-e\right)\boldsymbol{A}\left(t\right)$.
The time derivative $\hbar\dot{\boldsymbol{k}}_{t}=\left(-e\right)\left(-\dot{\boldsymbol{A}}\left(t\right)\right)$
immediately becomes Eq. (\ref{driving-E1}) of the main text. We now
derive Eq. (\ref{xdot-nra}) of the main text. The velocity operator
is defined by $\boldsymbol{v}\left(t\right):=\left(-i/\hbar\right)\left[\boldsymbol{r},H\left(t\right)\right]$
so the wave packet velocity is given by $\dot{\boldsymbol{x}}_{t}=\left\langle \Phi\left(t\right)\right\vert \boldsymbol{v}\left(t\right)\left\vert \Phi\left(t\right)\right\rangle $.
By the use of $\left\langle \boldsymbol{x},s\left\vert \Phi\left(t\right)\right.\right\rangle =\left(1/\sqrt{\mathcal{V}}\right)e^{i\boldsymbol{k}\cdot\boldsymbol{x}}\left\langle \boldsymbol{x},s\left\vert \chi_{t}\right.\right\rangle $
and the periodicity, it reads $\dot{\boldsymbol{x}}_{t}=\left\langle \chi_{t}\right\vert \boldsymbol{V}\left(\boldsymbol{k}_{t}\right)\left\vert \chi_{t}\right\rangle $
in which $\boldsymbol{V}\left(\boldsymbol{k}_{t}\right)$ is defined
by Eq. (\ref{vopt-def}) of the main text. To pronounce the geometric
aspects, we expand $\left\vert \chi_{t}\right\rangle =\sum_{n}\eta_{n}\left(t\right)e^{i\gamma_{n}\left(\boldsymbol{k}_{t}\right)}\left\vert u_{n}\left(\boldsymbol{k}_{t}\right)\right\rangle $
which then leads to $\dot{\boldsymbol{x}}_{t}=\sum_{n,m}\eta_{n}^{*}\left(t\right)e^{-i\gamma_{n}\left(\boldsymbol{k}_{t}\right)}\left[\hbar^{-1}\mathcal{D}_{\boldsymbol{k}_{t}},\mathcal{H}\left(\boldsymbol{k}_{t}\right)\right]_{n,m}\eta_{m}\left(t\right)e^{i\gamma_{m}\left(\boldsymbol{k}_{t}\right)}:=\left\langle \left[\hbar^{-1}\mathcal{D}_{\boldsymbol{k}_{t}},\mathcal{H}\left(\boldsymbol{k}_{t}\right)\right]\right\rangle $
as Eq. (\ref{xdot-nra}) of the main text. 

The more familiar expression for $\dot{\boldsymbol{x}}_{t}$ in which
the Berry curvature appears along with the anomalous velocity \cite{Xiao101959}
(as Eq. (\ref{nox-EOMx-ad1}) in the main text) can be derived straightforwardly
from the above by first segregating the bands $\left\{ \left\vert u_{n}\left(\boldsymbol{k}\right)\right\rangle \right\} _{n}$
into two sets, namely, the active and the remote manifolds, denoted
by $a$ and $r$ respectively. The definition is that during the driving
processes, the occupations on the remote manifold $\left\{ \left\vert u_{n}\left(\boldsymbol{k}\right)\right\rangle \right\} _{n\in r}$
are negligible while those on $\left\{ \left\vert u_{n}\left(\boldsymbol{k}\right)\right\rangle \right\} _{n\in a}$
are not. With the time-dependent basis of $\left\{ \left\vert u_{n}\left(\boldsymbol{k}_{t}\right)\right\rangle \right\} _{n}$,
the Schr\"{o}dinger equation $i\hbar\left\vert \dot{\chi}_{t}\right\rangle =\mathcal{H}\left(\boldsymbol{k}_{t}\right)\left\vert \chi_{t}\right\rangle $
is turned into $i\hbar\dot{\boldsymbol{\eta}}\left(t\right)=\bar{\mathcal{H}}\left(\boldsymbol{k}_{t},\boldsymbol{E}\left(t\right)\right)\boldsymbol{\eta}\left(t\right)$
in which the $n$th component of the complex vector $\boldsymbol{\eta}\left(t\right)$
is $\eta_{n}\left(t\right)$ and the moving-frame Hamiltonian $\bar{\mathcal{H}}\left(\boldsymbol{k}_{t},\boldsymbol{E}\left(t\right)\right)$
described below Eq. (\ref{incoh-rho-1}) of the main text naturally
appears. By taking the inter-manifold matrix elements $\bar{\mathcal{H}}_{n,m}\left(\boldsymbol{k}_{t},\boldsymbol{E}\left(t\right)\right)=-\left[\mathcal{\bar{\mathcal{R}}}_{\boldsymbol{k}_{t}}\right]_{n,m}\cdot\left(-e\right)\boldsymbol{E}\left(t\right)$
in which $n\in a$ and $m\in r$ or $n\in r$ and $m\in a$ as perturbation
to its first order, Eq. (\ref{xdot-nra}) of the main text becomes

\begin{align}
\dot{x}_{\alpha t}=\left\langle \left[\hbar^{-1}\mathcal{D}_{k_{\alpha t}},\mathcal{H}_{a}\left(\boldsymbol{k}_{t}\right)\right]\right\rangle _{a}-\sum_{\beta}\left\langle \mathcal{F}^{\alpha\beta}\left(\boldsymbol{k}_{t}\right)\right\rangle _{a}\dot{k}_{\beta t},\label{phsvelo-sep-2-x}
\end{align}
in which the active-manifold average $\left\langle \mathcal{M}\left(\boldsymbol{k}_{t}\right)\right\rangle _{a}:=\sum_{n,m\in a}\eta_{n}^{*}\left(t\right)e^{-i\gamma_{n}\left(\boldsymbol{k}_{t}\right)}\left[\mathcal{M}\left(\boldsymbol{k}_{t}\right)\right]_{nm}\eta_{m}\left(t\right)e^{i\gamma_{m}\left(\boldsymbol{k}_{t}\right)}$
has been defined for arbitrary matrices $\mathcal{M}\left(\boldsymbol{k}_{t}\right)$.
Here in Eq. (\ref{phsvelo-sep-2-x}), $\dot{x}_{\alpha t}$ and $\dot{k}_{\beta t}$
are respectively the $\alpha$th and the $\beta$th spatial components
of $\dot{\boldsymbol{x}}_{t}$ and $\dot{\boldsymbol{k}}_{t}$ and
the definition
\begin{equation}
\mathcal{F}^{\alpha\beta}\left(\boldsymbol{k}_{t}\right)=\frac{\partial\left[\mathcal{\mathcal{R}}_{k_{\beta t}}\right]}{\partial k_{\alpha t}}-\frac{\partial\left[\mathcal{\mathcal{R}}_{k_{\alpha t}}\right]}{\partial k_{\beta t}}-i\left[\mathcal{\mathcal{R}}_{k_{\alpha t}},\mathcal{\mathcal{R}}_{k_{\beta t}}\right],\label{nABCV-1_Mk}
\end{equation}
known as the non-Abelian Berry-curvature has been applied. The simpler
results Eqs. (\ref{nox-EOMx-ad1}) and (\ref{AbelianBrcv-defMk})
in the main text are obtained from Eqs. (\ref{phsvelo-sep-2-x}) and
(\ref{nABCV-1_Mk}) by assuming there is only one band indexed by
$n$ in the active manifold. Hereby we complete the derivation for
the equations of motion followed by a single wave packet present in
the main text.

\subsection{Dynamics of ensemble of wave packets }

\label{SI-wpt-deriv-ensemble}We now move to the second step of the
WPT in which we consider an ensemble of wave packets. We derive Eq.
(\ref{incoh-rho-1}) of the main text by extending the kinetic theory
readily framed in Refs. \cite{Ashcroft76book,Xiao101959} to account
for scattering effects. For being pedagogical, we discuss consecutively
three cases (a-c). (a): The band gaps and the external electric fields
are set to satisfy the adiabatic condition. (b): The band gaps and
the external electric fields are not set to satisfy the adiabatic
condition but the external electric fields are assumed to be static
at times of interests for steady states. (c): The external electric
fields are time-dependent and we are interested in the transient regime.
We start by describing the formulation in the kinetic theory for electron
gases, that is common to all the three cases. We then briefly review
the specification to situations (a) in Sec. \ref{KE-nac-SS} (well
established in Refs. \cite{Ashcroft76book,Xiao101959}) and (b) in
Sec. \ref{SI-wpt-deriv-ensemble-NADstatic} (detailed in Ref. \cite{Tu20045423}).
After this, how the specification to (c) can be done then becomes
obvious in Sec.~\ref{KE-nac-T}. Since the main focus of the present
research is on the time-dependent phenomena, in Sec.~\ref{nlqh},
we show that under periodic driving with small amplitudes such that
the adiabatic condition is valid, the second-order response calculated
using the density matrix given by Eq.~(\ref{incoh-rho-1}) reproduces
the well-established quantum nonlinear Hall effect of Ref. \cite{Sodemann15216806}.

The state of the ensemble is described by the density matrix $\varrho_{\boldsymbol{k}}\left(t\right)$
collected over all $\boldsymbol{k}$'s. Recall that previous discussions
for the dynamics of a single wave packet do not include any scattering
effects. So a single wave packet with CM momentum $\boldsymbol{k}=\boldsymbol{k}_{t_{0}}$
initially in an arbitrary pseudospin state $\left\vert \chi=\chi_{t_{0}}\right\rangle $
(collectively denoted as $\left(\boldsymbol{k},\chi\right)$) will
evolve to the state $\left(\boldsymbol{k}_{t},\chi_{t}\right)$ with
$\left\vert \chi_{t}\right\rangle :=U_{\boldsymbol{k}}\left(t,t_{0}\right)\left\vert \chi_{t_{0}}\right\rangle $
at any later time $t>t_{0}$. The matrix element $\left\langle \chi_{t_{0}}\right\vert \varrho_{\boldsymbol{k}}\left(t_{0}\right)\left\vert \chi_{t_{0}}\right\rangle $
gives the ensemble-averaged occupation on $\left\vert \chi_{t_{0}}\right\rangle $
from the initial ensemble $\varrho_{\boldsymbol{k}}\left(t_{0}\right)$.
Imposing scattering-free dynamics to all wave packets in the ensemble
results in $\left\langle \chi_{t}\right\vert \varrho_{\boldsymbol{k}}\left(t\right)\left\vert \chi_{t}\right\rangle =\left\langle \chi_{t_{0}}\right\vert \varrho_{\boldsymbol{k}}\left(t_{0}\right)\left\vert \chi_{t_{0}}\right\rangle $. 

We now consider how scattering events affect the occupation $\left\langle \chi_{t}\right\vert \varrho_{\boldsymbol{k}}\left(t\right)\left\vert \chi_{t}\right\rangle $.
Due to scattering, only a part of this occupation is actually contributed
by a wave packet that has been coherently driven to the state $\left(\boldsymbol{k}_{t},\chi_{t}\right)$.
The effects of scatterings are taken into account by the relaxation-time
approximation (RTA) which consists of two assumptions (i) and (ii)
that we will individually introduce in appropriate places. The assumption
(i) is that the scattering rate is independent of the form of the
electron distribution function. Using (i), we then denote by $1/\tau_{\boldsymbol{k}}$
as the scattering rate for a wave packet carrying momentum $\boldsymbol{k}$.
We denote by $P\left(\boldsymbol{k}_{t},\boldsymbol{k}_{t^{\prime}}\right)$
the probability that a wave packet in the state $\left(\boldsymbol{k}_{t^{\prime}},\chi_{t^{\prime}}\right)$
at time $t^{\prime}$ can evolve to the state $\left(\boldsymbol{k}_{t},\chi_{t}\right)$
at time $t\ge t^{\prime}$ without being scattered away. By a general
argument, $P\left(\boldsymbol{k}_{t},\boldsymbol{k}_{t^{\prime}}\right)$
is found to be given by Eq. (\ref{non-scatt-prob1}) of the main text
\cite{Ashcroft76book,footnoteNonScattProb}. Henceforth, the contribution
to $\left\langle \chi_{t}\right\vert \varrho_{\boldsymbol{k}}\left(t\right)\left\vert \chi_{t}\right\rangle $
by the above field-driven process is given by 
\begin{align}
g_{\chi_{t}}^{\text{drift}}\left(\boldsymbol{k}_{t},t\right)=\int_{t^{\prime}=t_{0}}^{t^{\prime}=t}\text{d}g_{\chi_{t^{\prime}}}\left(\boldsymbol{k}_{t^{\prime}},t^{\prime}\right)P\left(\boldsymbol{k}_{t},\boldsymbol{k}_{t^{\prime}}\right),\label{noscatt-stream}
\end{align}
where $\text{d}g_{\chi_{t^{\prime}}}\left(\boldsymbol{k}_{t^{\prime}},t^{\prime}\right)$
is the amount occupation gained into the state $\left(\boldsymbol{k}_{t^{\prime}},\chi_{t^{\prime}}\right)$
in an infinitesimal time interval of length $\text{d}t^{\prime}$
around time $t^{\prime}$. Note that at $t=t_{0}$, $g_{\chi}^{\text{drift}}\left(\boldsymbol{k},t_{0}\right)=0$
by the definition of Eq.~(\ref{noscatt-stream}). So if the initial
equilibrium readily has a nonzero occupation on $\left(\boldsymbol{k},\chi\right)$
(so it can be coherently driven to occupy $\left(\boldsymbol{k}_{t},\chi_{t}\right)$),
this source of contribution to the occupation is not accounted for
by Eq.~(\ref{noscatt-stream}). By including also this contribution,
we obtain 
\begin{align}
\left\langle \chi_{t}\right\vert \varrho_{\boldsymbol{k}}\left(t\right)\left\vert \chi_{t}\right\rangle =g_{\chi_{t}}^{\text{drift}}\left(\boldsymbol{k}_{t},t\right)+\left\langle \chi\right\vert \varrho_{\boldsymbol{k}}^{\text{eq}}\left\vert \chi\right\rangle P\left(\boldsymbol{k}_{t},\boldsymbol{k}\right)\label{fullOcc-t-cmm}
\end{align}
where the appearance of the factor $P\left(\boldsymbol{k}_{t},\boldsymbol{k}\right)$
in the second term in Eq.~(\ref{fullOcc-t-cmm}) accounts for the
fraction of the initial equilibrium occupation that has survived the
scattering.

The result Eq.~(\ref{fullOcc-t-cmm}) is common to all situations
(a),(b) and (c). If we are only interested in the steady states, we
then take $\lim_{t-t_{0}\rightarrow\infty}P\left(\boldsymbol{k}_{t},\boldsymbol{k}\right)=0$.
The occupation on $\left(\boldsymbol{k}_{t},\chi_{t}\right)$ simply
becomes 
\begin{align}
\left\langle \chi_{t}\right\vert \varrho_{\boldsymbol{k}}\left(t\right)\left\vert \chi_{t}\right\rangle =g_{\chi_{t}}^{\text{drift}}\left(\boldsymbol{k}_{t},t\right).\label{fullOcc-t-ab}
\end{align}
This is the starting point for the construction of the electron distribution
function in the kinetic theory framework in Ref. \cite{Ashcroft76book}
which is the basis also of Refs. \cite{Xiao101959,Tu20045423} for
exploring Berry curvature effects under adiabatic and non-adiabatic
conditions in the steady-state limits as the circumstances (a) and
(b) mentioned above.

To make further progresses from Eq.~(\ref{fullOcc-t-ab}) (which
amounts to further specifying the contents of Eq.~(\ref{noscatt-stream})),
we need the assumption (ii) of the RTA. This assumption is based on
the observation that in the absence of external fields, the collection
of scattering events maintains the electronic system to obey some
form of the distribution function, e.g., the Fermi-Dirac distribution
function that describes the thermal equilibrium of the electron gases.
Therefore, with the external field $\boldsymbol{E}$ being turned
on, we assume that there exists some form of distribution function
$g_{\chi_{t^{\prime}}}^{0}\left(\boldsymbol{k}_{t^{\prime}}\right)$
such that whenever the electron occupation at a momentum $\boldsymbol{k}_{t^{\prime}}$
arrives this form, the occupation number should not be changed by
the effects of scatterings. In other words, during the infinitesimal
interval of length $\text{d}t^{\prime}$ around time $t^{\prime}$,
the amount of electrons gained into the state $\left(\boldsymbol{k}_{t^{\prime}},\chi_{t^{\prime}}\right)$,
denoted by $\text{d}g_{\chi_{t^{\prime}}}\left(\boldsymbol{k}_{t^{\prime}},t^{\prime}\right)$,
should equal to the amount of electrons lost from the occupation specified
by $g_{\chi_{t^{\prime}}}^{0}\left(\boldsymbol{k}_{t^{\prime}}\right)$
due to scatterings, namely, 
\begin{align}
\text{d}g_{\chi_{t^{\prime}}}\left(\boldsymbol{k}_{t^{\prime}},t^{\prime}\right)=\frac{\text{d}t^{\prime}}{\tau_{\boldsymbol{k}_{t^{\prime}}}}g_{\chi_{t^{\prime}}}^{0}\left(\boldsymbol{k}_{t^{\prime}}\right).\label{eq-scatt-g}
\end{align}
Further explicating $g_{\chi_{t^{\prime}}}^{0}\left(\boldsymbol{k}_{t^{\prime}}\right)$
then requires the differentiation between the circumstance (a) (under
the adiabatic condition) and (b),(c) (not under the adiabatic condition).

\subsubsection{case (a): adiabatic drivings}

\label{KE-nac-SS}

If the circumstance (a) is fulfilled, then an electron wave packet
starting from one band $n$ at time $t^{\prime}$ (carrying momentum
$\boldsymbol{k}_{t^{\prime}}$) will remain at the same band $n$
under a different momentum $\boldsymbol{k}_{t}$ at a later time $t>t^{\prime}$.
It is thus sufficient to consider $\left\vert \chi_{t^{\prime}}\right\rangle =\left\vert u_{n}\left(\boldsymbol{k}_{t^{\prime}}\right)\right\rangle $,
and we denote $f_{n}\left(\boldsymbol{k}_{t^{\prime}},t^{\prime}\right):=\left.\left\langle \chi_{t^{\prime}}\right\vert \varrho_{\boldsymbol{k}}\left(t^{\prime}\right)\left\vert \chi_{t^{\prime}}\right\rangle \right\vert _{\left\vert \chi_{t^{\prime}}\right\rangle =\left\vert u_{n}\left(\boldsymbol{k}_{t^{\prime}}\right)\right\rangle }$
and $f_{n}^{0}\left(\boldsymbol{k}_{t^{\prime}}\right)=g_{\chi_{t^{\prime}}}^{0}\left(\boldsymbol{k}_{t^{\prime}}\right)$
to follow the notational convention in the literatures. Here $f_{n}^{0}\left(\boldsymbol{k}_{t^{\prime}}\right)=f_{FD}^{\mu,T}\left(\varepsilon_{n}\left(\boldsymbol{k}_{t^{\prime}}\right)\right)$
describes the statistical mixture of the original bands $\left\vert u_{n}\left(\boldsymbol{k}_{t^{\prime}}\right)\right\rangle $'s.
Given the scattering rates $\tau_{\boldsymbol{k}_{t^{\prime}}}^{-1}$,
the number of electrons lost/gained from/into the state $\left(\boldsymbol{k}_{t^{\prime}},\left\vert \chi_{t^{\prime}}\right\rangle =\left\vert u_{n}\left(\boldsymbol{k}_{t^{\prime}}\right)\right\rangle \right)$
for a differential time interval $\text{d}t^{\prime}$ for Eq.~(\ref{eq-scatt-g})
is then given by

\begin{align}
\text{d}g_{\chi_{t^{\prime}}}\left(\boldsymbol{k}_{t^{\prime}},t^{\prime}\right)=\frac{\text{d}t^{\prime}}{\tau_{\boldsymbol{k}_{t^{\prime}}}}f_{n}^{0}\left(\boldsymbol{k}_{t^{\prime}}\right).\label{eq-scatt-fm}
\end{align}
Substituting Eq.~(\ref{eq-scatt-fm}) into the right-hand side of
Eq.~(\ref{noscatt-stream}), and identifying $P\left(\boldsymbol{k}_{t},\boldsymbol{k}_{t^{\prime}}\right)/\tau_{\boldsymbol{k}_{t^{\prime}}}=\partial P\left(\boldsymbol{k}_{t},\boldsymbol{k}_{t^{\prime}}\right)/\partial t^{\prime}$
(see \cite{footnoteNonScattProb}), then the integral over $t^{\prime}$
there can be performed by using the integration-by-part technique.
The result for Eq. (\ref{fullOcc-t-ab}) with $f_{n}\left(\boldsymbol{k}_{t},t\right)=\left.\left\langle \chi_{t}\right\vert \varrho_{\boldsymbol{k}}\left(t\right)\left\vert \chi_{t}\right\rangle \right\vert _{\left\vert \chi_{t}\right\rangle =\left\vert u_{n}\left(\boldsymbol{k}_{t}\right)\right\rangle }$
is then $f_{n}\left(\boldsymbol{k}_{t},t\right)=f_{n}^{0}\left(\boldsymbol{k}_{t}\right)+\delta{f}_{n}\left(\boldsymbol{k}_{t},t\right)$
as given by Eq. (\ref{non-eq-ad-tt}) of the main text, omitting the
only band index $n$. The deviation from equilibrium appears more
often in the literature in the form of $\delta{f}_{n}\left(\boldsymbol{k}_{t},t\right)=-\tau_{\boldsymbol{k}_{t}}\left[\partial f_{n}^{0}\left(\boldsymbol{k}_{t}\right)/\partial\boldsymbol{k}_{t}\right]\cdot\left(-e/\hbar\right)\boldsymbol{E}\left(t\right)$
\cite{Ashcroft76book,Xiao101959,Chang951348,Chang967010,Deyo091917v1,Moore10026805,Sodemann15216806}.
This is arrived by further making the short-relaxation-time approximation
to Eq. (\ref{non-eq-ad-tt}) of the main text, namely, 
\begin{align}
P\left(\boldsymbol{k}_{t},\boldsymbol{k}_{t^{\prime}}\right)f\left(t^{\prime}\right)\approx e^{-\frac{t-t^{\prime}}{\tau_{\boldsymbol{k}_{t}}}}f\left(t\right)\label{short-RT}
\end{align}
for arbitrary function $f\left(t^{\prime}\right)$ \cite{footnote-shortRlxT}. 

\subsubsection{case (b): non-adiabatic drivings by time-independent electric fields}

\label{SI-wpt-deriv-ensemble-NADstatic}If the adiabatic condition
(a) is not fulfilled, then the external electric field can hybridise
the bands as we come to the circumstance (b). We now review such situation
but with $\boldsymbol{E}\left(t\right)=\boldsymbol{E}$ as a constant
in time. The hybridised bands, $\left\vert \mathfrak{u}_{i}\left(\boldsymbol{k}_{t^{\prime}}\right)\right\rangle $'s,
each as a superposition of the original bands $\left\vert u_{n}\left(\boldsymbol{k}_{t^{\prime}}\right)\right\rangle $'s,
depend non-perturbatively on $\boldsymbol{E}$ and are eigenfunctions
of the moving-frame Hamiltonian $\bar{\mathcal{H}}\left(\boldsymbol{k}_{t^{\prime}},\boldsymbol{E}\right)$
(see its definition below Eq.~(\ref{incoh-rho-1}) in the main text)
with eigenenergies $\mathcal{E}_{i}\left(\boldsymbol{k}_{t^{\prime}}\right)$'s.
The joint action of the decoherence and the electric field $\boldsymbol{E}$
results in a statistical mixture of the hybridised bands, leading
to 
\begin{align}
g_{\chi_{t^{\prime}}}^{0}\left(\boldsymbol{k}_{t^{\prime}}\right)=\sum_{i}\left\vert \left\langle \left.\mathfrak{u}_{i}\left(\boldsymbol{k}_{t^{\prime}}\right)\right\vert \chi_{t^{\prime}}\right\rangle \right\vert ^{2}g_{i}^{0}\left(\boldsymbol{k}_{t^{\prime}}\right),\label{inv-gSS}
\end{align}
where $g_{i}^{0}\left(\boldsymbol{k}_{t^{\prime}}\right)=f_{FD}^{\mu,T}\left(\mathcal{E}_{i}\left(\boldsymbol{k}_{t^{\prime}}\right)\right)$
\cite{Tu20045423,Tu20045004}. This casts Eq.~(\ref{eq-scatt-g})
to 
\begin{align}
\text{d}g_{\chi_{t^{\prime}}}\left(\boldsymbol{k}_{t^{\prime}},t^{\prime}\right)=\frac{\text{d}t^{\prime}}{\tau_{\boldsymbol{k}_{t^{\prime}}}}\sum_{i}\left\vert \left\langle \left.\mathfrak{u}_{i}\left(\boldsymbol{k}_{t^{\prime}}\right)\right\vert \chi_{t^{\prime}}\right\rangle \right\vert ^{2}g_{i}^{0}\left(\boldsymbol{k}_{t^{\prime}}\right).\label{eq-scatt-gSS}
\end{align}
By applying similar procedures leading from Eq.~(\ref{eq-scatt-fm})
to $f\left(\boldsymbol{k}_{t},t\right)$ of Eq. (\ref{non-eq-ad-tt})
in the main text, additionally with Eq. (\ref{short-RT}), here Eq.~(\ref{eq-scatt-gSS})
finally gives 
\begin{align}
g_{i}\left(\boldsymbol{k}_{t},t\right)=g_{i}^{0}\left(\boldsymbol{k}_{t}\right)+\delta g_{i}\left(\boldsymbol{k}_{t}\right),\label{neqgSS}
\end{align}
where 
\begin{align}
\delta g_{i}\left(\boldsymbol{k}_{t}\right)=-\tau_{\boldsymbol{k}_{t}}\frac{\partial g_{i}^{0}\left(\boldsymbol{k}_{t}\right)}{\partial\boldsymbol{k}_{t}}\cdot\left(\frac{-e}{\hbar}\right)\boldsymbol{E},\label{neqgSS-d1}
\end{align}
and we have notated $g_{i}\left(\boldsymbol{k}_{t},t\right)=\left\langle \mathfrak{u}_{i}\left(\boldsymbol{k}_{t}\right)\right\vert \varrho_{\boldsymbol{k}}\left(t\right)\left\vert \mathfrak{u}_{i}\left(\boldsymbol{k}_{t}\right)\right\rangle $.
How the result of Eq.~(\ref{neqgSS}) reduces to the well-known result
of (a) mentioned below Eq.~(\ref{eq-scatt-fm}) in the limit of weak
electric field has been detailed in Ref.~\cite{Tu20045423}. Together
with Eq.~(\ref{j-def0}) (in the main text) for computing the currents,
the main results of revealing the Berry curvature effects through
the currents in Ref.~\cite{Xiao101959} are also reproduced in Ref.~\cite{Tu20045423}
by taking the adiabatic limit. 

Note that this steady-state non-adiabatic extension applied to gapped
Dirac cone with two bands indexed as $c$ (conduction) and $v$ (valence)
in 2D yields a non-adiabatically renormalised valley Hall current
$\boldsymbol{j}^{H}\propto\boldsymbol{E}\times\int\text{d}^{2}\boldsymbol{k}\boldsymbol{\Omega}\left(\boldsymbol{k}\right)/\sqrt{1+4\left\vert r_{c,v}\left(\boldsymbol{k},\boldsymbol{E}\right)\right\vert }$
where $r_{c,v}\left(\boldsymbol{k},\boldsymbol{E}\right)$ is found
in Eq. (\ref{nonad-meas-def}) of the main text \cite{Tu20045004}.
It is non-perturbation in $\boldsymbol{E}$ since $r_{c,v}\left(\boldsymbol{k},\boldsymbol{E}\right)$
appears in the denominator. However, the geometric manifestation is
still through the Berry curvature $\boldsymbol{\Omega}\left(\boldsymbol{k}\right)$
in this steady-state limit. By going to the transient regime with
arbitrary time-dependent driving electric fields, it is possible to
have geometric manifestation even with $\boldsymbol{\Omega}\left(\boldsymbol{k}\right)=0$.

\subsubsection{case (c): non-adiabatic drivings by time-dependent electric fields}

\label{KE-nac-T}

We are now facilitated to generalise the above construction (b) to
the circumstance (c) with time-dependent electric field to obtain
$\varrho_{\boldsymbol{k}}\left(t\right)$ for $t$ not far away from
$t_{0}$. Since the circumstance (b) is a special case of (c) by taking
the electric field to be constant in time, we simply need to restore
$\boldsymbol{E}\rightarrow\boldsymbol{E}\left(t^{\prime}\right)$
in $\bar{\mathcal{H}}\left(\boldsymbol{k}_{t^{\prime}},\boldsymbol{E}\left(t^{\prime}\right)\right)$
for specifying $\left\vert \mathfrak{u}_{i}\left(\boldsymbol{k}_{t^{\prime}}\right)\right\rangle $
and $\mathcal{E}_{i}\left(\boldsymbol{k}_{t^{\prime}}\right)$ used
in Eq.~(\ref{eq-scatt-gSS}) and substitute it to the right-hand
side of Eq.~(\ref{noscatt-stream})) to get
\begin{align}
g_{\chi_{t}}^{\text{drift}}\left(\boldsymbol{k}_{t},t\right)=\int_{t_{0}}^{t}\text{d}t^{\prime}\left\langle \chi_{t^{\prime}}\right\vert \bar{\varrho}_{\boldsymbol{k}}\left(t^{\prime}\right)\left\vert \chi_{t^{\prime}}\right\rangle \frac{\partial P\left(\boldsymbol{k}_{t},\boldsymbol{k}_{t^{\prime}}\right)}{\partial t^{\prime}},\label{t-drift}
\end{align}
where $\bar{\varrho}_{\boldsymbol{k}}\left(t^{\prime}\right)=\sum_{i}g_{i}^{0}\left(\boldsymbol{k}_{t^{\prime}}\right)\left\vert \mathfrak{u}_{i}\left(\boldsymbol{k}_{t^{\prime}}\right)\right\rangle \left\langle \mathfrak{u}_{i}\left(\boldsymbol{k}_{t^{\prime}}\right)\right\vert $
(see also Eq.~(\ref{incoh-rho-1bar}) in the main text). Performing
similarly the integral of Eq.~(\ref{t-drift}) using the integration-by-part
method and applying $\left\vert \chi_{t^{\prime}}\right\rangle =U_{\boldsymbol{k}}(t^{\prime},t)\left\vert \chi_{t}\right\rangle $
for $t_{0}\le t^{\prime}\le t$ to turn the right-hand side of Eq.~(\ref{fullOcc-t-cmm})
to the form of $\left\langle \chi_{t}\right\vert \bullet\left\vert \chi_{t}\right\rangle $
so that the density matrix $\varrho_{\boldsymbol{k}}\left(t\right)$
can be identified with $\bullet$ in Eq.~(\ref{fullOcc-t-cmm}),
the result then is just Eq.~(\ref{incoh-rho-1}) of the main text. 

One can rewrite Eq. (\ref{incoh-rho-1}) of the main text via the
integration-by-part method to \begin{subequations}\label{jrho-ini-drf}
\begin{equation}
\varrho_{\boldsymbol{k}}\left(t\right)=\varrho_{\boldsymbol{k}}^{\text{ini}}\left(t\right)+\varrho_{\boldsymbol{k}}^{\text{drf}}\left(t\right),\label{rho-ini_fil_gen0-1}
\end{equation}
where
\begin{equation}
\varrho_{\boldsymbol{k}}^{\text{ini}}\left(t\right)=U_{\boldsymbol{k}}\left(t,t_{0}\right)\varrho_{\boldsymbol{k}}^{\text{eq}}U_{\boldsymbol{k}}\left(t_{0},t\right)P\left(\boldsymbol{k}_{t},\boldsymbol{k}\right).\label{rho-ini_gen0-1}
\end{equation}
and
\begin{equation}
\varrho_{\boldsymbol{k}}^{\text{drf}}\left(t\right)=\int_{t_{0}}^{t}\text{d}t^{\prime}\frac{P\left(\boldsymbol{k}_{t},\boldsymbol{k}_{t^{\prime}}\right)}{\tau_{\boldsymbol{k}_{t^{\prime}}}}U_{\boldsymbol{k}}\left(t,t^{\prime}\right)\bar{\varrho}_{\boldsymbol{k}}\left(t^{\prime}\right)U_{\boldsymbol{k}}\left(t^{\prime},t\right).\label{rho-fil_gen0-1}
\end{equation}
\end{subequations} If we let $\tau_{\boldsymbol{k}}\rightarrow\infty$,
then $\int_{t_{0}}^{t}\text{d}t^{\prime}\bar{\varrho}_{\boldsymbol{k}}^{U}\left(t,t^{\prime}\right)P\left(\boldsymbol{k}_{t},\boldsymbol{k}_{t^{\prime}}\right)/\tau_{\boldsymbol{k}_{t^{\prime}}}\rightarrow0$,
$P\left(\boldsymbol{k}_{t},\boldsymbol{k}\right)\rightarrow1$, and
by the use of $\bar{\varrho}_{\boldsymbol{k}}\left(t_{0}\right)=\varrho_{\boldsymbol{k}}^{\text{eq}}$,
we have $\varrho_{\boldsymbol{k}}\left(t\right)\rightarrow\varrho_{\boldsymbol{k}}^{\text{id}}\left(t\right)$.
So assuming the relaxation times to be infinite reduces the time-dependent
density matrix of Eq. (\ref{incoh-rho-1}) to the intrinsic limit
given by Eq. (\ref{ideal-dmtx}) in the main text (see also Eqs. (\ref{eq_1RDM-def})
and (\ref{intrinsic-stat_nk}) there). 

\subsection{Reproduction of Eq. (\ref{Jad-tdep0}) of the main text and second-order
responses under periodic drivings}

\label{nlqh}

We should verify Eq.~(\ref{incoh-rho-1}) by substituting it to Eq.~(\ref{j-def0})
in the main text to obtain the current for arbitrary time-dependent
electric field $\boldsymbol{E}\left(t\right)$ and compare it to the
known results in the literatures. We first show that under the adiabatic
condition, such current is reduced to Eq. (\ref{Jad-tdep0}) of the
main text, which has been applied to analyse non-perturbative intra-band
effects (see references below Eq. (\ref{Jad-tdep0}) of the main text).
If we further restrict the time-dependent electric fields to be time-periodic
at times far away from $t_{0}$ and consider the current to the second
order in the electric field, then it gives the Berry curvature dipole
that serves as the basis of quantum nonlinear Hall effect \cite{Sodemann15216806}.

Under the adiabatic condition, the weak applied field only weakly
hybridises the bands. The index $i$ for labelling a hybridised band
can thus be reduced to the label of an original band $n$, namely,
$\left\vert \mathfrak{u}_{n}\left(\boldsymbol{k}_{t}\right)\right\rangle \approx e^{i\gamma_{n}\left(\boldsymbol{k}_{t}\right)}\left\vert u_{n}\left(\boldsymbol{k}_{t}\right)\right\rangle +\left\vert \delta u_{n}\left(\boldsymbol{k}_{t}\right)\right\rangle $,
where $\left\vert \delta u_{n}\left(\boldsymbol{k}_{t}\right)\right\rangle $
is the perturbation to the first order in the electric field while
to the same order $\mathcal{E}_{n}\left(\boldsymbol{k}_{t}\right)\approx\varepsilon_{n}\left(\boldsymbol{k}_{t}\right)$
so $g_{i}^{0}\left(\boldsymbol{k}_{t}\right)\approx f_{n}^{0}\left(\boldsymbol{k}_{t}\right)$
reducing the first term of Eq.~(\ref{incoh-rho-1sum}) to 
\begin{equation}
\bar{\varrho}_{\boldsymbol{k}}\left(t\right)=\sum_{n}f_{n}^{0}\left(\boldsymbol{k}_{t}\right)\left\vert \mathfrak{u}_{n}\left(\boldsymbol{k}_{t}\right)\right\rangle \left\langle \mathfrak{u}_{n}\left(\boldsymbol{k}_{t}\right)\right\vert .\label{rho-bar-ad}
\end{equation}
With the adiabatic condition imposed to the dynamics of a single wave
packet, namely, $\left\vert \mathfrak{u}_{n}\left(\boldsymbol{k}_{t}\right)\right\rangle =U_{\boldsymbol{k}}\left(t,t^{\prime}\right)\left\vert \mathfrak{u}_{n}\left(\boldsymbol{k}_{t^{\prime}}\right)\right\rangle $,
rendering $\bar{\varrho}_{\boldsymbol{k}}^{U}\left(t,t^{\prime}\right)=\sum_{n}\left\vert \mathfrak{u}_{n}\left(\boldsymbol{k}_{t}\right)\right\rangle \left\langle \mathfrak{u}_{n}\left(\boldsymbol{k}_{t}\right)\right\vert f_{n}^{0}\left(\boldsymbol{k}_{t^{\prime}}\right)$
(see Eq. (\ref{incoh-rho-1evm})), the second term of Eq.~(\ref{incoh-rho-1sum})
becomes 
\begin{align}
 & \delta\varrho_{\boldsymbol{k}}\left(t\right)=-\sum_{n}\left\vert \mathfrak{u}_{n}\left(\boldsymbol{k}_{t}\right)\right\rangle \left\langle \mathfrak{u}_{n}\left(\boldsymbol{k}_{t}\right)\right\vert \delta{f}_{n\boldsymbol{k}}^{\left(1\right)}\left(t\right)\label{rho-del-ad}
\end{align}
where 
\begin{align}
\delta{f}_{n\boldsymbol{k}}^{\left(1\right)}\left(t\right)=\int_{t_{0}}^{t}\text{d}t^{\prime}P\left(\boldsymbol{k}_{t},\boldsymbol{k}_{t^{\prime}}\right)\frac{\partial f_{n}^{0}\left(\boldsymbol{k}_{t^{\prime}}\right)}{\partial\boldsymbol{k}_{t^{\prime}}}\cdot\left(\frac{-e}{\hbar}\right)\boldsymbol{E}\left(t^{\prime}\right),\label{delfm-1E}
\end{align}
in which the term $\left(\partial f_{n}^{0}\left(\boldsymbol{k}_{t^{\prime}}\right)/\partial\boldsymbol{k}_{t^{\prime}}\right)\cdot\left(-e/\hbar\right)\boldsymbol{E}\left(t^{\prime}\right)$
comes from $\text{d}f_{n}^{0}\left(\boldsymbol{k}_{t^{\prime}}\right)/\text{d}t^{\prime}$.
The expectation values of the velocity operator (see its definition
in Eq.~(\ref{vopt-def}) of the main text) on the weakly hybridised
bands are given by, 
\begin{align}
\left\langle \mathfrak{u}_{n}\left(\boldsymbol{k}_{t}\right)\right\vert \boldsymbol{V}\left(\boldsymbol{k}_{t}\right)\left\vert \mathfrak{u}_{n}\left(\boldsymbol{k}_{t}\right)\right\rangle =\boldsymbol{v}_{n}^{b}\left(\boldsymbol{k}_{t}\right)+\boldsymbol{v}_{n}^{r}\left(\boldsymbol{k}_{t},\boldsymbol{E}\left(t\right)\right),\label{adia-velo}
\end{align}
in which we have explained the meanings of $\boldsymbol{v}_{n}^{b}\left(\boldsymbol{k}_{t}\right)$
and $\boldsymbol{v}_{n}^{r}\left(\boldsymbol{k}_{t},\boldsymbol{E}\left(t\right)\right)$
in the main text below Eq. (\ref{nox-EOMx-ad1}) and one immediately
identifies $\dot{\boldsymbol{x}}_{t}^{n}=\left\langle \mathfrak{u}_{n}\left(\boldsymbol{k}_{t}\right)\right\vert \boldsymbol{V}\left(\boldsymbol{k}_{t}\right)\left\vert \mathfrak{u}_{n}\left(\boldsymbol{k}_{t}\right)\right\rangle $.
Taking the first line of Eq.~(\ref{incoh-rho-1}) and putting it
into the definition Eq.~(\ref{j-def0}) in the main text, one obtains
$\boldsymbol{j}\left(t\right)=\bar{\boldsymbol{j}}\left(t\right)+\delta\boldsymbol{j}\left(t\right)$
in which $\bar{\boldsymbol{j}}\left(t\right)=-e\int\text{d}^{D}\boldsymbol{k}\text{Tr}\left(\boldsymbol{V}\left(\boldsymbol{k}_{t}\right)\bar{\varrho}_{\boldsymbol{k}}\left(t\right)\right)$
and $\delta{\boldsymbol{j}}\left(t\right)=-e\int\text{d}^{D}\boldsymbol{k}\text{Tr}\left(\boldsymbol{V}\left(\boldsymbol{k}_{t}\right)\delta{\varrho}_{\boldsymbol{k}}\left(t\right)\right)$.
Specifying $\bar{\varrho}_{\boldsymbol{k}}\left(t\right)$ and $\delta{\varrho}_{\boldsymbol{k}}\left(t\right)$
by Eqs. (\ref{rho-bar-ad}) and (\ref{rho-del-ad}) respectively together
with the use of Eq. (\ref{adia-velo}) and focusing on one band, one
then straightforwardly obtains Eq. (\ref{Jad-tdep0}) of the main
text.

We now turn to the second-order current. Note that $\bar{\boldsymbol{j}}\left(t\right)$
is only to the first order of $\boldsymbol{E}\left(t\right)$ which
gives the usual linear-response Hall current well reviewed in Ref.~\cite{Xiao101959}.
The second-order response then must be contained in $\delta{\boldsymbol{j}}\left(t\right)$.
One therefore has $\delta{\boldsymbol{j}}\left(t\right)=\delta{\boldsymbol{j}}^{\left(1\right)}\left(t\right)+\delta{\boldsymbol{j}}^{\left(2\right)}\left(t\right)$
where $\delta{\boldsymbol{j}}^{\left(1\right)}\left(t\right)=-e\sum_{n}\int\text{d}^{D}\boldsymbol{k}\boldsymbol{v}_{n}^{b}\left(\boldsymbol{k}_{t}\right)\delta{f}_{n\boldsymbol{k}}^{\left(1\right)}\left(t\right)$
is to the first order in the electric field and $\delta{\boldsymbol{j}}^{\left(2\right)}\left(t\right)=-e\sum_{n}\int\text{d}^{D}\boldsymbol{k}\boldsymbol{v}_{n}^{a}\left(\boldsymbol{k}_{t}\right)\delta{f}_{n\boldsymbol{k}}^{\left(1\right)}\left(t\right)$
accounts for the second-order response. We further assume that $\tau_{\boldsymbol{k}^{\prime}}=\tau$
(single-relaxation time) for all possible $\boldsymbol{k}^{\prime}$.
We also assume that $\tau$ is short so that Eq.~(\ref{short-RT})
can be applied. We only apply Eq.~(\ref{short-RT}) to the part $P\left(\boldsymbol{k}_{t},\boldsymbol{k}_{t^{\prime}}\right)\left(\partial f_{n}^{0}\left(\boldsymbol{k}_{t^{\prime}}\right)/\partial\boldsymbol{k}_{t^{\prime}}\right)$
of the integrand in the integral of Eq.~(\ref{delfm-1E}), while
the part of the integrand given by $\boldsymbol{E}\left(t^{\prime}\right)$
is left intact \cite{footnote-partshortRlxT}. We now impose the time-periodic
driving $\boldsymbol{E}\left(t^{\prime}\right)=\text{Re}\left\{ \tilde{\boldsymbol{E}}e^{i\omega t^{\prime}}\right\} $
where $\tilde{\boldsymbol{E}}$ is the time-independent complex electric-field
amplitude and take the limit $t_{0}\rightarrow-\infty$, we then end
up with 
\begin{align}
\delta{\boldsymbol{j}}^{\left(2\right)}\left(t\right) & =-\sum_{n\alpha}\int\text{d}^{D}\boldsymbol{k}f_{n}^{0}\left(\boldsymbol{k}\right)\frac{\partial\boldsymbol{\Omega}_{n}\left(\boldsymbol{k}\right)}{\partial k_{\alpha}}\times\text{Re}\left\{ \frac{\tau e^{3}}{2\hbar^{2}\left(1+i\omega\tau\right)}\left(\tilde{E}_{\alpha}\boldsymbol{\tilde{E}}^{*}+e^{2i\omega t}\tilde{E}_{\alpha}\boldsymbol{\tilde{E}}\right)\right\} ,\label{j2qnh}
\end{align}
where the index $\alpha$ runs over the spatial components and in
the intermediate procedure, we have identified $\text{d}^{D}\boldsymbol{k}$
with $\text{d}^{D}\boldsymbol{k}_{t}$ and afterwards rewritten $\boldsymbol{k}_{t}$
to $\boldsymbol{k}$ for the integral. Here the Berry curvature $\boldsymbol{\Omega}_{n}\left(\boldsymbol{k}\right)$
is given by Eq. (\ref{AbelianBrcv-defMk}) of the main text and the
Berry curvature dipole is given by the components of $\partial\boldsymbol{\Omega}_{n}\left(\boldsymbol{k}\right)/\partial k_{\alpha}$.
By focusing on one active band $n$, Eq.~(\ref{j2qnh}) then reproduces
the second-order response describing the quantum nonlinear Hall effect
in accordance with Ref.~\cite{Sodemann15216806}. 

\subsection{The post-pulse current rates }

\label{SI-wpt-deriv-ppcrntrt}

In the above, we have derived the main formulation of NADT-WPT developed
here as Eq. (\ref{incoh-rho-1}) of the main text and showed its reproductions
of well known results of the AD-WPT. We now discuss the non-adiabatic
transient consequences that one can find with such NADT-WPT. We first
analyse the current rate $\dot{\boldsymbol{j}}\left(t\right)$ induced
by a pulse of finite time width (subjecting to Eq. (\ref{pulse-Eft})
of the main text) whose relation to manifesting geometric properties
of the bands has been discussed in the main text Sec. \ref{WPT-basics-QnAccHistory}.
Taking the time derivative of Eq.~(\ref{j-def0}) gives \begin{subequations}\label{djovdt-def}
\begin{equation}
\dot{j}_{\alpha}\left(t\right)=\dot{j}_{\alpha}^{drv}\left(t\right)+\dot{j}_{\alpha}^{rp}\left(t\right),\label{djovdt-decomp1}
\end{equation}
where
\begin{equation}
\dot{j}_{\alpha}^{drv}\left(t\right)=\frac{\left(-e\right)^{2}}{\hbar}\boldsymbol{E}\left(t\right)\cdot\int\text{d}^{D}\boldsymbol{k}\text{Tr}\left(\frac{\partial V_{\alpha}\left(\boldsymbol{k}_{t}\right)}{\partial\boldsymbol{k}_{t}}\varrho_{\boldsymbol{k}}\left(t\right)\right),\label{djovdt-Kcomp}
\end{equation}
and
\begin{equation}
\dot{j}_{\alpha}^{rp}\left(t\right)=-e\int\text{d}^{D}\boldsymbol{k}\text{Tr}\left(V_{\alpha}\left(\boldsymbol{k}_{t}\right)\dot{\varrho}_{\boldsymbol{k}}\left(t\right)\right),\label{djovdt-gcomp}
\end{equation}
\end{subequations} in which $\dot{j}_{\alpha}\left(t\right)$ and
$V_{\alpha}\left(\boldsymbol{k}_{t}\right)$ are the $\alpha$th spatial
component of $\dot{\boldsymbol{j}}\left(t\right)$ and $\boldsymbol{V}\left(\boldsymbol{k}_{t}\right)$
respectively. Here the superscript ``$drv$'' for $\dot{\boldsymbol{j}}^{drv}\left(t\right)$
indicates that this component instantaneously reflects the driving
field $\boldsymbol{E}\left(t\right)$ and the superscript ``$rp$''
for $\dot{\boldsymbol{j}}^{rp}\left(t\right)$ means that this component
behaves in response to the driving field. Noticeably from setting
$\boldsymbol{E}\left(t>t_{\text{off}}\right)=0$ to Eq. (\ref{djovdt-def}),
we see $\dot{\boldsymbol{j}}^{drv}\left(t>t_{\text{off}}\right)=0$.
The post-pulse current rate then is only contributed by $\dot{\boldsymbol{j}}\left(t>t_{\text{off}}\right)=\dot{\boldsymbol{j}}^{rp}\left(t\right)$. 

We first clarify the intrinsic limit in which the scattering rates
are zero $\tau_{\boldsymbol{k}}^{-1}\rightarrow0$ such that $\varrho_{\boldsymbol{k}}\left(t\right)\rightarrow\varrho_{\boldsymbol{k}}^{id}\left(t\right)$
(see Eq. (\ref{ideal-dmtx}) of the main text). The extrinsic effects
due to non-vanishing scattering rates will be discussed later. Taking
the intrinsic limit of Eq. (\ref{incoh-rho-1}) leads to $\dot{\varrho}_{\boldsymbol{k}}\left(t\right)=\left(-i/\hbar\right)\left[\mathcal{H}\left(\boldsymbol{k}_{t}\right),\varrho_{\boldsymbol{k}}\left(t\right)\right]$
which in turn casts Eq. (\ref{djovdt-gcomp}) to $\dot{\boldsymbol{j}}^{rp}\left(t\right)=\dot{\boldsymbol{j}}^{geo}\left(t\right)$
as Eq. (\ref{jdt-qcc}) of the main text, where quantum acceleration
$\vec{\mathbb{A}}\left(\boldsymbol{k}_{t}\right)$, other than the
anomalous velocity, appears and dominates the manifestation of geometric
effects. We now explain why the manifestation of geometric properties
of bands in the current rate after switching off the laser pulse is
exclusive to non-adiabatic drivings but not adiabatic drivings during
the in-pulse processes. 

\subsubsection{Exclusive effects of non-adiabatic drivings on post-pulse wave-packet
dynamics}

\label{SI-wpt-deriv-ppcrntrt-postpulse}

For adiabatic dynamics of a single wave packet, the expectation value
$\left\langle \phi_{n,\boldsymbol{k}}^{ad}\left(t\right)\right\vert O\left\vert \phi_{n,\boldsymbol{k}}^{ad}\left(t\right)\right\rangle $
of any observable $O$ at any moment of time $t$ depends on $t$
only through $\boldsymbol{k}_{t}$. As the driving ceases for $t>t_{\text{off}}$,
Eq. (\ref{driving-E1}) of the main text dictates that $\boldsymbol{k}_{t}$
no longer changes with time, namely, $\left.\boldsymbol{k}_{t}\right\vert _{t>t_{\text{off}}}=\boldsymbol{k}+\boldsymbol{\mathcal{K}}_{t_{\text{off}}}\equiv\boldsymbol{k}_{t_{\text{off}}}$.
The expectation value also stops to change with time so its time derivative
remains zero after the driving is switched off which corresponds to
$\dot{\boldsymbol{j}}\left(t>t_{\text{off}}\right)=0$ in the above
discussion. 

In contrast, non-adiabatically, the time-dependent expectation value
$\left\langle \phi_{n,\boldsymbol{k}}\left(t,t_{0}\right)\right\vert O\left\vert \phi_{n,\boldsymbol{k}}\left(t,t_{0}\right)\right\rangle $
of an arbitrary observable $O$ depends on $t$ through the whole
driving history up to time $t$ more than just $\boldsymbol{k}_{t}$
at that moment $t$. As the non-adiabatic drivings enable $\left\vert \phi_{n,\boldsymbol{k}}\left(t_{\text{off}},t_{0}\right)\right\rangle $
to acquire non-negligible occupations on bands other than the starting
band $n$, the state $\left\vert \phi_{n,\boldsymbol{k}}\left(t_{\text{off}},t_{0}\right)\right\rangle $
becomes a driving-history dependent superposition of eigenstates of
$\mathcal{H}\left(\boldsymbol{k}_{t_{\text{off}}}\right)$ as a genuine
non-adiabatic multiple-band effect. Consequently, as the wave packet
continues to evolve to the post-pulse time region as $\left\vert \phi_{n,\boldsymbol{k}}\left(t>t_{\text{off}},t_{0}\right)\right\rangle =\exp\left\{ -\left(i/\hbar\right)\left(t-t_{\text{off}}\right)\mathcal{H}\left(\boldsymbol{k}_{t_{\text{off}}}\right)\right\} \left\vert \phi_{n,\boldsymbol{k}}\left(t_{\text{off}},t_{0}\right)\right\rangle $,
the time derivative of the observables, e.g., the quantum acceleration,
$\vec{\mathbb{A}}\left(\boldsymbol{k}_{t}\right)$, for $t>t_{\text{off}}$
does not vanish. 

Explicitly, such non-adiabatic effects from transient drivings $\boldsymbol{E}\left(t_{0}\le t\le t_{\text{off}}\right)\ne0$
can be explicated in the post-pulse time region $t>t_{\text{off}}$
in which $\boldsymbol{E}\left(t>t_{\text{off}}\right)=0$ from two
complementary views. First, with a fixed pulse whose peak amplitude
is large enough to activate non-adiabatic regime, two different band
structures with the same band gaps but differentiated only by their
geometric connections can leave such geometric differences imprinted
to the band-pseudospin state $\left\vert \phi_{n,\boldsymbol{k}}\left(t_{\text{off}},t_{0}\right)\right\rangle $
at the end of the pulse and manifested through the driving-free dynamics
governed by $\mathcal{H}\left(\boldsymbol{k}_{t_{\text{off}}}\right)$
in the post-pulse time region. Second, two different pulses are expected
to induce different in-pulse dynamics as usual. Therefore, the post-pulse
differences due to different pulses applied before are then evidences
of history dependencies accumulated to $\left\vert \phi_{n,\boldsymbol{k}}\left(t_{\text{off}},t_{0}\right)\right\rangle $
by non-adiabatic in-pulse drivings. These microscopic effects on the
dynamics of the band pseudospin of a wave packet is then manifested
in $\dot{\boldsymbol{j}}\left(t>t_{\text{off}}\right)$ as macroscopic
observable through the quantum acceleration mechanism as discussed
around Eqs. (\ref{jdt-qcc}) and (\ref{jdt-qcc-bndrep}) in the main
text.

\subsubsection{Inclusion of extrinsic scattering effects}

\label{SI-wpt-deriv-ppcrntrt-extrinsic}

We now conclude the general discussion of post-pulse current rate
by taking into account the extrinsic scattering effects. In the post-pulse
time region, we always have $\dot{\boldsymbol{j}}^{drv}\left(t>t_{\text{off}}\right)=0$
due to its direct proportionality to the electric field at time $t$
as Eq. (\ref{djovdt-Kcomp}) regardless of whether the scattering
effects are ignored. For convenience of analysing the post-pulse current,
instead of using Eq. (\ref{incoh-rho-1}) for the density matrix,
here we substitute its alternative expression Eq. (\ref{jrho-ini-drf})
to Eq. (\ref{djovdt-gcomp}) and obtain

\begin{equation}
\dot{\boldsymbol{j}}^{rp}\left(t\right)=\dot{\boldsymbol{j}}^{rp,ini}\left(t\right)+\dot{\boldsymbol{j}}^{rp,drf}\left(t\right),\label{jdot-init-drf}
\end{equation}
where $\dot{j}_{\alpha}^{rp,ini}\left(t\right)=-e\int\text{d}^{D}\boldsymbol{k}\text{Tr}\left(V_{\alpha}\left(\boldsymbol{k}_{t}\right)\text{d}\varrho_{\boldsymbol{k}}^{\text{ini}}\left(t\right)/\text{d}t\right)$
and $\dot{j}_{\alpha}^{rp,drf}\left(t\right)=-e\int\text{d}^{D}\boldsymbol{k}\text{Tr}\left(V_{\alpha}\left(\boldsymbol{k}_{t}\right)\text{d}\varrho_{\boldsymbol{k}}^{\text{drf}}\left(t\right)/\text{d}t\right)$
are the $\alpha$th component of the vectors $\dot{\boldsymbol{j}}^{rp,ini}\left(t\right)$
and $\dot{\boldsymbol{j}}^{rp,drf}\left(t\right)$ respectively. More
explicitly, \begin{subequations}\label{jdot-rp-ini}
\begin{equation}
\dot{\boldsymbol{j}}^{rp,ini}\left(t\right)=\dot{\boldsymbol{j}}_{vel}^{rp,ini}\left(t\right)+\dot{\boldsymbol{j}}_{acc}^{rp,ini}\left(t\right),\label{jdot-rp-ini-sum}
\end{equation}
where
\begin{equation}
\dot{\boldsymbol{j}}_{vel}^{rp,ini}\left(t\right)=-e\int\text{d}^{D}\boldsymbol{k}\left[-\frac{P\left(\boldsymbol{k}_{t},\boldsymbol{k}\right)}{\tau_{\boldsymbol{k}}}\right]\text{Tr}\left(\boldsymbol{V}\left(\boldsymbol{k}_{t}\right)\varrho_{\boldsymbol{k}}^{id}\left(t\right)\right),\label{jdot-rp-ini-loss}
\end{equation}
\begin{equation}
\dot{\boldsymbol{j}}_{acc}^{rp,ini}\left(t\right)=-e\int\text{d}^{D}\boldsymbol{k}P\left(\boldsymbol{k}_{t},\boldsymbol{k}\right)\text{Tr}\left(\vec{\mathbb{A}}\left(\boldsymbol{k}_{t}\right)\varrho_{\boldsymbol{k}}^{id}\left(t\right)\right),\label{jdot-rp-ini-acc}
\end{equation}
\end{subequations} in which $\varrho_{\boldsymbol{k}}^{id}\left(t\right)$
is found from Eqs. (\ref{ideal-dmtx}) and (\ref{intrinsic-stat_nk}),
$\boldsymbol{V}\left(\boldsymbol{k}_{t}\right)$ is defined in Eq.
(\ref{vopt-def}) and $\vec{\mathbb{A}}\left(\boldsymbol{k}_{t}\right)$
is still the quantum acceleration defined in Eq. (\ref{q-acc-opt-def})
in the main text respectively.

To take the time derivative of $\varrho_{\boldsymbol{k}}^{\text{drf}}\left(t\right)$
to obtain $\dot{\boldsymbol{j}}^{rp,drf}\left(t\right)$, we first
note that there exists a time scale $\tau_{m}\gtrsim\text{max}\left\{ \tau_{\boldsymbol{k}}\right\} _{\boldsymbol{k}}$
such that $P\left(\boldsymbol{k}_{t},\boldsymbol{k}_{t^{\prime}}\right)\approx0$
for $t-t^{\prime}\ge\tau_{m}$ so the integral $\int_{t_{0}}^{t}\text{d}t^{\prime}$
in Eq. (\ref{rho-fil_gen0-1}) is only contributed by its integrand
at $t^{\prime}$ that satisfies $t-\tau_{m}\le t^{\prime}\le t$.
By further realising that $\bar{\varrho}_{\boldsymbol{k}}\left(t^{\prime}>t_{\text{off}}\right)=\varrho_{\boldsymbol{k}_{t_{\text{off}}}}^{eq}$
and $U_{\boldsymbol{k}}\left(t,t^{\prime}>t_{\text{off}}\right)=\exp\left\{ -\left(i/\hbar\right)\left(t-t^{\prime}\right)\mathcal{H}\left(\boldsymbol{k}_{t_{\text{off}}}\right)\right\} $
so $U_{\boldsymbol{k}}\left(t,t^{\prime}\right)\bar{\varrho}_{\boldsymbol{k}}\left(t^{\prime}\right)=\bar{\varrho}_{\boldsymbol{k}}\left(t^{\prime}\right)U_{\boldsymbol{k}}\left(t,t^{\prime}\right)$
for $t\ge t^{\prime}>t_{\text{off}}$ in Eq. (\ref{rho-fil_gen0-1}),
it becomes 
\begin{equation}
\varrho_{\boldsymbol{k}}^{\text{drf}}\left(t>t_{\text{off}}+\tau_{m}\right)\approx\left[\int_{t-\tau_{m}}^{t}\text{d}t^{\prime}\frac{P\left(\boldsymbol{k}_{t},\boldsymbol{k}_{t^{\prime}}\right)}{\tau_{\boldsymbol{k}_{t^{\prime}}}}\right]\varrho_{\boldsymbol{k}_{t_{\text{off}}}}^{eq}.\label{rho-drf-off}
\end{equation}
The integral over $t^{\prime}$ is found to be $\int_{t-\tau_{m}}^{t}\text{d}t^{\prime}P\left(\boldsymbol{k}_{t},\boldsymbol{k}_{t^{\prime}}\right)/\tau_{\boldsymbol{k}_{t^{\prime}}}=\int_{t-\tau_{m}}^{t}\text{d}t^{\prime}\partial P\left(\boldsymbol{k}_{t},\boldsymbol{k}_{t^{\prime}}\right)/\partial t^{\prime}=1-P\left(\boldsymbol{k}_{t},\boldsymbol{k}_{t-\tau_{m}}\right)$
by the use of integration by part. Noting for $t\ge t^{\prime}>t_{\text{off}}$,
one has $P\left(\boldsymbol{k}_{t},\boldsymbol{k}_{t^{\prime}}\right)=\exp\left\{ -\int_{t^{\prime}}^{t}\text{d}t^{\prime\prime}\tau_{\boldsymbol{k}_{t^{\prime\prime}}}^{-1}\right\} =\exp\left\{ -\int_{t^{\prime}}^{t}\text{d}t^{\prime\prime}\tau_{\boldsymbol{k}_{t_{\text{off}}}}^{-1}\right\} =\exp\left\{ -\left(t-t^{\prime}\right)/\tau_{\boldsymbol{k}_{t_{\text{off}}}}\right\} $.
Therefore, $P\left(\boldsymbol{k}_{t},\boldsymbol{k}_{t-\tau_{m}}\right)=\exp\left\{ -\tau_{m}/\tau_{\boldsymbol{k}_{t_{\text{off}}}}\right\} $
which then gives $\varrho_{\boldsymbol{k}}^{\text{drf}}\left(t>t_{\text{off}}+\tau_{m}\right)\approx\left[1-\exp\left\{ -\tau_{m}/\tau_{\boldsymbol{k}_{t_{\text{off}}}}\right\} \right]\varrho_{\boldsymbol{k}_{t_{\text{off}}}}^{eq}$
leading to $\left.\text{d}\varrho_{\boldsymbol{k}}^{\text{drf}}\left(t\right)/\text{d}t\right\vert _{t>t_{\text{off}}+\tau_{m}}=0$
and $\dot{\boldsymbol{j}}^{rp,drf}\left(t>t_{\text{off}}+\tau_{m}\right)=0$.
Consequently, by Eq. (\ref{jdot-init-drf}) with $\dot{\boldsymbol{j}}^{rp,drf}\left(t>t_{\text{off}}+\tau_{m}\right)=0$,
we obtain 
\begin{equation}
\dot{\boldsymbol{j}}\left(t>t_{\text{off}}+\tau_{m}\right)\approx\dot{\boldsymbol{j}}^{rp,ini}\left(t\right),\label{crnt-rt-extsM}
\end{equation}
where $\dot{\boldsymbol{j}}^{rp,ini}\left(t\right)$ has been found
as Eq. (\ref{jdot-rp-ini}). Comparing Eq. (\ref{jdot-rp-ini-acc})
for $\dot{\boldsymbol{j}}_{acc}^{rp,ini}\left(t\right)$ with Eq.
(\ref{jdt-qcc}) of the main text for $\dot{\boldsymbol{j}}^{geo}\left(t\right)$,
we see that the post-pulse current rate $\dot{\boldsymbol{j}}\left(t>t_{\text{off}}+\tau_{m}\right)$
given by Eq. (\ref{crnt-rt-extsM}) still retain the manifestation
of geometric effects through the quantum acceleration, albeit it is
damped by the factor $0\le P\left(\boldsymbol{k}_{t},\boldsymbol{k}\right)<1$
due to scattering. If adiabatic drivings had been applied in the time
region $t_{0}\le t\le t_{\text{off}}$, then $\dot{\boldsymbol{j}}_{acc}^{rp,ini}\left(t>t_{\text{off}}\right)=0$
by the reason that makes $\dot{\boldsymbol{j}}^{geo}\left(t\right)=0$
pointed out above. For longer relaxation times, both the damping factor
$P\left(\boldsymbol{k}_{t},\boldsymbol{k}\right)$ and the contamination
of this manifestation by $\dot{\boldsymbol{j}}_{vel}^{rp,ini}\left(t\right)$
can be reduced. 

\section{Wave-packet spin dynamics and SOC-geometric effects}

\label{SI-spin-basics}

In Secs. \ref{WPT-basics-spntxt} and \ref{sec-examples-SOC-isc}
of the main text, we have used the concept that driving-induced intrinsic
spin coherence represents SOC geometric effects to illustrate the
prospects of macroscopically manifesting microscopic SOC-geometric
effects through CISP and spin-mediated part of the photocurrent. Here
in Sec. \ref{SI-spin-basics-txtcoh}, we explain this basic concept
from the wave packet point of view. With this conceptual picture in
mind, we then continue in Sec. \ref{SI-spin-basics-ppdys} to study
characteristics of driven spin-texture dynamics in the post-pulse
time region. This also explicates the first abstract point about transient
non-adiabatic manifestation of band-geometric effects discussed in
Sec. \ref{WPT-basics-QnAccHistory} of the main text that relates
such effects embedded in the quantum acceleration to the post-pulse
current rate. In Sec. \ref{SI-spin-basics-adiaCISP}, we recall the
well-established AD-WPT applied to CISP effects to provide relevant
reference results to be compared with non-adiabatic CISP effects addressed
in Secs. \ref{WPT-basics-spntxt} and \ref{sec-examples-Rashba} of
the main text.

\subsection{Spin geometric connections for intrinsic spin coherences}

\label{SI-spin-basics-txtcoh}

We now inspect dynamics of the spin DOF of the wave packet under external
drivings with the general Hamiltonian given by Eq. (\ref{SOC-H-gDef})
of the main text. Let us consider a driving that pushes the CM of
the wave packet in the BZ from $\boldsymbol{k}_{t_{0}}=\boldsymbol{k}$
at time $t_{0}$ to $\boldsymbol{k}_{t}$ at time $t$ along a path
$\Gamma_{t_{0}}^{t}$. We have introduced in Sec. \ref{WPT-basics-spntxt}
of the main text the driven spin texture $\left\langle \boldsymbol{\sigma}\right\rangle _{n}\left(\boldsymbol{k},t\right)$
which includes effects of drivings so that $\left\langle \boldsymbol{\sigma}\right\rangle _{n}\left(\boldsymbol{k},t\right)=\left\langle \phi_{n,\boldsymbol{k}}\left(t,t_{0}\right)\right\vert \boldsymbol{\sigma}\left\vert \phi_{n,\boldsymbol{k}}\left(t,t_{0}\right)\right\rangle $
is expected to deviate from $\left\langle \boldsymbol{\sigma}\right\rangle _{n}^{0}\left(\boldsymbol{k}_{t}\right)$
(the native spin texture, defined by Eq. (\ref{spintxt-native-def})
in the main text). Consequently, the spin DOF of the wave packet described
by the state $\left\vert \phi_{n,\boldsymbol{k}}\left(t,t_{0}\right)\right\rangle $
becomes a coherent superposition between $\left\vert u_{-}\left(\boldsymbol{k}_{t}\right)\right\rangle $
and $\left\vert u_{+}\left(\boldsymbol{k}_{t}\right)\right\rangle $
showing intrinsic spin coherence. The native spin texture along the
path $\Gamma_{t_{0}}^{t}$ determines how such driving induction of
the intrinsic spin coherence can be incurred. Below we explain this
for two opposite situations, namely, $\partial\left\langle \boldsymbol{\sigma}\right\rangle _{n}^{0}\left(\boldsymbol{k}_{t^{\prime}}\right)/\partial\boldsymbol{k}_{t^{\prime}}\ne0$
and $\partial\left\langle \boldsymbol{\sigma}\right\rangle _{n}^{0}\left(\boldsymbol{k}_{t^{\prime}}\right)/\partial\boldsymbol{k}_{t^{\prime}}=0$
for $t_{0}\le t^{\prime}\le t$.

We first consider a path that fulfils $\partial\left\langle \boldsymbol{\sigma}\right\rangle _{n}^{0}\left(\boldsymbol{k}_{t^{\prime}}\right)/\partial\boldsymbol{k}_{t^{\prime}}\ne0$
for $t_{0}\le t^{\prime}\le t$ so that $\left[\mathcal{\mathcal{R}}_{\boldsymbol{k}_{t^{\prime}}}\right]_{n,\bar{n}}\ne0$
(indicated by Eq. (\ref{sptxt-geoct-r1}) of the main text). For general
non-adiabatic drivings, we have $\left(\hbar/2\right)\left\langle \dot{\boldsymbol{\sigma}}\right\rangle _{n}\left(\boldsymbol{k},t\right)=\boldsymbol{\Lambda}_{so}\left(\boldsymbol{k}_{t}\right)\times\left\langle \boldsymbol{\sigma}\right\rangle _{n}\left(\boldsymbol{k},t\right)$.
This describes the spin precession dynamics in an effective time-dependent
magnetic field $\boldsymbol{\Lambda}_{so}\left(\boldsymbol{k}_{t}\right)$
whose orientation is determined by the native spin texture $\boldsymbol{\Lambda}_{so}\left(\boldsymbol{k}_{t}\right)/\left\vert \boldsymbol{\Lambda}_{so}\left(\boldsymbol{k}_{t}\right)\right\vert =\pm\left\langle \boldsymbol{\sigma}\right\rangle _{\pm}^{0}\left(\boldsymbol{k}_{t}\right)$.
The driving of the CM of the wave packet in the BZ along the path
$\Gamma_{t_{0}}^{t}$ characterised by $\partial\left\langle \boldsymbol{\sigma}\right\rangle _{n}^{0}\left(\boldsymbol{k}_{t^{\prime}}\right)/\partial\boldsymbol{k}_{t^{\prime}}\ne0$
for $\boldsymbol{k}_{t^{\prime}}\in\Gamma_{t_{0}}^{t}$ keeps the
orientation of the effective magnetic field $\boldsymbol{\Lambda}_{so}\left(\boldsymbol{k}_{t}\right)$
changing in time which causes the driven spin texture $\left\langle \boldsymbol{\sigma}\right\rangle _{n}\left(\boldsymbol{k},t\right)$
to deviate from $\left\langle \boldsymbol{\sigma}\right\rangle _{n}^{0}\left(\boldsymbol{k}_{t}\right)$.
This thus arrives a coherent superposition between $\left\vert u_{-}\left(\boldsymbol{k}_{t}\right)\right\rangle $
and $\left\vert u_{+}\left(\boldsymbol{k}_{t}\right)\right\rangle $
for the spin DOF of the wave packet. In case of adiabatic drivings,
the spin state of the wave packet is simplified to $\left\vert \phi_{n,\boldsymbol{k}}\left(t,t_{0}\right)\right\rangle \rightarrow\left\vert \phi_{n,\boldsymbol{k}}^{ad}\left(t\right)\right\rangle $
given by Eq. (\ref{n-ad-uni}) of the main text which obviously shows
spin coherence at $\boldsymbol{k}_{t}$ as superposition between $\left\vert u_{n}\left(\boldsymbol{k}_{t}\right)\right\rangle $
and $\left\vert u_{\bar{n}}\left(\boldsymbol{k}_{t}\right)\right\rangle $
because $r_{n,\bar{n}}\left(\boldsymbol{k}_{t},\boldsymbol{E}\left(t\right)\right)\ne0$
by $\left[\mathcal{\mathcal{R}}_{\boldsymbol{k}_{t}}\right]_{n,\bar{n}}\ne0$.
The deviation of $\left\langle \boldsymbol{\sigma}\right\rangle _{n}\left(\boldsymbol{k},t\right)$
from $\left\langle \boldsymbol{\sigma}\right\rangle _{n}^{0}\left(\boldsymbol{k}_{t}\right)$
is on the first order of the driving field $\boldsymbol{E}\left(t\right)$.
Therefore, regardless whether the driving is adiabatic or non-adiabatic,
driving-induced intrinsic spin coherence is attained due to the effects
of the non-vanishing inter-spin geometric connection $\left[\mathcal{\mathcal{R}}_{\boldsymbol{k}_{t^{\prime}}}\right]_{n,\bar{n}}\ne0$
associated with the native spin texture $\partial\left\langle \boldsymbol{\sigma}\right\rangle _{n}^{0}\left(\boldsymbol{k}_{t^{\prime}}\right)/\partial\boldsymbol{k}_{t^{\prime}}\ne0$.
Such driving-induced intrinsic spin coherence thus faithfully represents
the SOC-geometric effects.

We now consider the opposite case that the native spin texture along
the path $\Gamma_{t_{0}}^{t}$ instead fulfills $\partial\left\langle \boldsymbol{\sigma}\right\rangle _{n}^{0}\left(\boldsymbol{k}_{t^{\prime}}\right)/\partial\boldsymbol{k}_{t^{\prime}}=0$
for $t_{0}\le t^{\prime}\le t$. This implies that the direction of
$\boldsymbol{\Lambda}_{so}\left(\boldsymbol{k}_{t^{\prime}}\right)$
is a time constant given by $\left\langle \boldsymbol{\sigma}\right\rangle _{\pm}^{0}\left(\boldsymbol{k}_{t_{0}}=\boldsymbol{k}\right)$.
With the initial value of the driven spin texture set by $\left\langle \boldsymbol{\sigma}\right\rangle _{n}\left(\boldsymbol{k},t_{0}\right)=\left\langle \boldsymbol{\sigma}\right\rangle _{n}^{0}\left(\boldsymbol{k}\right)$,
we have $\left.\left\langle \dot{\boldsymbol{\sigma}}\right\rangle _{n}\left(\boldsymbol{k},t^{\prime}\right)\right\vert _{t^{\prime}=t_{0}}\propto\left\langle \boldsymbol{\sigma}\right\rangle _{n}^{0}\left(\boldsymbol{k}\right)\times\left\langle \boldsymbol{\sigma}\right\rangle _{n}\left(\boldsymbol{k},t_{0}\right)=0$
such that $\left\langle \boldsymbol{\sigma}\right\rangle _{n}\left(\boldsymbol{k},t_{0}+\delta t\right)=\left\langle \boldsymbol{\sigma}\right\rangle _{n}\left(\boldsymbol{k},t_{0}\right)$
for small enough $\delta t>0$. This can be iterated throughout the
path $\Gamma_{t_{0}}^{t}$ to see that $\left\langle \boldsymbol{\sigma}\right\rangle _{n}\left(\boldsymbol{k},t_{0}\le t^{\prime}\le t\right)=\left\langle \boldsymbol{\sigma}\right\rangle _{n}\left(\boldsymbol{k},t_{0}\right)=\left\langle \boldsymbol{\sigma}\right\rangle _{n}^{0}\left(\boldsymbol{k}\right)$.
Along such path on which the native spin texture fulfills $\partial\left\langle \boldsymbol{\sigma}\right\rangle _{n}^{0}\left(\boldsymbol{k}^{\prime}\right)/\partial\boldsymbol{k}^{\prime}=0$
for $\boldsymbol{k}^{\prime}\in\Gamma_{t_{0}}^{t}$, the CM of the
wave packet and its spin are essentially decoupled so the geometric
properties of the bands are not manifested in wave-packet dynamics
even if the inter-spin connections can be nonzero elsewhere. This
is why the persistent spin texture, characterised by $\partial\left\langle \boldsymbol{\sigma}\right\rangle _{n}^{0}\left(\boldsymbol{k}^{\prime}\right)/\partial\boldsymbol{k}^{\prime}=0$,
excludes itself from our interests. In contrast, for the Rashba-Dresselhaus
spin texture, one easily finds driving paths \textbf{$\Gamma_{t_{0}}^{t}$}
endowed with the geometric property $\partial\left\langle \boldsymbol{\sigma}\right\rangle _{n}^{0}\left(\boldsymbol{k}^{\prime}\right)/\partial\boldsymbol{k}^{\prime}\ne0$
for $\boldsymbol{k}^{\prime}\in\Gamma_{t_{0}}^{t}$ that allows driving-induced
intrinsic spin coherence.

\subsection{Post-pulse driven spin texture dynamics linked to current rate}

\label{SI-spin-basics-ppdys}

We have explained that the driven spin texture can show the intrinsic
spin coherence induced by drivings to manifest underlying SOC-geometric
effects. Such microscopic manifestation goes without distinguishing
non-adiabatic and adiabatic drivings. We have discussed in Sec. \ref{WPT-basics-QnAccHistory}
of the main text that the geometric manifestation of the non-adiabatic
drivings can be exclusively captured in the post-pulse time region
by the current rate as a macroscopic observable. We therefore should
link the microscopic driven spin texture to the macroscopic current
rate.

With the general Hamiltonian given by Eq. (\ref{SOC-H-gDef}) of the
main text, the quantum acceleration given by Eq. (\ref{jdt-qcc-bndrep})
of the main text reads 
\begin{equation}
\mathbb{A}_{\alpha}\left(\boldsymbol{k}_{t}\right)=2\hbar^{-2}\left[\frac{\partial\boldsymbol{\Lambda}_{so}\left(\boldsymbol{k}_{t}\right)}{\partial k_{\alpha t}}\times\boldsymbol{\Lambda}_{so}\left(\boldsymbol{k}_{t}\right)\right]\cdot\boldsymbol{\sigma},\label{qAccTwoBndfm}
\end{equation}
and consequently

\begin{equation}
\dot{j}_{\alpha}^{geo}\left(t\right)=2\left(-e\right)\hbar^{-2}\int\text{d}^{D}\boldsymbol{k}\left[\frac{\partial\boldsymbol{\Lambda}_{so}\left(\boldsymbol{k}_{t}\right)}{\partial k_{\alpha t}}\times\boldsymbol{\Lambda}_{so}\left(\boldsymbol{k}_{t}\right)\right]\cdot\left\langle \boldsymbol{\sigma}\right\rangle \left(\boldsymbol{k},t\right),\label{rtgeo-twobndFm}
\end{equation}
where $k_{\alpha t}$ denotes the $\alpha$th component of $\boldsymbol{k}_{t}$.
The above discussed geometric effects realised through the driven
(and therefore the non-equilibrium) spin texture, namely, $\left\langle \boldsymbol{\sigma}\right\rangle \left(\boldsymbol{k},t\right)=\sum_{n}f_{FD}^{\mu,T}\left(\varepsilon_{n}\left(\boldsymbol{k}\right)\right)\left\langle \boldsymbol{\sigma}\right\rangle _{n}\left(\boldsymbol{k},t\right)$,
is then concretely manifested in the post-pulse current rate Eq. (\ref{rtgeo-twobndFm})
that does not vanish only for non-adiabatic driving during the pulse.
Below we will discuss the properties of the dynamics of the driven
spin texture $\left\langle \boldsymbol{\sigma}\right\rangle _{n}\left(\boldsymbol{k},t\right)$
in more detail. In Sec. \ref{sec-examples-SSH} of the main text,
we have noted that the SSH system is obtained by simply replacing
the unspecified $\boldsymbol{\Lambda}_{so}\left(\boldsymbol{k}\right)$
by $\boldsymbol{\Lambda}_{SSH}\left(k\right)$ specified in the main
text. Therefore, by corresponding the spin to the sublattice pseudospin,
the discussions below about the properties of the driven spin texture
and their implications for the current rates are also valid for the
SSH system as illustrated in Sec. \ref{sec-examples-SSH} of the main
text.

To better understand the non-adiabatic driving effects on the spin
texture, we consider a pulsing scenario in which $\left.\boldsymbol{\mathcal{K}}_{t}\right\vert _{t>t_{\text{off}}}\rightarrow0$
in Eq. (\ref{k-shift}) of the main text. We find it convenient to
define $\left\langle \boldsymbol{\sigma}\right\rangle _{n,m}^{0}\left(\boldsymbol{k}\right):=\left\langle u_{n}\left(\boldsymbol{k}\right)\right\vert \boldsymbol{\sigma}\left\vert u_{m}\left(\boldsymbol{k}\right)\right\rangle $.
Noting $\left\langle \boldsymbol{\sigma}\right\rangle _{n,n}^{0}\left(\boldsymbol{k}\right)=\left\langle \boldsymbol{\sigma}\right\rangle _{n}^{0}\left(\boldsymbol{k}\right)$,
the diagonals/off-diagonals of $\left\langle \boldsymbol{\sigma}\right\rangle _{n,m}^{0}\left(\boldsymbol{k}\right)$
are called the native intra-band/inter-band spin texture. As $\boldsymbol{k}_{t_{\text{off}}}\rightarrow\boldsymbol{k}$,
the driven spin texture in the post-pulse time region is universally
given by \begin{subequations}\label{spintxt-id_toff-sum}
\begin{equation}
\left\langle \boldsymbol{\sigma}\right\rangle _{n}\left(\boldsymbol{k},t>t_{\text{off}}\right)=\left\langle \boldsymbol{\sigma}\right\rangle _{n}^{\text{off},dc}\left(\boldsymbol{k}\right)+\left\langle \boldsymbol{\sigma}\right\rangle _{n}^{\text{off},ac}\left(\boldsymbol{k},t\right).\label{spintxt-id_toff-1}
\end{equation}
It contains a time-independent contribution, $\left\langle \boldsymbol{\sigma}\right\rangle _{n}^{\text{off},dc}\left(\boldsymbol{k}\right)=\sum_{m}\left\vert \varphi_{m\boldsymbol{k}}\left(t_{\text{off}}\right)\right\vert ^{2}\left\langle \boldsymbol{\sigma}\right\rangle _{m}^{0}\left(\boldsymbol{k}\right)$,
in which $\varphi_{m\boldsymbol{k}}\left(t\right)=\left\langle u_{m}\left(\boldsymbol{k}\right)\left\vert \phi_{n,\boldsymbol{k}}\left(t,t_{0}\right)\right.\right\rangle $
and a time-oscillating contribution, 
\begin{equation}
\left\langle \boldsymbol{\sigma}\right\rangle _{n}^{\text{off},ac}\left(\boldsymbol{k},t\right)=\sum_{m}\sum_{m^{\prime}\ne m}e^{-\left(i/\hbar\right)\left(t-t_{\text{off}}\right)\left(\varepsilon_{m}-\varepsilon_{m^{\prime}}\right)\left(\boldsymbol{k}\right)}\varphi_{m^{\prime}\boldsymbol{k}}^{*}\left(t_{\text{off}}\right)\left\langle \boldsymbol{\sigma}\right\rangle _{m^{\prime},m}^{0}\left(\boldsymbol{k}\right)\varphi_{m\boldsymbol{k}}\left(t_{\text{off}}\right).\label{spintxt-id_toff-ac1}
\end{equation}
\end{subequations} In Eq. (\ref{spintxt-id_toff-sum}), we thus see
clearly how the different band-structure factors (spin texture and
band gaps) and the external-driving scenarios $\boldsymbol{E}\left(t_{0}\le t\le t_{\text{off}}\right)$
come together to determine the behaviours of the dc and the ac components
of the driven spin texture in the post-pulse time region. The dc component
$\left\langle \boldsymbol{\sigma}\right\rangle _{n}^{\text{off},dc}\left(\boldsymbol{k}\right)$
is solely characterised by its magnitude as a time-independent offset.
The dynamics of the ac component is characterised by three attributes,
namely, the amplitude, the period and the phase of the ac oscillation.
The result of Eq. (\ref{spintxt-id_toff-sum}) says that the native
intra-band spin texture $\left\langle \boldsymbol{\sigma}\right\rangle _{m}^{0}\left(\boldsymbol{k}\right)$
determines the magnitudes of the dc component while the native inter-band
spin texture $\left.\left\langle \boldsymbol{\sigma}\right\rangle _{m^{\prime},m}^{0}\left(\boldsymbol{k}\right)\right\vert _{m^{\prime}\ne m}$
is one factor behind the amplitude of the ac component. The band gaps
$\left(\varepsilon_{m}-\varepsilon_{m^{\prime}}\right)\left(\boldsymbol{k}\right)$
determine the periods of the ac oscillations in time. 

We now expound on the interplay between the in-pulse non-adiabatic
drivings and the geometric properties of the bands. We address how
such interplay manifests itself in the post-pulse driven spin texture
dynamics and the current rate in correspondence to the points (i)
and (ii) abstracted in Sec. \ref{WPT-basics-QnAccHistory} of the
main text. Certainly, the band occupation $\left\vert \eta_{m\boldsymbol{k}}\left(t_{\text{off}}\right)\right\vert ^{2}$
and inter-band coherence $\eta_{m^{\prime}\boldsymbol{k}}^{*}\left(t_{\text{off}}\right)\eta_{m\boldsymbol{k}}\left(t_{\text{off}}\right)$
of the wave-packet band pseudospin at the end of the pulse are results
of CM-band-pseudospin coupled dynamics that combined effects from
both the band-geometric properties and the drivings as explained before.
Here, more specifically for the post-pulse driven spin textures, the
band occupation and the inter-band coherence at the end of the pulse
respectively affect the magnitudes of the dc and the amplitudes of
the ac components as told by Eq. (\ref{spintxt-id_toff-sum}). This
illustrates the point (i) about how the in-pulse drivings allow the
geometric properties of the bands to leave imprints on the post-pulse
current rate $\dot{\boldsymbol{j}}\left(t>t_{\text{off}}\right)=\dot{\boldsymbol{j}}^{geo}\left(t\right)$.
As indicated by Eq. (\ref{rtgeo-twobndFm}) such imprints are realised
via the underlying driven spin texture $\left\langle \boldsymbol{\sigma}\right\rangle \left(\boldsymbol{k},t>t_{\text{off}}\right)$. 

For the point (ii), we should consider two different non-adiabatic
driving scenarios $A$ and $B$ to inspect the geometric significance
of history-dependence effects. Different in-pulse drivings are not
expected to result in different periods of the post-pulse ac oscillations
because the period is determined by the band gaps $\left(\varepsilon_{m}-\varepsilon_{m^{\prime}}\right)\left(\boldsymbol{k}\right)$,
in accordance of Eq. (\ref{spintxt-id_toff-sum}). Effects of different
driving scenarios can be better envisaged by assuming that the two
pulses have the same duration $t_{\text{off}}-t_{0}$ but leading
to different inter-band coherence at the end of the pulse, namely,
$\left.\eta_{m^{\prime}\boldsymbol{k}}^{*}\left(t_{\text{off}}\right)\eta_{m\boldsymbol{k}}\left(t_{\text{off}}\right)\right\vert _{A}\ne\left.\eta_{m^{\prime}\boldsymbol{k}}^{*}\left(t_{\text{off}}\right)\eta_{m\boldsymbol{k}}\left(t_{\text{off}}\right)\right\vert _{B}$.
We now take the time $t$ in Eq. (\ref{rtgeo-twobndFm}) to be in
the post-pulse time region, namely, $\dot{j}_{\alpha}^{geo}\left(t>t_{\text{off}}\right)=2\left(-e\right)\hbar^{-2}\int\text{d}^{D}\boldsymbol{k}\left(\partial\boldsymbol{\Lambda}_{so}\left(\boldsymbol{k}\right)/\partial k_{\alpha}\right)\times\boldsymbol{\Lambda}_{so}\left(\boldsymbol{k}\right)\cdot\left\langle \boldsymbol{\sigma}\right\rangle \left(\boldsymbol{k},t>t_{\text{off}}\right)$.
Here we have applied $\left.\boldsymbol{\Lambda}_{so}\left(\boldsymbol{k}_{t}\right)\right\vert _{t>t_{\text{off}}}=\boldsymbol{\Lambda}_{so}\left(\boldsymbol{k}_{t_{\text{off}}}\right)\rightarrow\boldsymbol{\Lambda}_{so}\left(\boldsymbol{k}\right)$.
In terms of the microscopic driven spin texture, it is clear that
$\left.\left\langle \boldsymbol{\sigma}\right\rangle \left(\boldsymbol{k},t>t_{\text{off}}\right)\right\vert _{A}$
and $\left.\left\langle \boldsymbol{\sigma}\right\rangle \left(\boldsymbol{k},t>t_{\text{off}}\right)\right\vert _{B}$
share the same oscillation period while the situation $\left.\eta_{m^{\prime}\boldsymbol{k}}^{*}\left(t_{\text{off}}\right)\eta_{m\boldsymbol{k}}\left(t_{\text{off}}\right)\right\vert _{A}\ne\left.\eta_{m^{\prime}\boldsymbol{k}}^{*}\left(t_{\text{off}}\right)\eta_{m\boldsymbol{k}}\left(t_{\text{off}}\right)\right\vert _{B}$
caused by different drivings only result in differences in the ac
oscillation amplitudes between $\left.\left\langle \boldsymbol{\sigma}\right\rangle \left(\boldsymbol{k},t>t_{\text{off}}\right)\right\vert _{A}$
and $\left.\left\langle \boldsymbol{\sigma}\right\rangle \left(\boldsymbol{k},t>t_{\text{off}}\right)\right\vert _{B}$,
as already discussed with Eq. (\ref{spintxt-id_toff-sum}) above.
Macroscopically, however, for the current rate, the above microscopic
$\boldsymbol{k}$-dependent amplitude difference can give rise to
different oscillation phases between $\left.\dot{j}_{\alpha}^{geo}\left(t>t_{\text{off}}\right)\right\vert _{A}$
and $\left.\dot{j}_{\alpha}^{geo}\left(t>t_{\text{off}}\right)\right\vert _{B}$
after the integration over $\boldsymbol{k}$. Therefore, different
in-pulse non-adiabatic drivings can give rise to different oscillation
phases rooted to the band-geometric effects. 

\subsection{Properties of adiabatic CISP}

\label{SI-spin-basics-adiaCISP}

We have discussed the motivations of studying CISP from the perspective
of SOC-geometric effects in Sec. \ref{WPT-basics-spntxt} of the main
text. With Rashba SOC, we have explored their salient transient non-adiabatic
manifestations in Sec. \ref{sec-examples-Rashba} of the main text.
Here we supplement information about CISP under adiabatic drivings
first without specifying the SOC to be Rashba. This readily allows
one to see the cancellation of the intrinsic CISP. We then specify
to the setting of Rashba SOC to investigate properties of the remained
CISP effects involving extrinsic mechanisms more explicitly.

\subsubsection{Cancellation of intrinsic spin coherence effects in CISP}

\label{sec-examples-SOC-adnoCISP}

Based on Eq. (\ref{SOC-H-gDef}) of the main text, we first recap
that the adiabatically induced CISP in the linear response regime
does not retain properties of the intrinsic spin coherence in the
resulting macroscopic spin polarisation. This goes without any specification
of the spin-orbit field $\boldsymbol{\Lambda}_{so}\left(\boldsymbol{k}\right)$
except that it satisfies 
\begin{equation}
\boldsymbol{\Lambda}_{so}\left(\boldsymbol{k}\right)=-\boldsymbol{\Lambda}_{so}\left(-\boldsymbol{k}\right),\label{SO-fld-TRS}
\end{equation}
as a result of respecting TRS . 

In the adiabatic regime to the first order in $\boldsymbol{E}\left(t\right)$,
the driven spin texture given by $\left\langle \boldsymbol{\sigma}\right\rangle _{n}^{ad}\left(\boldsymbol{k},t\right):=\left\langle \phi_{n,\boldsymbol{k}}^{ad}\left(t\right)\right\vert \boldsymbol{\sigma}\left\vert \phi_{n,\boldsymbol{k}}^{ad}\left(t\right)\right\rangle $
with $\left\vert \phi_{n,\boldsymbol{k}}^{ad}\left(t\right)\right\rangle $
found in Eq. (\ref{n-ad-uni}) of the main text leads to \begin{subequations}\label{ad-spntxt-E1}
\begin{equation}
\left\langle \boldsymbol{\sigma}\right\rangle _{n}^{ad}\left(\boldsymbol{k},t\right)=\left\langle \boldsymbol{\sigma}\right\rangle _{n}^{0}\left(\boldsymbol{k}_{t}\right)+\left\langle \boldsymbol{\sigma}\right\rangle _{n}^{r}\left(\boldsymbol{k}_{t},\boldsymbol{E}\left(t\right)\right)\label{ad-spntxt-sum}
\end{equation}
where $\left\langle \boldsymbol{\sigma}\right\rangle _{n}^{r}\left(\boldsymbol{k}_{t},\boldsymbol{E}\left(t\right)\right)$
is linear in $\boldsymbol{E}\left(t\right)$, i.e.,
\begin{equation}
\left\langle \sigma_{\beta}\right\rangle _{n}^{r}\left(\boldsymbol{k}_{t},\boldsymbol{E}\left(t\right)\right)=\sum_{\alpha}\chi_{n\alpha}^{\beta}\left(\boldsymbol{k}_{t}\right)\left(-eE_{\alpha}\left(t\right)\right),\label{ad-spntxt-rXp}
\end{equation}
with linear-response coefficients: 
\begin{equation}
\chi_{n\alpha}^{\beta}\left(\boldsymbol{k}_{t}\right)=\frac{\left\langle u_{n}\left(\boldsymbol{k}_{t}\right)\right\vert \sigma_{\beta}\left\vert u_{\bar{n}}\left(\boldsymbol{k}_{t}\right)\right\rangle \left[\mathcal{\mathcal{R}}_{k_{\alpha t}}\right]_{\bar{n},n}}{\varepsilon_{\bar{n}}\left(\boldsymbol{k}_{t}\right)-\varepsilon_{n}\left(\boldsymbol{k}_{t}\right)}+\text{c.c.},\label{ad-spntxt-rCoefR}
\end{equation}
\end{subequations} where $\bar{n}$ stands for the spin opposite
of $n$, e.g., $n=+$ then $\bar{n}=-$ and vice versa. Here in Eq.
(\ref{ad-spntxt-E1}), $\left\langle \boldsymbol{\sigma}\right\rangle _{n}^{r}\left(\boldsymbol{k}_{t},\boldsymbol{E}\left(t\right)\right)$
represents the part of the driven spin texture that contains effects
from the adiabatically-induced intrinsic spin coherence. As shown
in Eq. (\ref{ad-spntxt-rCoefR}), this is enabled by $\left[\mathcal{\mathcal{R}}_{k_{\alpha t}}\right]_{\bar{n},n}\ne0$
as a geometric effect. Without restricting the driving field to the
adiabatic regime, the driven spin texture $\left\langle \boldsymbol{\sigma}\right\rangle _{n}\left(\boldsymbol{k},t\right)$
in general develops complicated time-dependent field configurations
over the domain of $\boldsymbol{k}$. In contrast, under the adiabatic
limit, $\left\langle \boldsymbol{\sigma}\right\rangle _{n}\left(\boldsymbol{k},t\right)\rightarrow\left\langle \boldsymbol{\sigma}\right\rangle _{n}^{ad}\left(\boldsymbol{k},t\right)$,
the description of the spin-texture field is simplified to Eq. (\ref{ad-spntxt-E1})
in which the time development of $\boldsymbol{E}\left(t\right)$ is
factored out. Effectively, the driven spin texture is characterised
by $\chi_{n\alpha}^{\beta}\left(\boldsymbol{k}_{t}\right)$'s as static
field configurations over the domain of $\boldsymbol{k}_{t}$. By
the TRS of Eq. (\ref{SO-fld-TRS}), one sees that $\chi_{n\alpha}^{\beta}\left(\boldsymbol{k}_{t}\right)=-\chi_{\bar{n}\alpha}^{\beta}\left(-\boldsymbol{k}_{t}\right)$
and therefore 
\begin{equation}
\left\langle \sigma_{\beta}\right\rangle _{n}^{r}\left(\boldsymbol{k}_{t},\boldsymbol{E}\left(t\right)\right)=-\left\langle \sigma_{\beta}\right\rangle _{\bar{n}}^{r}\left(-\boldsymbol{k}_{t},\boldsymbol{E}\left(t\right)\right).\label{spinLR-TRS}
\end{equation}
This similarity to the anti-symmetric character of the native spin
texture is a signature of the SOC-geometric effects during adiabatic
driving processes.

We can calculate the non-equilibrium spin polarisation due to the
adiabatic drivings in the linear response regime by fully ignoring
band hybridisation in Eq.~(\ref{incoh-rho-1}) for the density matrix
(so $\bar{\varrho}_{\boldsymbol{k}}\left(t\right)=\sum_{n}f_{n}^{0}\left(\boldsymbol{k}_{t}\right)\left\vert u_{n}\left(\boldsymbol{k}_{t}\right)\right\rangle \left\langle u_{n}\left(\boldsymbol{k}_{t}\right)\right\vert $)
but maintaining the non-equilibrium deviation Eq. (\ref{rho-del-ad})
caused by the external field. We further make the approximation given
by Eq. (\ref{short-RT}). The resulting density matrix substituted
to Eq. (\ref{St-neq-0}) (of the main text) as the definition for
spin polarisation (here renotated to $\boldsymbol{S}\left(t\right)\rightarrow\boldsymbol{S}^{ad}\left(t\right)$
for adiabatic drivings) then gives \begin{subequations}\label{ad-spPolar-E1}
\begin{equation}
\boldsymbol{S}^{ad}\left(t\right)=\bar{\boldsymbol{S}}\left(t\right)+\delta\boldsymbol{S}\left(t\right),\label{ad-spPolar-E1-tot}
\end{equation}
where
\begin{equation}
\bar{\boldsymbol{S}}\left(t\right)=\frac{\hbar}{2}\sum_{n}\int\text{d}^{D}\boldsymbol{k}\left\langle \boldsymbol{\sigma}\right\rangle _{n}^{r}\left(\boldsymbol{k},\boldsymbol{E}\left(t\right)\right)f_{n}^{0}\left(\boldsymbol{k}\right),\label{ad-spPolar-E1-in}
\end{equation}
and
\begin{equation}
\delta\boldsymbol{S}\left(t\right)=-\tau\frac{\hbar}{2}\sum_{n}\int\text{d}^{D}\boldsymbol{k}\left\langle \boldsymbol{\sigma}\right\rangle _{n}^{0}\left(\boldsymbol{k}\right)\left[\frac{\partial f_{n}^{0}\left(\boldsymbol{k}\right)}{\partial\varepsilon_{n}\left(\boldsymbol{k}\right)}\frac{\partial\varepsilon_{n}\left(\boldsymbol{k}\right)}{\partial\hbar\boldsymbol{k}}\cdot(-e)\boldsymbol{E}\left(t\right)\right],\label{ad-spPolar-E1-ex}
\end{equation}
\end{subequations} in which we have used $\text{d}^{D}\boldsymbol{k}_{t}=\text{d}^{D}\boldsymbol{k}$.
The intrinsic spin-split-bands' geometric effects appear in the macroscopic
spin polarisation via $\bar{\boldsymbol{S}}\left(t\right)$ while
$\delta\boldsymbol{S}\left(t\right)$ in proportion to the relaxation
time $\tau$ is coupled with extrinsic scattering. By Eq. (\ref{spinLR-TRS})
and noting $f_{n}^{0}\left(\boldsymbol{k}\right)=f_{\bar{n}}^{0}\left(-\boldsymbol{k}\right)$
due to the Kramer degeneracy given by the TRS, we have
\begin{equation}
\bar{\boldsymbol{S}}\left(t\right)=0.\label{ad-spPolar-inzero}
\end{equation}
The adiabatic driving thus limits the nonzero spin polarisation arising
from the CISP effect to contain only the extrinsic contribution $\boldsymbol{S}^{ad}\left(t\right)=\delta\boldsymbol{S}\left(t\right)$,
reproducing the results in \cite{Edelstein90233,Tao21085438}. 

\subsubsection{Restrictions on the CISP orientation for Rashba SOC under adiabatic
drivings}

\label{sec-examples-Rashba-adCISP} We now specify $\boldsymbol{\Lambda}_{so}\left(\boldsymbol{k}\right)$
to be the spin-orbit field for the anisotropic Rashba SOC defined
in Eq.(\ref{sorb-fld-1}) of the main text. Under adiabatic drivings,
the driven spin texture from Eq. (\ref{ad-spntxt-E1}) becomes 
\begin{equation}
\left\langle \sigma_{z}\right\rangle _{n}^{r}\left(\boldsymbol{k}_{t},\boldsymbol{E}\left(t\right)\right)=\frac{\left(\hat{\boldsymbol{z}}\times\boldsymbol{k}_{t}\right)\cdot\hat{\boldsymbol{e}}}{2\alpha_{R}\left\vert \mathcal{X}_{K}\boldsymbol{k}_{t}\right\vert ^{3}}\left(-eE\left(t\right)\right),\label{ad-spntxt-E1-r}
\end{equation}
while $\left\langle \sigma_{x/y}\right\rangle _{n}^{r}\left(\boldsymbol{k}_{t},\boldsymbol{E}\left(t\right)\right)=0$.
The antisymmetry of $\left\langle \sigma_{z}\right\rangle _{n}^{r}\left(\boldsymbol{k}_{t},\boldsymbol{E}\left(t\right)\right)$
in $\boldsymbol{k}_{t}$ (as discussed with Eq. (\ref{spinLR-TRS}))
is made explicit here by Eq. (\ref{ad-spntxt-E1-r}). We already know
from the discussion for Eq. (\ref{ad-spPolar-inzero}) that under
the adiabatic driving, the CISP is solely produced by the extrinsic
contribution $\boldsymbol{S}^{ad}\left(t\right)=\delta\boldsymbol{S}\left(t\right)$.
By the nature of Eq. (\ref{sorb-fld-1}) of the Rashba spin-orbit
field, the extrinsic CISP given by $\delta\boldsymbol{S}\left(t\right)$
of Eq. (\ref{ad-spPolar-E1-ex}), in which $\left\langle \boldsymbol{\sigma}\right\rangle _{n}^{0}\left(\boldsymbol{k}\right)\cdot\hat{\boldsymbol{z}}=0$,
is limited to be in-plane which in turn supports that an out-of-plane
CISP should reveal the intrinsic properties. Under the isotropic condition,
namely, $r_{A}=1$, the native spin texture in Eq. (\ref{ad-spPolar-E1-ex})
becomes $\left\langle \boldsymbol{\sigma}\right\rangle _{\pm}^{0}\left(\boldsymbol{k}\right)=\pm\boldsymbol{k}\times\hat{\boldsymbol{z}}/\left\vert \boldsymbol{k}\right\vert $.
As a result, the orientation of the CISP is simplified to $\boldsymbol{S}^{ad}\left(t\right)\parallel\hat{\boldsymbol{e}}\times\hat{\boldsymbol{z}}$,
as also found in \cite{Edelstein90233,Tao21085438}. Therefore the
adiabatic driving to the linear response regime limits the spin polarisation
orientation raised by the CISP effect to 
\begin{equation}
\hat{\boldsymbol{z}}\cdot\boldsymbol{S}^{ad}\left(t\right)=0,\label{z-S-ad}
\end{equation}
\begin{equation}
\left.\boldsymbol{S}^{ad}\left(t\right)\cdot\hat{\boldsymbol{e}}\right\vert _{r_{A}=1}=0.\label{perp-S-iso-ad}
\end{equation}
Even though the intrinsic spin coherence evidenced by Eq. (\ref{ad-spntxt-E1-r})
is not manifested through CISP, it has been shown to be manifested
through the spin Hall current (not the ordinary electronic current
Eq. (\ref{j-def0}) defined in the main text) and serves as the basis
of the universal intrinsic spin Hall effect (see Ref. \cite{Sinova04126603}
in which the isotropic limit $\mathcal{X}_{K}\boldsymbol{k}=\boldsymbol{k}$
is considered). We will show later that the intrinsic spin coherence
effect induced by non-adiabatic drivings can also be manifested directly
in the electronic current and in the CISP under appropriate circumstances
(see discussions below).

\section{Supporting information on the Rashba-SOC system}

\label{SI-RSOC-dys}

In this section, we supplement information on the spin polarisation
and photocurrent dynamics to support the statements about the anisotropic
Rashba system briefed in Sec. \ref{sec-examples-Rashba} of the main
text.

\subsection{Non-adiabatic dynamics of driven spin texture and CISP}

\label{sec-examples-Rashba-nadCISP}In the non-adiabatic regime, to
more conveniently study the driven spin texture, which is generally
asymmetric, we define $\overline{\left\langle \boldsymbol{\sigma}\right\rangle }\left(\boldsymbol{k},t\right):=\left\langle \boldsymbol{\sigma}\right\rangle _{n_{-}}\left(\boldsymbol{k},t\right)+\left\langle \boldsymbol{\sigma}\right\rangle _{\bar{n}_{-}}\left(-\boldsymbol{k},t\right)$
by the virtue that before the electric field is turned on it is zero,
namely, $\overline{\left\langle \boldsymbol{\sigma}\right\rangle }\left(\boldsymbol{k},t_{0}\right)=0$
for all $\boldsymbol{k}$'s, due to the TRS of the native spin texture.
Here $n_{-}$ labels the lower energy band for a given $\boldsymbol{k}$
and $\left(\bar{n}_{-},-\boldsymbol{k}\right)$ labels the Kramer
partner of $\left(n_{-},\boldsymbol{k}\right)$. We assume the initial
equilibrium is given by filling up only the lower-energy band. The
intrinsic spin polarisation is then given by $\boldsymbol{S}\left(t\right)=\left(\hbar/2\right)\int_{\text{half FS}}\text{d}^{2}\boldsymbol{k}\overline{\left\langle \boldsymbol{\sigma}\right\rangle }\left(\boldsymbol{k},t\right)$,
in which the subscript ``half FS'' for the integral means half of
the fermi surface. Extrinsic effects will be discussed too alongside
with intrinsic effects. 

At a time $t_{1}=t_{0}+\delta t$ that is shortly away from the initial
equilibrium, complicated deviation from the native texture readily
appears in the driven spin texture which reads \cite{FootnoteIsoSpnTxtSdys}
\begin{subequations}\label{t1nadspntxt-aniso} 

\begin{align}
 & \overline{\left\langle \sigma_{\perp}\right\rangle }\left(\boldsymbol{k},t_{1}\right)=\overline{\delta s_{//}}\mathcal{G}_{0}\left(\theta,\varphi_{e}\right),\label{in-pern}
\end{align}

\begin{equation}
\overline{\left\langle \sigma_{\parallel}\right\rangle }\left(\boldsymbol{k},t_{1}\right)=\overline{\delta s_{//}}\mathcal{G}_{1}\left(\sqrt{r_{A}},\theta,\varphi_{e}\right),\label{in-paral}
\end{equation}
and 
\begin{equation}
\overline{\left\langle \sigma{}_{z}\right\rangle }\left(\boldsymbol{k},t_{1}\right)=\overline{\delta s_{z}}\mathcal{G}_{1}\left(r_{A},\theta,\varphi_{e}\right),\label{z-t4}
\end{equation}
\end{subequations} where
\begin{align}
 & \mathcal{G}_{0}\left(\theta,\varphi_{e}\right)=2\frac{\bar{r}\left(\cos\left[2\left(\theta-\varphi_{e}\right)\right]-1\right)+\delta r\left(\cos\left(2\theta\right)-\cos\left(2\varphi_{e}\right)\right)}{\sqrt{r_{X}^{2}\cos^{2}\theta+r_{Y}^{2}\sin^{2}\theta}},\label{sptxt-anig0}
\end{align}
and 
\begin{align}
 & \mathcal{G}_{1}\left(x,\theta,\varphi_{e}\right)=\frac{-\left(x+x^{-1}\right)\sin\left[2\left(\theta-\varphi_{e}\right)\right]+\left(x-x^{-1}\right)\left[\sin\left(2\varphi_{e}\right)-\sin\left(2\theta\right)\right]}{\sqrt{r_{X}^{2}\cos^{2}\theta+r_{Y}^{2}\sin^{2}\theta}},\label{sptxt-anig1}
\end{align}
in which 
\begin{equation}
\bar{r}=\left(r_{Y}+r_{X}\right)/2,\label{bar-r}
\end{equation}
 
\begin{equation}
\delta r=\frac{\left(r_{Y}-r_{X}\right)}{2}.\label{del-r}
\end{equation}
Recall from the main text that $r_{X}=\sqrt{m_{y}/m_{x}}$, $r_{Y}=\sqrt{m_{x}/m_{y}}$.
Here we have used the polar coordinate $\boldsymbol{k}=\left(\hat{\boldsymbol{x}}\cos\theta+\hat{\boldsymbol{y}}\sin\theta\right)k$.
Noticeably, under a non-adiabatic driving, regardless of the anisotropy/isotropy
of the Rashba SOC, the two components, $\overline{\left\langle \sigma_{\parallel}\right\rangle }\left(\boldsymbol{k},t_{1}\right)$
and $\overline{\left\langle \sigma_{z}\right\rangle }\left(\boldsymbol{k},t_{1}\right)$,
of the driven spin texture, share similar characters in the field
configuration over the angular part of $\boldsymbol{k}$ (see Eqs.
(\ref{in-paral}) and (\ref{z-t4}) with the common factor of $\mathcal{G}_{1}\left(x,\theta,\varphi_{e}\right)$),
which is distinct from $\mathcal{G}_{0}\left(\theta,\varphi_{e}\right)$
in the component $\overline{\left\langle \sigma_{\perp}\right\rangle }\left(\boldsymbol{k},t_{1}\right)$,
Eq. (\ref{in-pern}). This correlation between the parallel-to-field
in-plane component and out-of-plane component of the driven spin texture
is not found in the adiabatic driven spin texture. Below we further
investigate the consequences on the CISP effect from such driven spin
textures. 

\subsubsection{Properties of the CISP orientation with isotropic Rashba SOC}

\label{sec-examples-Rashba-nadCISP-iso}

Under the isotropic Rashba SOC, namely, $r_{A}=1$, the driven spin
texture is simplified to \begin{subequations}\label{t1nadspntxt-iso}
\begin{align}
 & \overline{\left\langle \sigma_{\perp}\right\rangle }\left(\boldsymbol{k},t_{1}\right)=2\overline{\delta s_{//}}\left[\cos\left(2\theta-2\varphi_{e}\right)-1\right],\label{in-pern-iso1}
\end{align}
\begin{equation}
\overline{\left\langle \sigma_{\parallel}\right\rangle }\left(\boldsymbol{k},t_{1}\right)=-2\overline{\delta s_{//}}\sin\left(2\theta-2\varphi_{e}\right),\label{in-paral-1}
\end{equation}
and 
\begin{equation}
\overline{\left\langle \sigma{}_{z}\right\rangle }\left(\boldsymbol{k},t_{1}\right)=-2\overline{\delta s_{z}}\sin\left(2\theta-2\varphi_{e}\right).\label{z-t4-1}
\end{equation}
\end{subequations}By $\int_{0}^{\pi}\text{d}\theta\sin\left(2\theta-2\varphi_{e}\right)=\int_{0}^{\pi}\text{d}\theta\cos\left(2\theta-2\varphi_{e}\right)=0$,
we find $S^{\parallel}\left(t_{1}\right)=S^{z}\left(t_{1}\right)=0$
and $S^{\perp}\left(t_{1}\right)\ne0$. This result is summarised
as
\begin{equation}
\left.S^{z}\left(t\right)\right\vert _{r_{A}=1}=0,\label{z-S-iso-nad}
\end{equation}
\begin{equation}
\left.\boldsymbol{S}\left(t\right)\cdot\hat{\boldsymbol{e}}\right\vert _{r_{A}=1}=0.\label{perp-S-iso-nad}
\end{equation}

\begin{figure}[h] \includegraphics[width=15cm, height=5.4cm]{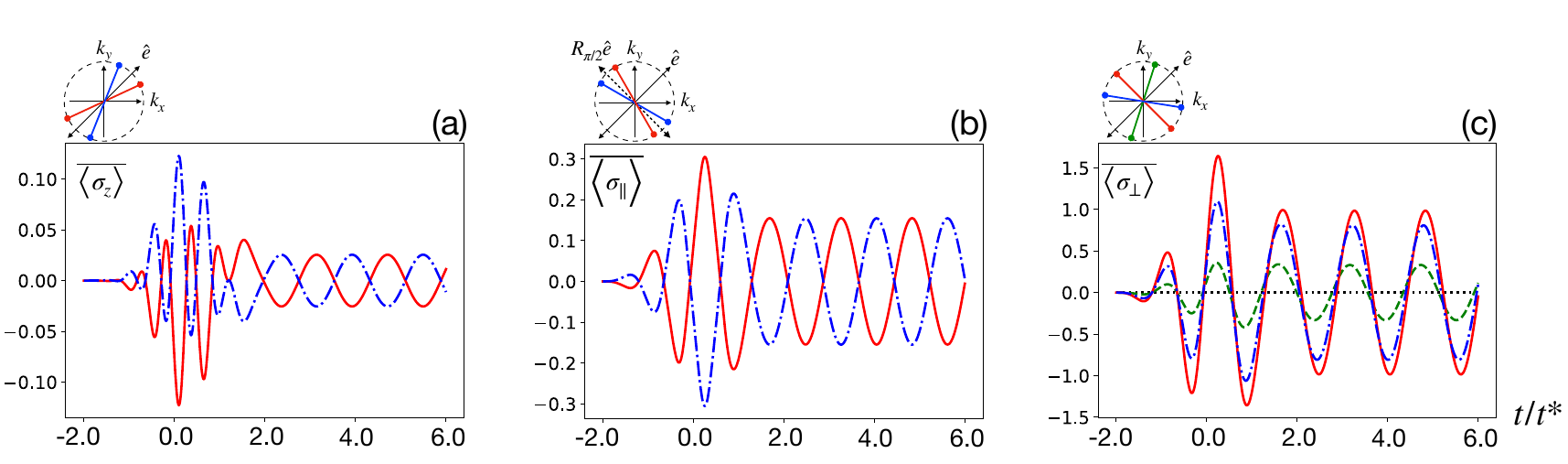} 
\caption{ Dynamics of driven spin texture components for selected $\boldsymbol{k}$'s. Writing in the polar coordinate $\boldsymbol{k}=(k,\theta)$, we fix  $k=k_{\text{so}}^{*}$ (see captions in Fig. \ref{CISP-Szpp} of the main text).  The applied electric field is polarised at $\varphi_{e}=0.25\pi$.  The directions of selected $\boldsymbol{k}$'s relative to the polarisation direction $\hat{\boldsymbol{e}}$ are illustrated on the upper-left corner of each plot. In (a) and (b),  the anti-symmetry in $\theta$ with respect to $\varphi_{r}$ for the $z$- and $\parallel$-components are exemplified by choosing $\theta=\varphi_{r}+\delta\theta$ (blue dashed) and $\theta=\varphi_{r}-\delta\theta$ (red solid)  where we set $\delta\theta=0.05\pi$ and $\varphi_{r}=\varphi_{e}$ for (a) and $\varphi_{r}=\varphi_{e}+\pi/2$ for (b).  In (c),  $\theta$'s are given by $\theta=\varphi_{e}$ (black dotted) and $\theta=\varphi_{e}+\delta\theta$ with $\delta\theta=0.15\pi$ for the green dashed,  $\delta\theta=0.5\pi$ for the red solid and $\delta\theta=0.7\pi$ for the blue dash-dot lines. The same pulse as that used in Fig. \ref{CISP-Szpp} of the main text is applied here.
} 
\label{drnspntxt} 
\end{figure}

Here we should explain that the reason behind the vanishing of $S^{\parallel/z}\left(t>t_{0}\right)$
under the isotropic condition, $r_{A}=1$, is quite different from
zero spin polarisation $\boldsymbol{S}_{\text{eq}}=\boldsymbol{S}\left(t\le t_{0}\right)=0$
given by the equilibrium spin texture regardless of the value of $r_{A}$.
The latter is simply due to $\overline{\left\langle \boldsymbol{\sigma}\right\rangle }\left(\boldsymbol{k},t\le t_{0}\right)=0$
by the TRS while the driven spin texture generally satisfies $\overline{\left\langle \boldsymbol{\sigma}\right\rangle }\left(\boldsymbol{k},t>t_{0}\right)\ne0$.
Noticeably from Eqs. (\ref{in-paral-1}) and (\ref{z-t4-1}), we find
$\left.\overline{\left\langle \sigma_{\parallel/z}\right\rangle }\left(\boldsymbol{k},t_{1}\right)\right\vert _{\theta=\varphi_{r}+\delta\theta}=-\left.\overline{\left\langle \sigma_{\parallel/z}\right\rangle }\left(\boldsymbol{k},t_{1}\right)\right\vert _{\theta=\varphi_{r}-\delta\theta}$
in which $\varphi_{r}$ specifies the direction of the unit vector
$\hat{\boldsymbol{r}}_{S}$ given by $\hat{\boldsymbol{r}}_{S}\parallel\hat{\boldsymbol{e}}$
or $\hat{\boldsymbol{r}}_{S}\parallel R_{\pi/2}\hat{\boldsymbol{e}}$
where $R_{\pi/2}$ is a rotation by $\pi/2$ and $\delta\theta$ is
the angle between $\boldsymbol{k}$ and $\hat{\boldsymbol{r}}_{S}$.
So the isotropy of the Rashba SOC forces the driven spin texture in
the $\parallel$ and $z$ components to become anti-symmetric in $\theta$
with respect to $\theta=\varphi_{r}$ causing $S^{\parallel}\left(t_{1}\right)=S^{z}\left(t_{1}\right)=0$.
Quite differently, we see from Eq. (\ref{in-pern-iso1}) that $\left\vert \overline{\left\langle \sigma_{\perp}\right\rangle }\left(\boldsymbol{k},t_{1}\right)\right\vert $
becomes larger when $\theta$ gets away from $\varphi_{e}$ and it
is maximised at $\theta=\varphi_{e}\pm\pi/2$. The above points manifest
that the external field along $\hat{\boldsymbol{e}}$ creates an anisotropic
feature in the driven spin texture despite the Rashba SOC is isotropic.
By numerically calculating the components of the driven spin texture
under the same electric-field pulse used in the main text, we confirm
that the above conclusion is held also to any later time $t>t_{0}$.
The anti-symmetric character of the components $\overline{\left\langle \sigma_{z}\right\rangle }\left(\boldsymbol{k},t\right)$
and $\overline{\left\langle \sigma_{\parallel}\right\rangle }\left(\boldsymbol{k},t\right)$
of driven spin texture with respect to the driving field orientation
$\hat{\boldsymbol{e}}$ is demonstrated in Fig. \ref{drnspntxt}(a)
and Fig. \ref{drnspntxt}(b) respectively. The dynamics of $\overline{\left\langle \sigma_{\perp}\right\rangle }\left(\boldsymbol{k},t\right)$
is exemplified in Fig. \ref{drnspntxt}(c). As expected from isotropic
condition, we verified from the numerical results that the non-vanishing
magnitude of $S^{\perp}\left(t\right)$ is independent of the direction
of $\hat{\boldsymbol{e}}$. Therefore, under the isotropic condition,
the orientational properties, Eqs. (\ref{z-S-iso-nad}) and (\ref{perp-S-iso-nad})
of CISP, caused by the non-adiabatic driving is not different from
those arising from adiabatic drivings, namely, Eqs. (\ref{z-S-ad})
and (\ref{perp-S-iso-ad}). Our findings here are consistent with
Ref. \cite{Vignale16035310} in which a time-independent electric
field is applied to study the time development of the spin polarisation
in isotropic Rashba systems in the intrinsic limit.

\subsubsection{Out-of-plane CISP induced by transient mirror-symmetry breaking }

\label{sec-examples-Rashba-nadCISP-mrbrk}

\begin{figure}[h] \includegraphics[width=15cm, height=4.4cm]{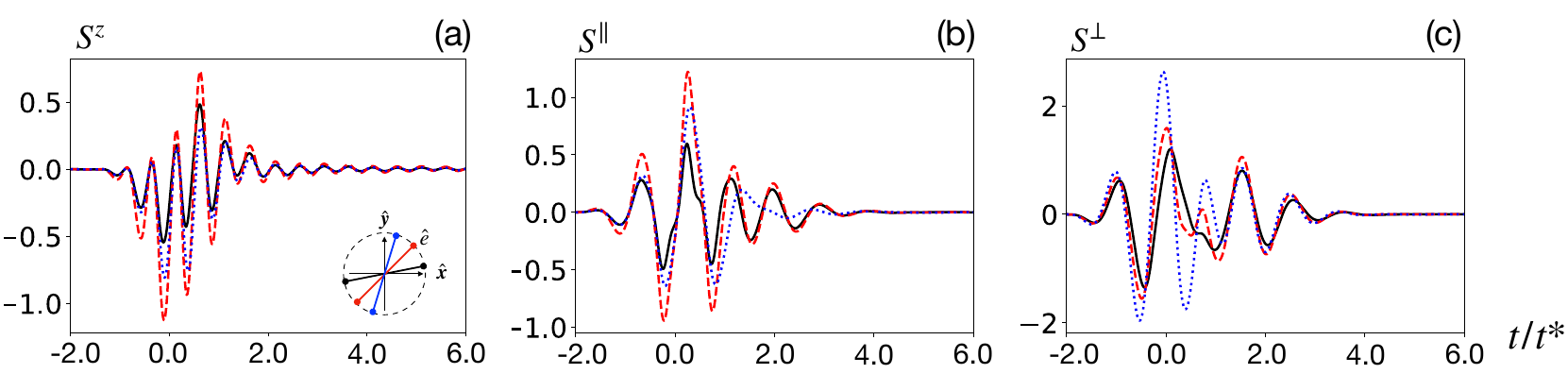} 
\caption{ Intrinsic CISP dynamics for anisotropic Rashba SOC.  The demonstrated polarisation angles $\varphi_{e}$'s and all other parameters are the same as those used in Fig. \ref{CISP-Szpp} of the main text except here all scattering rates are zero.   The directions of applied field's polarisation $\hat{\boldsymbol{e}}$ with respect to the mirror lines $\hat{\boldsymbol{x}}$ and $\hat{\boldsymbol{y}}$ are illustrated as the inset of (a).  When $\hat{\boldsymbol{e}}$ lies in the middle between $\hat{\boldsymbol{x}}$ and $\hat{\boldsymbol{y}}$ corresponding to $\varphi_{e}=0.25\pi$ (the red rod in the inset),  the applied field maximally breaks the mirror symmetry and thus gives the highest peak values in $S_{z}(t)$ and $S_{\parallel}(t)$ but not in $S_{\perp}(t)$ (see the red dashed lines in comparison to the other two lines in all three plots).
} 
\label{iCISP} 
\end{figure}

By lifting up the isotropy and restoring the anisotropy in the Rashba
SOC which establishes the mirror symmetry as its native structure,
one finds that the non-adiabatic drivings can yield results qualitatively
different from the adiabatic drivings. The anisotropic Rashba SOC
featured by $r_{A}\ne1$ yields $S^{\parallel/z}\left(t_{1}\right)\propto\sin\left(2\varphi_{e}\right)$
\cite{FootnoteAniSPzp}, which is nonzero when $\hat{\boldsymbol{e}}$
is not parallel to either of two the mirror lines $\hat{\boldsymbol{x}}$
and $\hat{\boldsymbol{y}}$. In other words, by allowing anisotropy
in Rashba SOC so the system's symmetry is featured by the two mirrors,
a non-adiabatic driving field can be oriented to break this mirror
symmetry and induces nonzero spin polarisation in both $S^{\parallel}\left(t\right)$
and $S^{z}\left(t\right)$ components. In contrast, in the adiabatic
regime, even when the electric field is not applied along the mirror
lines, it cannot induce an out-of-plane spin polarisation $\hat{\boldsymbol{z}}\cdot\boldsymbol{S}^{ad}\left(t\right)=0$.
We have demonstrated the induction of the out-of-plane CISP by transient
mirror-symmetry breaking without excluding extrinsic scatterings in
Fig. \ref{CISP-Szpp}(a) of the main text. Here, we illustrate this
intrinsic mechanism in the intrinsic limit in Fig. \ref{iCISP}(a)
and (b). It testifies that the more the external field's polarisation
is away from the mirror lines, the higher the peak values of $S^{\parallel/z}\left(t\right)$
can be attained (see the larger magnitudes in the red dashed lines
relative to other lines there). The CISP component $S^{\perp}\left(t\right)$
which is not induced due to breaking of the mirror symmetry by the
applied field is supplemented in Fig. \ref{iCISP}(c). The black solid
and blue dash-dot lines in all plots of Fig. \ref{iCISP} are obtained
by deviating $\hat{\boldsymbol{e}}$ from $\hat{\boldsymbol{x}}$
and $\hat{\boldsymbol{y}}$ by the same angle. Their difference attests
the anisotropy between $x$- and $y$- directions. This anisotropic
effect is then present in all three spin polarisation components as
shown in Fig. \ref{iCISP}.

\subsection{Extraction of spin-mediated processes in the photocurrents by mirror-symmetry
breaking}

\label{sec-examples-Rashba-mrr}

Having discussed the pulse-induced dynamics of the spin polarisation,
here we provide accompanied studies of the photocurrent. In the main
text, we have suggested that the spin-mediated part of the photocurrent
may contain SOC-geometric effects in Sec. \ref{sec-examples-SOC-isc}
without specifying the SOC type. Here we discuss such manifestation
with the Rashba SOC explicated in Sec. \ref{sec-examples-Rashba}
of the main text.

Substituting Eqs. (\ref{ani-kinetic-1}) and (\ref{sorb-fld-1}) into
Eqs. (\ref{JtoKnL-defK}) and (\ref{JtoKnL-defL}) from the main text,
we obtain

\begin{equation}
\boldsymbol{j}_{K}\left(t\right)=-e\mathcal{X}_{K}\int\text{d}^{2}\boldsymbol{k}\text{\ensuremath{\frac{\hbar\boldsymbol{k}_{t}}{m_{r}}}}N_{\boldsymbol{k}}\left(t\right),\label{j-K-1}
\end{equation}
\begin{equation}
\boldsymbol{j}_{so}\left(t\right)=\left(\frac{-e}{\hbar/2}\right)\frac{\hbar k_{\text{so}}^{*}}{m_{r}}\hat{\boldsymbol{z}}\times\mathcal{X}_{S}\boldsymbol{S}\left(t\right),\label{jSO-1}
\end{equation}
where $\mathcal{X}_{S}=\left(\begin{array}{cc}
r_{Y} & 0\\
0 & r_{X}
\end{array}\right)$, $m_{r}=\sqrt{m_{x}m_{y}}$, $N_{\boldsymbol{k}}\left(t\right)=\text{tr}\left(\varrho_{\boldsymbol{k}}\left(t\right)\right)$
and we have defined $\hbar k_{\text{so}}^{*}/m_{r}=\alpha_{R}/\hbar$
with $\alpha_{R}$ being the Rashba coefficient in Eq. (\ref{sorb-fld-1})
of the main text. The SOC-geometric effects are thus carried by $\boldsymbol{j}_{so}\left(t\right)$
when the spin polarisation, $\boldsymbol{S}\left(t\right)\ne0$, can
be attributed to the intrinsic spin coherence induced by the external
field. Previous discussion already shows that using the asymmetry
offered by $r_{A}\ne1,\hat{\boldsymbol{e}}\nparallel\hat{\boldsymbol{x}}/\hat{\boldsymbol{y}}$,
one can tell, by the qualitative difference between $S^{z}\left(t\right)=0$
and $S^{z}\left(t\right)\ne0$ (the out-of-plane macroscopic spin
polarisation), whether or not the macroscopic current $\boldsymbol{j}_{so}\left(t\right)$
(which is coupled to the in-plane spin polarisation $S^{\parallel/\perp}\left(t\right)$)
contains microscopic intrinsic spin coherence effects. We now investigate
the implications of $r_{A}\ne1,\hat{\boldsymbol{e}}\nparallel\hat{\boldsymbol{x}}/\hat{\boldsymbol{y}}$
for the extraction of spin-mediated processes in the transient photocurrent
$\boldsymbol{j}\left(t,\varphi_{e}\right)=\hat{\boldsymbol{e}}j^{\parallel}\left(t,\varphi_{e}\right)+\left(\hat{\boldsymbol{z}}\times\hat{\boldsymbol{e}}\right)j^{\perp}\left(t,\varphi_{e}\right)$
in which $j^{\parallel}\left(t,\varphi_{e}\right)$ and $j^{\perp}\left(t,\varphi_{e}\right)$
are called the longitudinal and the transverse components respectively.
We make the dependence of the photocurrent on laser polarisation $\varphi_{e}$
explicit. 

\subsubsection{Transverse component as a mirror-symmetry-breaking signature}

\label{Rashba-mrr-tvsl}

Explicitly, the the longitudinal and the transverse components of
$\boldsymbol{j}_{K}\left(t\right)$ from Eq. (\ref{j-K-1}) read \begin{subequations}\label{j-K-fr}
\begin{equation}
j_{K}^{\parallel}\left(t,\varphi_{e}\right)=-e\frac{\hbar\mathcal{K}_{t}}{m_{r}}\bar{n}_{f}\left[\bar{r}-\delta r\cos\left(2\varphi_{e}\right)\right],\label{j-K-frPL}
\end{equation}
\begin{equation}
j_{K}^{\perp}\left(t,\varphi_{e}\right)=-e\bar{n}_{f}\frac{\hbar\mathcal{K}_{t}}{m_{r}}\delta r\sin\left(2\varphi_{e}\right),\label{j-K-frPN}
\end{equation}
\end{subequations} in ignorance of scattering effects, $\varrho_{\boldsymbol{k}}\left(t\right)\rightarrow\varrho_{\boldsymbol{k}}^{\text{id}}\left(t\right)$.
Here $\bar{n}_{f}=\int\text{d}^{2}\boldsymbol{k}\varrho_{\boldsymbol{k}}^{eq}$
is the carrier density. Meanwhile, the components of $\boldsymbol{j}_{so}\left(t\right)$
from Eq. (\ref{jSO-1}) can be written as \begin{subequations}\label{jSO-eC0}

\begin{equation}
j_{so}^{\parallel}\left(t,\varphi_{e}\right)=\left(\frac{-e}{\hbar/2}\right)\frac{\hbar k_{\text{so}}^{*}}{m_{r}}\left\{ \delta r\left[\sin\left(2\varphi_{e}\right)S^{\parallel}\left(t,\varphi_{e}\right)+\cos\left(2\varphi_{e}\right)S^{\perp}\left(t,\varphi_{e}\right)\right]-\bar{r}S^{\perp}\left(t,\varphi_{e}\right)\right\} ,\label{jSO-eC0-S}
\end{equation}
\begin{align}
j_{so}^{\perp}\left(t,\varphi_{e}\right)=\left(\frac{-e}{\hbar/2}\right)\frac{\hbar k_{\text{so}}^{*}}{m_{r}}\left\{ \delta r\left[-\sin\left(2\varphi_{e}\right)S^{\perp}\left(t,\varphi_{e}\right)+\cos\left(2\varphi_{e}\right)S^{\parallel}\left(t,\varphi_{e}\right)\right]+\bar{r}S^{\parallel}\left(t,\varphi_{e}\right)\right\}  & .\label{jSO-eCL}
\end{align}
\end{subequations} Clearly, by setting $\delta r=0$ (so $r_{A}=1$)
in Eqs. (\ref{j-K-fr}) and (\ref{jSO-eC0}), the transverse components
vanish $j_{K/so}^{\perp}\left(t,\varphi_{e}\right)=0$ (recalling
that $\left.S^{\parallel}\left(t,\varphi_{e}\right)\right\vert _{r_{A}=1}=0$)
and only the longitudinal components remain. When $\delta r\ne0$
and at the same time $\sin\left(2\varphi_{e}\right)\ne0$, which is
equivalent to $r_{A}\ne1,\hat{\boldsymbol{e}}\nparallel\hat{\boldsymbol{x}}/\hat{\boldsymbol{y}}$,
then the possibility $j_{K/so}^{\perp}\left(t,\varphi_{e}\right)\ne0$
arises. The transverse component $j^{\perp}\left(t,\varphi_{e}\right)$
thus is a signature of breaking the mirror symmetry. Note that $j_{K/so}^{\parallel}\left(t,\varphi_{e}\right)$
is invariant under the mirror reflection operations $\varphi_{e}\rightarrow-\varphi_{e}$
and $\varphi_{e}\rightarrow\pi-\varphi_{e}$.

\subsubsection{Differentiating spin- and bond-mediated processes via the longitudinal
and transverse photocurrents with anisotropy }

\label{Rashba-mrr-stdm}

The mirror symmetry here is associated with the anisotropy in the
Rashba SOC by $r_{A}\ne1$. The efficacy of the anisotropy in principle
can be quantified by the difference between the results obtained under
$\hat{\boldsymbol{e}}\parallel\hat{\boldsymbol{x}}$ and that obtained
under $\hat{\boldsymbol{e}}\parallel\hat{\boldsymbol{y}}$. However,
since the transverse components of the current need to be induced
by deviating $\hat{\boldsymbol{e}}$ from $\hat{\boldsymbol{x}}$
and $\hat{\boldsymbol{y}}$, we instead measure the efficacy of the
anisotropy by the difference between $\hat{\boldsymbol{e}}\parallel R_{z}\left(\delta\varphi\right)\hat{\boldsymbol{x}}$
and $\hat{\boldsymbol{e}}\parallel R_{z}\left(-\delta\varphi\right)\hat{\boldsymbol{y}}$.
Here $R_{z}\left(\delta\varphi\right)\hat{\boldsymbol{x}}$ rotates
$\hat{\boldsymbol{x}}$ by an angle $\delta\varphi$ with $0<\delta\varphi\le\pi/4$
toward $\hat{\boldsymbol{y}}$ (so $\varphi_{e}=\delta\varphi$) and
correspondingly $R_{z}\left(-\delta\varphi\right)\hat{\boldsymbol{y}}$
rotates $\hat{\boldsymbol{y}}$ by the same amount of angle $\delta\varphi$
toward $\hat{\boldsymbol{x}}$ ($\varphi_{e}=\pi/2-\delta\varphi$).
This leads to the definition of anisotropic asymmetry as Eq. (\ref{ani-asym-def})
in the main text. We now study the anisotropic asymmetry in the longitudinal
($\alpha=\parallel$) and transverse ($\alpha=\perp$) components
of the bond- and spin-mediated parts of the photocurrents given by
Eqs. (\ref{j-K-fr}) and (\ref{jSO-eC0}) one by one. 

To simplify the analysis of the anisotropic asymmetry, we considered
$\delta\varphi\gtrsim0$ such that $\sin\left(2\delta\varphi\right)\approx0$
and $\cos\left(2\delta\varphi\right)\approx1$. Then, applying Eq.
(\ref{ani-asym-def}) to Eqs. (\ref{j-K-fr}) and (\ref{jSO-eC0})
and taking $\alpha=\parallel$, we obtain \begin{subequations}\label{aniasym-j-K-fr}
\begin{equation}
\Delta_{\delta\varphi}j_{K}^{\parallel}\left(t\right)\approx e\frac{\hbar\mathcal{K}_{t}}{m_{r}}\left(r_{Y}-r_{X}\right)\bar{n}_{f},\label{aniasym-Kpara}
\end{equation}
\begin{equation}
\Delta_{\delta\varphi}j_{so}^{\parallel}\left(t\right)\approx\left(-e\frac{\hbar k_{\text{so}}^{*}}{m_{r}}\right)\left(\frac{1}{\hbar/2}\right)\left[r_{Y}S^{\perp}\left(t,\pi/2-\delta\varphi\right)-r_{X}S^{\perp}\left(t,\delta\varphi\right)\right].\label{aniasym-jso-para}
\end{equation}
\end{subequations} Note here that in Eq. (\ref{aniasym-j-K-fr}),
the lack of dependence on $\bar{r}=\left(r_{X}+r_{Y}\right)/2$ in
Eq. (\ref{aniasym-Kpara}) is not due to the small-$\delta\varphi$
approximation but an exact result from Eq. (\ref{j-K-frPL}). The
limitation of the spin $\left\vert \left\langle \sigma_{\alpha}\right\rangle \left(\boldsymbol{k},t\right)\right\vert \le1$
imposes an upper bound on $\left\vert S^{\perp}\left(t,\varphi_{e}\right)\right\vert \lesssim\left(\hbar/2\right)\bar{n}_{f}$.
So with an estimation of $\left\vert S^{\perp}\left(t,\varphi_{e}\right)\right\vert \lesssim\left(\hbar/2\right)\bar{n}_{f}$,
the difference between the factor $e\left[r_{Y}S^{\perp}\left(t,\pi/2-\delta\varphi\right)-r_{X}S^{\perp}\left(t,\delta\varphi\right)\right]/\left(\hbar/2\right)$
in Eq. (\ref{aniasym-jso-para}) and $e\left(r_{Y}-r_{X}\right)\bar{n}_{f}$
in Eq. (\ref{aniasym-Kpara}), both in unit of the charge density,
does not inform clear separation between the magnitudes of $\left\vert \Delta_{\delta\varphi}j_{so}^{\parallel}\left(t\right)\right\vert $
and $\left\vert \Delta_{\delta\varphi}j_{K}^{\parallel}\left(t\right)\right\vert $.
Nevertheless, as $\mathcal{K}_{t}$ is the tunable amplitude of the
externally applied field, it can be tuned to satisfy $\mathcal{K}_{t}\gg k_{\text{so}}^{*}$
such that
\begin{equation}
\left\vert \Delta_{\delta\varphi}j^{\parallel}\left(t\right)\approx\Delta_{\delta\varphi}j_{K}^{\parallel}\left(t\right)\right\vert \gg\left\vert \Delta_{\delta\varphi}j_{so}^{\parallel}\left(t\right)\right\vert .\label{AniAsym-LgDom-1}
\end{equation}
This separation between the two parts of the photocurrent points to
the fundamental difference between charge and spin in response to
the electric field. The applied field $\mathcal{K}_{t}$ directly
couples to charge motion along the direction of the field and contributes
to the longitudinal current. The spin-mediated part of the photocurrent,
however, is limited by the efficiency of the SOC characterised by
$k_{\text{so}}^{*}$. 

Substituting Eq. (\ref{ani-asym-def}) to Eqs. (\ref{j-K-fr}) and
(\ref{jSO-eC0}) and taking $\alpha=\perp$ then yield \begin{subequations}\label{aniasym-jSO-eC0}
\begin{equation}
\Delta_{\delta\varphi}j_{K}^{\perp}\left(t\right)=0,\label{aniasym-Kzero}
\end{equation}

\begin{equation}
\Delta_{\delta\varphi}j_{so}^{\perp}\left(t\right)\approx\left(-e\frac{\hbar k_{\text{so}}^{*}}{m_{r}}\right)\left(\frac{1}{\hbar/2}\right)\left[r_{Y}S^{\parallel}\left(t,\delta\varphi\right)-r_{X}S^{\parallel}\left(t,\pi/2-\delta\varphi\right)\right].\label{aniasym-jso-perp}
\end{equation}
\end{subequations} Note here that Eq. (\ref{aniasym-Kzero}) is an
exact consequence of Eq. (\ref{j-K-frPN}) without any use of approximations.
The approximation $\delta\varphi\gtrsim0$ has been applied to simplify
Eq. (\ref{jSO-eCL}) to arrive Eq. (\ref{aniasym-jso-perp}). In general
for $\delta\varphi\ne0$ and $\delta r\ne0$ we have non-vanishing
$\Delta_{\delta\varphi}j_{so}^{\perp}\left(t\right)\ne0$. The sharp
contrast between the spin-mediated part $\Delta_{\delta\varphi}j_{so}^{\perp}\left(t\right)\ne0$
and the bond-mediated part $\Delta_{\delta\varphi}j_{K}^{\perp}\left(t\right)=0$,
and therefore $\left\vert \Delta_{\delta\varphi}j^{\perp}\left(t\right)=\Delta_{\delta\varphi}j_{so}^{\perp}\left(t\right)\right\vert \gg\left\vert \Delta_{\delta\varphi}j_{K}^{\perp}\left(t\right)\right\vert =0$,
here obtained under the intrinsic limit, is directly linked to the
difference between $\boldsymbol{V}^{K}\left(\boldsymbol{k}_{t}\right)$
and $\boldsymbol{V}^{so}\left(\boldsymbol{k}_{t}\right)$ in the definition
of the velocity matrix Eq. (\ref{Vdecompose}) in the main text. The
inclusion of scattering effects does not affect the contents of $\boldsymbol{V}^{K}\left(\boldsymbol{k}_{t}\right)$
and $\boldsymbol{V}^{so}\left(\boldsymbol{k}_{t}\right)$ at all.
It only modifies the dynamics of $\varrho_{\boldsymbol{k}}\left(t\right)$
so one could still have 
\begin{equation}
\left\vert \Delta_{\delta\varphi}j^{\perp}\left(t\right)\approx\Delta_{\delta\varphi}j_{so}^{\perp}\left(t\right)\right\vert \gg\left\vert \Delta_{\delta\varphi}j_{K}^{\perp}\left(t\right)\right\vert ,\label{AniAsym-TvDom-1}
\end{equation}
with $\Delta_{\delta\varphi}j_{K}^{\perp}\left(t\right)$ being not
strictly zero, as verified from numerical calculations shown in Fig.
\ref{CISP-Szpp}(c) of the main text.

Combining Eqs. (\ref{AniAsym-TvDom-1}) and (\ref{AniAsym-LgDom-1}),
we see that the spin/bond-mediated part of the photocurrent is then
largely contained in its transverse/longitudinal component of anisotropic
asymmetry. Noticeably, $\mathcal{K}_{t}\gg k_{\text{so}}^{*}$ for
holding (\ref{AniAsym-LgDom-1}) is another indication of the activation
of the non-adiabatic regime of spin dynamics. So the distinction between
bond-mediated and the spin-mediated parts of the photocurrent manifested
by the distinction between the longitudinal and the transverse components
of the photocurrents, in terms of the anisotropic characters, is crucially
facilitated by non-adiabatic drivings.

\section{Supporting information for implications on experiments}

\label{SI-EXP} Implications of NADT-WPT to experiments with ultrafast
lasers applied to quantum materials have been discussed in Sec. \ref{concl-EXPIM}
of the main text in two parts. In the first part, we point out the
general existence of links between the present theoretical framework
to three kinds of experiments without going to particular points specific
to each type of experiments. These particular points are discussed
in more details for each of the three types of experimental detections
in Sec. \ref{SI-EXP-broadcmmts} here. In the second part of Sec.
\ref{concl-EXPIM} of the main text, we summarised theoretical implications
relevant for exploiting SOC anisotropic effects in experiments. For
this, we have performed THz-emission experiments on SnSe. Therefore,
in Sec. \ref{SI-EXP-SnSe} here, we describe experimental details
and how the consistency between theoretical implications and experimental
observations is reached. 

\subsection{General considerations of relating NADT-WPT to experiments with ultrafast
lasers }

\label{SI-EXP-broadcmmts}

\subsubsection{Time-resolved ARPES (Tr-ARPES)}

The first type concerns the Tr-ARPES. With subcycle time resolution
\cite{Reimann18396}, the Tr-ARPES is capable of tackling what we
call the in-pulse time region here. From the time-dependent data of
occupation distributed over energy and momentum, the group velocities
of the bands and the dynamics of Bloch-momentum-resolved intra-band
processes can be tracked. These capabilities of Tr-ARPES have been
demonstrated to manifest the intrinsic properties of the linearly
dispersed surface Dirac cones in experiments \cite{Reimann18396}.
Here we seek for temporal dynamics of momentum-resolved inter-band
coherence rooted to inter-band geometric connections. The challenge,
therefore lies in getting the transient inter-band coherence and not
just the occupation. With the advancement of spin-resolved ARPES \cite{Cacho15097401,Bussolotti23032001},
it may be possible to extract information about spin coherence from
the spin-resolved-ARPES data to microscopically address the SOC-geometric
effects in real time. 

\subsubsection{Time-averaged electronic currents}

The second type is implemented by attaching electrodes to measure
the laser-excited currents. Given the limitation on the response time
of the electronic devices, the time-averaged currents, integrated
on a time scale larger than the microscopic time scale of the excited
electron dynamics, are often obtained. The information contained in
such currents is further complicated by laser-spot-to-collector diffusions
as well as the interfaces between the sample quantum material and
the electrodes. Nevertheless, through a number of studies on graphene
under strong-field drivings where the non-adiabatic dynamics is relevant,
an important experimental observation has been confirmed. That is,
the time-averaged currents carry the imprints of the subcycle properties
of the driving fields \cite{Higuchi17224,Heide21023103,Boolakee22251,Weitz24206901}.
In understanding these results, the current is formulated in terms
of the asymptotic conduction population and the group velocity \cite{Higuchi17224,Weitz24206901}.
The development of NADT-WPT made here, without prior references to
the above works, is unexpectedly found to be related to the above
results when we tried to look for implications of NADT-WPT on experiments.
We analytically showed that the post-pulse current has a history dependence
on the in-pulse non-adiabatic dynamics arising from the geometric
connections among the bands. The final result is akin to the dependence
of post-pulse time-averaged currents on the parameters of the in-pulse
drivings observed in experiments which can be literally interpreted
also as history-dependence effects. The current, from a wave-packet
point of view, however, is not only contributed by the group velocities
of bands but also by the anomalous velocities supported by inter-band
coherence. To which extent these two different sources of currents
survive the time-averaging processes and carry history information
of in-pulse-drivings despite of spot-to-contact diffusion and other
complications remain to be investigated in the future. 

\subsubsection{Time-resolved THz emissions}

The third type measures the THz-emission associated with the laser-excited
photocurrents. In principle, one considers transient currents in this
case, instead of time-integrated currents. Within the framework of
electromagnetism without any explicit band structure consideration,
the electric fields of the emitted signals $\boldsymbol{E}_{em}\left(t\right)$
have been shown to be proportional to $\text{d}\boldsymbol{j}\left(t\right)/\text{d}t$
\cite{Shan04book}. This seems to directly relate experimental signals
to the present theory which computes $\boldsymbol{j}\left(t\right)$
and $\text{d}\boldsymbol{j}\left(t\right)/\text{d}t$. Nevertheless,
excitations other than the sought-for geometrically-rooted intrinsic
inter-band coherence can be induced. Detected transient signals in
the time domain can be time-delayed from the photoexcitations due
to inelastic relaxations. The resulting emission then contains more
information than just $\text{d}\boldsymbol{j}\left(t\right)/\text{d}t$
produced from the theory (even when it includes finite scattering
rates) used here. Therefore with these caveats in mind, to relate
NADT-WPT to THz emissions, one shall not directly compare the detailed
time dependent profiles of experimentally produced $\boldsymbol{E}_{em}\left(t\right)\sim\text{d}\boldsymbol{j}\left(t\right)/\text{d}t$
with that of a clean theory result. Instead, it is more suitable to
relate the transient dynamics of currents given by NADT-WPT to that
given by Tr-ARPES because the latter detection is targeted directly
at the electrons themselves other than electromagnetic waves associated
with electronic motions.

Crucially, one purpose of NADT-WPT here is to manifest band-geometric
effects via macroscopic observables not entirely through the time-resolved
patterns in the transient processes but through other more robust
features. We have theoretically demonstrated in Sec. \ref{sec-examples-Rashba}
of the main text such scenarios using the anisotropic characters as
a more robust feature without looking into detailed structure of time
dependent profiles of the dynamics of observables, e.g., spin polarisation
and the two differentiated parts of the photocurrents. This gives
a more suitable way to relate views of NADT-WPT to THz emissions data
as we demonstrate below.

\subsection{THz-emission experiments on $\text{SnSe}$ viewed from NADT-WPT}

\label{SI-EXP-SnSe}

We have collected THz-emission data as a result of applying ultrafast
laser pulses on single crystal SnSe. The experimental methods are
described in Sec. \ref{SI-EXP-exp}. Analysis of the experimental
data in line with implications from NADT-WPT summarised in Sec. \ref{conclude-sec}
of the main text is conducted in Secs. \ref{SI-EXP-planeforSOC} and
\ref{SI-EXP-qualitative}. There, we have employed particular data
extraction procedures to separate away the signal components relevant
to our investigation purpose from those uninterested components. These
fitting procedures guided by NADT-WPT were detailed in Sec. \ref{SI-EXP-fit}.

\subsubsection{Sample preparation and experimental methods}

\label{SI-EXP-exp}

The SnSe single crystals were prepared using the Bridgman method.
Tin rods (99.99\%) and Selenium beads (99.999\%) were used as starting
materials. Stoichiometric amounts of raw elements were loaded into
a small quartz tube and sealed into a second ampoule to avoid cracking
upon cooling. The charged ampoule was heated in a vertical furnace
to 1183 K in 12 hr. and kept at 1183 K for 24-48 hr. for homogenisation.
After that, the ampoule was slowly cooled to 973 K at a rate of 1
K/h and kept for 50 hr. Millimetre-sized crystals with metallic luster
were obtained. The crystal structure was confirmed by X-ray diffraction
(XRD) with Cu K$\text{\ensuremath{\alpha}}$ radiation.

\begin{figure}[h] \includegraphics[width=12cm, height=11.0cm]{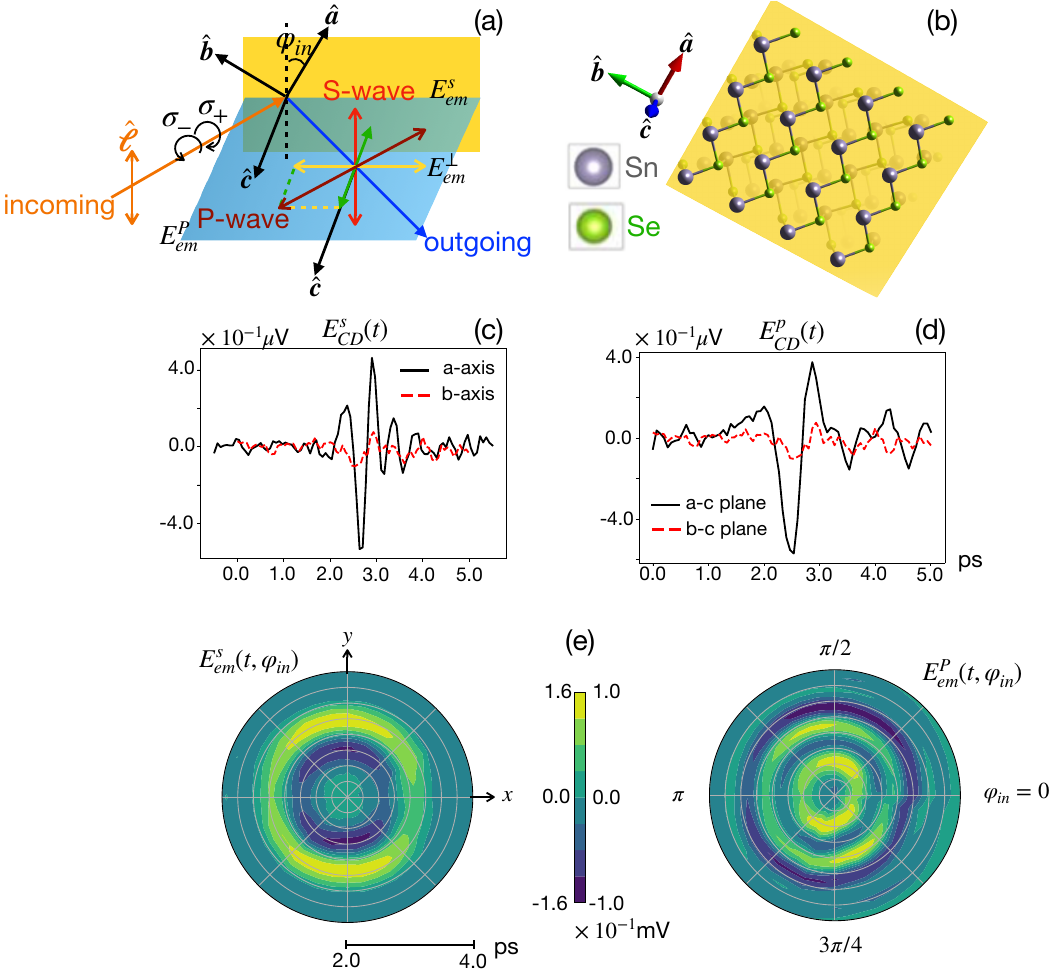} 
\caption{THz-emission experiments on SnSe. (a): Schematics of layouts for detecting emitted THz electric fields generated by LP/CP pulses.  The S-wave (the double-headed red arrow) is polarised along $\hat{\boldsymbol{\ell}}$,  the lab-fixed direction,  while the electric fields of the P-wave (the double-headed brown arrow) contain both an uninterested out-of-plane (c-axis) component (the double-headed green arrow) and the interested in-plane component (the double-headed yellow arrow) perpendicular to $\hat{\boldsymbol{\ell}}$.  (b): Atomic orthorhombic structure of the SnSe crystal.   (c)/(d): CD signals of S/P-wave THz emissions show strong anisotropy between the a- and b- axes (see legends). 
(e): The LP-pulses-induced S-wave $E^{s}_{em}(t,\varphi_{in})$ (the left panel) and P-wave $E^{p}_{em}(t,\varphi_{in})$ (the right panel) data collected by scanning the angle $\varphi_{in}$ over one round.  The radial direction corresponds to the delay time (see the legend).  
} 
\label{exp-phi-scan} 
\end{figure}

A Ti:sapphire regenerative amplifier (Astrella UPS, Coherent, Inc.)
was employed to perform the THz emission experiments. An optical pulse
train with a central wavelength of 800 nm (\ensuremath{\sim}1.55eV),
a pulse duration of \ensuremath{\sim}35 fs, and a repetition rate
of 1 kHz was split into a pump beam and a probe beam. The pump beam
(average power \ensuremath{\sim} 1.5 mW) was used to illuminate the
surface of SnSe single crystals to generate THz electromagnetic radiation.
The probe beam was used in electro-optic sampling to depict the THz
waveform in the time domain by changing the optical path difference
between the pump and probe beams. A wire-grid polariser was used to
purify the polarisation of the emitted THz radiation from samples
in vertical (S-wave) or horizontal directions (P-wave). See detection
layouts in Fig. \ref{exp-phi-scan}(a), where the unit vectors along
the three crystal axes, namely, a-, b- and c-axes are denoted by $\hat{\boldsymbol{a}}$,
$\hat{\boldsymbol{b}}$ and $\hat{\boldsymbol{c}}$ respectively.
Corresponding atomic structure of the single crystal SnSe is schematically
shown in Fig. \ref{exp-phi-scan}(b).

We apply laser pulses linearly polarised (LP) in the lab-referenced
fixed direction $\hat{\boldsymbol{\ell}}$, perpendicular to the incident
plane (shallow blue) spanned by the incoming and outgoing propagation
vectors (the orange and the deep-blue solid arrows respectively) of
the electromagnetic waves (see Fig. \ref{exp-phi-scan}(a)). The incident
electric fields of the LP pulses parallel to $\hat{\boldsymbol{\ell}}$
thus lie on the a-b plane of the sample (coloured in yellow, see Fig.
\ref{exp-phi-scan}(a)). The plane spanned by the a- and b- crystal
axes is called the sample plane throughout this work. The angle $\varphi_{in}$
between the crystal a-axis and the fixed direction $\hat{\boldsymbol{\ell}}$
can be varied by rotating the sample plane around the c-axis. The
electric fields of the emitted S-wave are along the direction $\hat{\boldsymbol{\ell}}$
(see the double-headed red arrow in Fig. \ref{exp-phi-scan}(a)).
The electric fields of the emitted P-wave (see the double-headed brown
arrow in Fig. \ref{exp-phi-scan}(a)) are perpendicular to $\hat{\boldsymbol{\ell}}$.
The LP-pulse induced S-wave and P-wave with the sample rotation angle
$\varphi_{in}$ are denoted respectively by $E_{em}^{s}\left(t,\varphi_{in}\right)$
and $E_{em}^{p}\left(t,\varphi_{in}\right)$. We also apply right-/left-hand
circularly polarised (CP) pulses $\sigma_{+}/\sigma_{-}$. The corresponding
THz emissions as S-wave and P-wave are denoted by $E_{em}^{s}\left(t,\sigma_{\pm}\right)$
and $E_{em}^{p}\left(t,\sigma_{\pm}\right)$ respectively and the
circular-dichroism (CD) signals are defined by $E_{CD}^{s/p}\left(t\right):=E_{em}^{s/p}\left(t,\sigma_{+}\right)-E_{em}^{s/p}\left(t,\sigma_{-}\right)$. 

\subsubsection{The a-b plane of SnSe single crystals as the playground for the in-plane
SOC}

\label{SI-EXP-planeforSOC}

We first confirm that the plane on which the SOC plays a major role
is the a-b plane of SnSe single crystals. Two different methods have
been used to reach the same conclusion that the c-axis signals are
not important with regards to our interests here. The first method
uses CD data only. The second method uses data from LP excitations
only with auxiliary analysis guided by NADT-WPT.

The S-wave CD data $E_{CD}^{s}\left(t\right)$ obtained with $\varphi_{in}=0$
and $\varphi_{in}=\pi/2$ represent the SOC effects encoded in the
emitted waves along the a-axis and b-axis respectively. They are shown
respectively as the black solid (for a-axis) and the red dashed (for
b-axis) lines in Fig. \ref{exp-phi-scan}(c). In accompanying to this,
the P-wave CD data $E_{CD}^{p}\left(t\right)$ obtained with $\varphi_{in}=0$
and $\varphi_{in}=\pi/2$ show the CD emission waves on the b-c plane
and the a-c plane, respectively plotted as the red dashed and the
black solid lines in Fig. \ref{exp-phi-scan}(d). By comparing the
overall magnitudes of the a-axis signal in Fig. \ref{exp-phi-scan}(c)
with that of the a-c plane signal in Fig. \ref{exp-phi-scan}(d) (both
as black solid lines), we find the involvement of the c-axis signal
does little to affect the observed CD magnitudes contributed by the
signal fields along the a-axis. Similar conclusions can be drawn from
the comparison between the b-axis signal in Fig. \ref{exp-phi-scan}(c)
with that of the b-c plane in Fig. \ref{exp-phi-scan}(d) (both as
red dashed lines). Therefore, the SOC-related electronic motion is
mainly on the a-b plane, which is also called the sample plane here.
We extract the peak-to-peak values from the time sequences of $E_{CD}^{s}\left(t\right)$
obtained for the a-axis and b-axis respectively denoted as $\bar{E}_{CD}^{a/b}$.
They exhibit obvious anisotropy, namely, $\bar{E}_{CD}^{a}\approx1\mu\text{V}>\bar{E}_{CD}^{b}\approx0.2\mu\text{V}$. 

The time-resolved THz emissions induced by LP pulses applied on the
sample plane with the sample rotation angle $\varphi_{in}$ scanned
over a round are documented in Fig. \ref{exp-phi-scan}(e). From these
data we deduce the in-plane (on the sample-plane) longitudinal and
transverse components of the emitted THz electric fields in line with
the NADT-WPT treatment of the photocurrent induced by LP pulses in
terms of its spatial components $\parallel$ and $\perp$ to $\hat{\boldsymbol{e}}$
in the main text. The in-plane anisotropy here is furnished by two
orthogonal and non-equivalent axes, namely, the a- and b-axes, showing
similarity to Sec. \ref{sec-examples-Rashba}. We henceforth tried
Eq. (\ref{mirror-perp-pp}) of the main text to construct mirror-symmetric
and mirror-asymmetric functions as fitting functions for the emitted
in-plane electric fields that vary longitudinally and transversely
to $\hat{\boldsymbol{\ell}}$ respectively. Details are in Sec. \ref{SI-EXP-fit}.
The longitudinal components are given by the electric fields of the
S-wave $E_{em}^{s}\left(t,\varphi_{in}\right)$ since they extend
in parallel to $\hat{\boldsymbol{\ell}}$. The suitability of mirror-symmetric
fitting $E_{em}^{\parallel}\left(t,\varphi_{in}\right)$ for $E_{em}^{s}\left(t,\varphi_{in}\right)$
is demonstrated in Fig. \ref{mirror-nonmirror-expthe}(a). The directly
detectable P-wave $E_{em}^{p}\left(t,\varphi_{in}\right)$ contains
both an uninterested c-axis component (the double-headed green arrow
in parallel with $\hat{\boldsymbol{c}}$ in Fig. \ref{exp-phi-scan}(a))
and the targeted component $E_{em}^{\perp}\left(t,\varphi_{in}\right)$
(the double-headed yellow arrow in Fig. \ref{exp-phi-scan}(a)). A
particular procedure (see Sec. \ref{SI-EXP-fit}) is thus applied
to fix this c-axis component with direct fitting to $E_{em}^{p}\left(t,\varphi_{in}\right)$
by either mirror-symmetric $f_{S}^{p}\left(t,\varphi_{in}\right)$
or mirror-asymmetric $f_{A}^{p}\left(t,\varphi_{in}\right)$ functions.
From the above procedure, then $E_{em}^{\perp}\left(t,\varphi_{in}\right)$
is extracted. An improvement using the mirror-asymmetric function
$f_{A}^{p}\left(t,\varphi_{in}\right)$ over the mirror-symmetric
function $f_{S}^{p}\left(t,\varphi_{in}\right)$ to fit $E_{em}^{p}\left(t,\varphi_{in}\right)$
is demonstrated in Fig. \ref{mirror-nonmirror-expthe}(b). The mirror-symmetric/asymmetric
contrast $E_{em}^{\parallel}\left(t,\varphi_{in}\right)$/$E_{em}^{\perp}\left(t,\varphi_{in}\right)$
is showcased in Fig. \ref{mirror-nonmirror-expthe}(c) along with
a reference of similar contrast shown in Fig. \ref{mirror-nonmirror-expthe}(d)
from the theoretical calculation of the Rashba system of Sec. \ref{sec-examples-Rashba}
in the main text. 

From the time sequences of $E_{em}^{s}\left(t,\varphi_{in}\right)$,
$E_{em}^{p}\left(t,\varphi_{in}\right)$ and $E_{em}^{\perp}\left(t,\varphi_{in}\right)$,
we extract their peak-to-peak values at each angle $\varphi_{in}$
and they are denoted by $\bar{E}_{em}^{\parallel}\left(\varphi_{in}\right)$,
$\bar{E}_{em}^{p}\left(\varphi_{in}\right)$ and $\bar{E}_{em}^{\perp}\left(\varphi_{in}\right)$
respectively. These peak-to-peak values are documented in Fig. \ref{mirror-nonmirror-expthe}(e).
The closeness of $\bar{E}_{em}^{\perp}\left(\varphi_{in}\right)$
(green triangles) to $\bar{E}_{em}^{p}\left(\varphi_{in}\right)$
(black crosses) in Fig. \ref{mirror-nonmirror-expthe}(e) indicates
that the c-axis components contained in $\bar{E}_{em}^{p}\left(\varphi_{in}\right)$
are relatively small. This conclusion about the insignificance of
the c-axis component, obtained purely from the LP data aided by fittings
in accordance with NADT-WPT, is in line with the similar conclusion
that is purely obtained from the CD data above.

\subsubsection{Supports on the theoretical implications briefed in the main text
without specification of the type of SOC}

\label{SI-EXP-qualitative}

We are now facilitated with enough information to inspect the theoretical
points (i-iii) abstracted in Sec. \ref{concl-EXPIM} of the main text.
Theoretically the logic usually flows as (i)+(ii)$\Rightarrow$(iii),
started by the physical picture of (i) depicting the spin-mediation
and the bond-mediation of electronic currents as fundamentally different
processes. However, from the perspective of data availability in experiments
where it is only the full signal (spin-mediated plus bond-mediated
ones) that is accessed directly, it is operationally more convenient
to first verify (ii) and (iii). Then by plain logic, (i) directly
follows from (ii)+(iii). Below we discuss how the consistency between
the points (ii-iii) and the experimental data is observed. For tackling
(ii), we study the differentiation between longitudinally and transversely
excited THz emissions under LP pulses, in terms of their respective
anisotropic properties. This can be approached with or without the
involvement of the CD data.

\begin{figure}[h] \includegraphics[width=13cm, height=7.0cm]{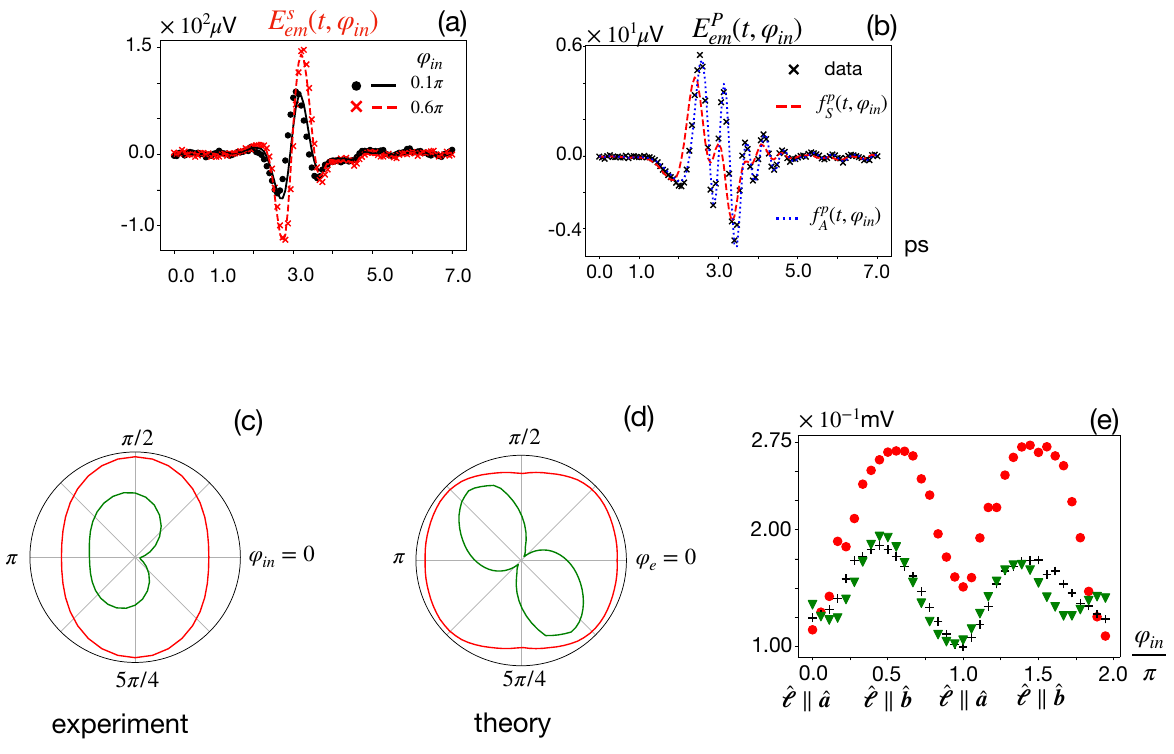} 
\caption{Differentiation between $\parallel$ and $\perp$ components of THz emissions.  (a): Detected S-wave $E^{s}_{em}(t,\varphi_{in})$ data points (black circle and red cross marks) and fitting to them by mirror-symmetric functions $E^{\parallel}_{em}(t,\varphi_{in})$ (black solid and red dashed lines) (see legends for selected values of $\varphi_{in}$).  (b):  Fitting to  data of $E^{p}_{em}(t,\varphi_{in})$ (black cross marks)  by mirror-symmetric function $f^{p}_{S}(t,\varphi_{in})$ (red dashed line) is improved by mirror-asymmetric function  $f^{p}_{A}(t,\varphi_{in})$ (blue dotted line) illustrated with $\varphi_{in}=0.9\pi$.  
(c): A mirror-symmetric/asymmetric contrast between $E^{\parallel}_{em}(t,\varphi_{in})$ (the red line) and $E^{\perp}_{em}(t,\varphi_{in})$ (the green line) from fittings to the  experimental data is snapshotted at $t=3.1$ps as a polar plot over varied $\varphi_{in}$.  (d): Similar mirror-symmetric/asymmetric contrast is seen in the polar plot of $\varphi_{e}$-dependencies of $j^{\parallel}(t,\varphi_{e})$ (the red line) and $j^{\perp}(t,\varphi_{e})$ (the green line) computed theoretically using the Hamiltonian in Sec. \ref{sec-examples-Rashba} of the main text. This is snapshotted at $t-t_{0}=2.4t^{*}$ and with the same parameters used for the calculations of Fig. \ref{CISP-Szpp} in the main text.   In both (c) and (d),  the zero and maximum radial lengths correspond to $-0.8$ and $1.0$ in their respective units $0.1$mV for (c) and $\left(-e/\hbar\right)k_{\text{so}}^{*}\varepsilon_{\text{so}}^{*}$ for (d).   (e): Peak-to-peak THz emitted signals as a function of $\varphi_{in}$.   The red  disks,  black crosses and green triangles mark respectively $\bar{E}^{\parallel}(\varphi_{in})$,  $\bar{E}^{p}(\varphi_{in})$ and $\bar{E}^{\perp}(\varphi_{in})$.  
} 
\label{mirror-nonmirror-expthe} 
\end{figure}

We first go without consulting the CD data by directly inspecting
the $\varphi_{in}$-dependence patterns displayed in the transient
processes and in the peak-to-peak extractions. For the longitudinal
component, both the transient snapshots $E_{em}^{\parallel}\left(t,\varphi_{in}\right)$
and the peak-to-peak values $\bar{E}^{\parallel}\left(\varphi_{in}\right)$
exhibit visually identifiable mirror-symmetric pattern (see the red
line and the red disks in Fig. \ref{mirror-nonmirror-expthe}(c) and
(e) respectively). In contrast, for the transverse component, only
the peak-to-peak extractions $\bar{E}^{\perp}\left(\varphi_{in}\right)$
show a more mirror-symmetric pattern which is apparently not visible
in the snapshots $E_{em}^{\perp}\left(t,\varphi_{in}\right)$ (see
the green triangles and the green line in Fig. \ref{mirror-nonmirror-expthe}(e)
and (c) respectively). This observation of the contrast between the
longitudinally and transversely induced signals corroborates that
longitudinal and transverse currents are excited by underlying different
mechanisms, abstracted as the point (ii) for Sec. \ref{conclude-sec}
in the main text. 

We now describe how the CD data can be useful. In parallel to the
theoretical view of Sec. \ref{sec-examples-SOC-isc} of the main text,
the emitted electric fields lying on the sample plane are also decomposed
to two parts respectively attributed to the underlying bond-mediated
(with subscript $K$) and the spin-mediated (with subscript $so$)
processes behind the excited photocurrents, namely, 
\begin{equation}
\bar{E}_{em}^{\parallel/\perp}\left(\hat{\boldsymbol{\ell}}\right)=\bar{E}_{K}^{\parallel/\perp}\left(\hat{\boldsymbol{\ell}}\right)+\bar{E}_{so}^{\parallel/\perp}\left(\hat{\boldsymbol{\ell}}\right).\label{EM-Kso-d1}
\end{equation}
In Eq. (\ref{EM-Kso-d1}), we replace $\varphi_{in}$ that appear
in the arguments of the LP-pulse-induced signals by $\hat{\boldsymbol{\ell}}$
for convenience of later discussions. The previously obtained anisotropic
relation in CD signals,$\bar{E}_{CD}^{a}>\bar{E}_{CD}^{b}$, is first
used to deduce similar anisotropic relations for the spin-mediated
parts of the LP-induced emissions according to the link between CD
and SOC \cite{Ganichev03R935}. Then together with Eq. (\ref{EM-Kso-d1}),
we can attempt to infer possible anisotropic relations for the bond-mediated
parts to deduce further differentiation between the spin-mediated
and the bond-mediated contributions in LP-induced THz emissions.

The CD detection is a common experimental mean to certify the existence
of SOC \cite{Ganichev03R935} that also underlies the LP-induced signals
$\bar{E}_{so}^{\parallel/\perp}\left(\hat{\boldsymbol{\ell}}\right)$.
The result $\bar{E}_{CD}^{a}>\bar{E}_{CD}^{b}$ implies that the SOC
effect is stronger for electronic motion along the direction of $\hat{\boldsymbol{a}}$
than that in the direction of $\hat{\boldsymbol{b}}$. Therefore,
for longitudinally induced signals, we have
\begin{equation}
\bar{E}_{so}^{\parallel}\left(\hat{\boldsymbol{\ell}}\parallel\hat{\boldsymbol{a}}\right)>\bar{E}_{so}^{\parallel}\left(\hat{\boldsymbol{\ell}}\parallel\hat{\boldsymbol{b}}\right).\label{a-b_ani-soE}
\end{equation}
For the transverse response, note that $\bar{E}_{so/K}^{\perp}\left(\hat{\boldsymbol{\ell}}\parallel\hat{\boldsymbol{a}}\right)$
is associated with currents that run in the direction of $\hat{\boldsymbol{b}}$
and $\bar{E}_{so/K}^{\perp}\left(\hat{\boldsymbol{\ell}}\parallel\hat{\boldsymbol{b}}\right)$
is emitted by current that runs along $\hat{\boldsymbol{a}}$. The
implication from $\bar{E}_{CD}^{a}>\bar{E}_{CD}^{b}$ for transversely
excited THz emissions by LP pulses should then be read as 
\begin{equation}
\bar{E}_{so}^{\perp}\left(\hat{\boldsymbol{\ell}}\parallel\hat{\boldsymbol{b}}\right)>\bar{E}_{so}^{\perp}\left(\hat{\boldsymbol{\ell}}\parallel\hat{\boldsymbol{a}}\right).\label{trv-absoE}
\end{equation}
With Eqs. (\ref{a-b_ani-soE}) and (\ref{trv-absoE}) for the spin-mediated
parts and knowing also the full signals $\bar{E}_{em}^{\parallel/\perp}\left(\hat{\boldsymbol{\ell}}\right)=\bar{E}_{K}^{\parallel/\perp}\left(\hat{\boldsymbol{\ell}}\right)+\bar{E}_{so}^{\parallel/\perp}\left(\hat{\boldsymbol{\ell}}\right)$
in Fig. \ref{mirror-nonmirror-expthe}(e), we should be able to deduce
corresponding relation between $\bar{E}_{K}^{\parallel}\left(\hat{\boldsymbol{\ell}}\parallel\hat{\boldsymbol{a}}\right)$
and $\bar{E}_{K}^{\parallel}\left(\hat{\boldsymbol{\ell}}\parallel\hat{b}\right)$
as well as the relation between $\bar{E}_{K}^{\perp}\left(\hat{\boldsymbol{\ell}}\parallel\hat{\boldsymbol{a}}\right)$
and $\bar{E}_{K}^{\perp}\left(\hat{\boldsymbol{\ell}}\parallel\hat{b}\right)$.
For the longitudinal component, the full signals satisfy $\bar{E}_{em}^{\parallel}\left(\hat{\boldsymbol{\ell}}\parallel\hat{\boldsymbol{a}}\right)<\bar{E}_{em}^{\parallel}\left(\hat{\boldsymbol{\ell}}\parallel\hat{\boldsymbol{b}}\right)$
(see red circles in Fig. \ref{mirror-nonmirror-expthe}(e)). We thus
deduce from Eq. (\ref{a-b_ani-soE}) that
\begin{equation}
\bar{E}_{K}^{\parallel}\left(\hat{\boldsymbol{\ell}}\parallel\hat{\boldsymbol{a}}\right)<\bar{E}_{K}^{\parallel}\left(\hat{\boldsymbol{\ell}}\parallel\hat{\boldsymbol{b}}\right).\label{a-b_ani-KE}
\end{equation}
On the contrary, for the transverse components for which the full
signals satisfy $\bar{E}^{\perp}\left(\hat{\boldsymbol{\ell}}\parallel\hat{\boldsymbol{a}}\right)=\bar{E}_{so}^{\perp}\left(\hat{\boldsymbol{\ell}}\parallel\hat{\boldsymbol{a}}\right)+\bar{E}_{K}^{\perp}\left(\hat{\boldsymbol{\ell}}\parallel\hat{\boldsymbol{a}}\right)<\bar{E}^{\perp}\left(\hat{\boldsymbol{\ell}}\parallel\hat{\boldsymbol{b}}\right)=\bar{E}_{so}^{\perp}\left(\hat{\boldsymbol{\ell}}\parallel\hat{\boldsymbol{b}}\right)+\bar{E}_{K}^{\perp}\left(\hat{\boldsymbol{\ell}}\parallel\hat{\boldsymbol{b}}\right)$
(see green triangles in Fig. \ref{mirror-nonmirror-expthe}(e)) no
definite relation between the bond-mediated parts $\bar{E}_{K}^{\perp}\left(\hat{\boldsymbol{\ell}}\parallel\hat{\boldsymbol{b}}\right)$
and $\bar{E}_{K}^{\perp}\left(\hat{\boldsymbol{\ell}}\parallel\hat{\boldsymbol{a}}\right)$
similar to Eq. (\ref{a-b_ani-KE}) can be deduced from the spin-mediated
parts Eq. (\ref{trv-absoE}). 

The above contrast between $\parallel$ and $\perp$ components of
the THz emissions attributed to underlying photocurrents is briefly
interpreted as the following. The information deduction from the spin-mediated
signals to the bond-mediated signals works in one way for the longitudinal
components while it works in other ways for the transverse components.
This is another side evidence for the fundamental difference between
longitudinal and transverse responses as point (ii). Since such difference
between $\parallel$ and $\perp$ components is facilitated by the
differentiation between spin-mediated and bond-mediated processes,
it is equivalent to the point (iii), namely, these two different photocurrent
components respectively encode the spin/bond-mediated processes in
quite different ways. Hereby we find the observed data consistent
with the points (ii) and (iii) briefed in Sec. \ref{concl-EXPIM}
of the main text. The point (i) there is a logical consequence of
(ii) and (iii).

\subsubsection{Analysis of anisotropic characters under the excitation of LP pulses}

\label{SI-EXP-fit}

Here we analyse the characters of the anisotropic dependence of THz
signals $E_{em}^{s}\left(t,\varphi_{in}\right)$ and $E_{em}^{p}\left(t,\varphi_{in}\right)$
on the linear polarisation angle $\varphi_{in}$ or simply denoted
as $\varphi$. The reference axis from which $\varphi$ is measured
is re-named as the $x$-axis. The axis orthogonal to the $x$-axis
is of course called the $y$-axis as usual. Taking these two axes
as the mirror lines, we first note from the angular scans of $E_{em}^{s/p}\left(t,\varphi\right)$
that the mirror symmetric character in $E_{em}^{s}\left(t,\varphi\right)$
is more visible than that in $E_{em}^{p}\left(t,\varphi\right)$ (see
Fig. \ref{exp-phi-scan}(e)). More explicitly, a mirror-symmetric
function $f_{S}\left(t,\varphi\right)$ is defined by satisfying

\begin{equation}
f_{S}\left(t,-\varphi\right)=f_{S}\left(t,\varphi\right),f_{S}\left(t,\pi-\varphi\right)=f_{S}\left(t,\varphi\right),\label{C2v-fitreq}
\end{equation}
in which the first/second of Eq. (\ref{C2v-fitreq}) describes reflection
about the $x$/$y$-axis. According to Sec. \ref{sec-examples-Rashba}
of the main text, the analytical form of the bond-mediated part of
longitudinal current Eq. (\ref{j-K-frPL}), which does not involve
any form of SOC, is simply a result of generic quadratic dispersion.
Its dependence on $\varphi$ given by $\cos\left(2\varphi\right)$
satisfies the mirror-symmetric property of Eq. (\ref{C2v-fitreq}).
In order to have the fitting parameters to be optimised, then according
to $\cos\left(2\varphi\right)=\cos^{2}\varphi-\sin^{2}\varphi$ we
assume 
\begin{equation}
f_{S}\left(t,\varphi\right)=\alpha_{0}\left(t\right)\cos^{2}\varphi+\alpha_{1}\left(t\right)\sin^{2}\varphi,\label{sfit-f1}
\end{equation}
in which $\alpha_{0}\left(t\right)$ and $\alpha_{1}\left(t\right)$
are treated as the fitting parameters. For mirror-asymmetric functions
$f_{A}\left(t,\varphi\right)$, the transverse components of Eqs.
(\ref{j-K-fr}) and (\ref{jSO-eC0}) suggest the use of $\cos\left(2\varphi\right)$
and $\sin\left(2\varphi\right)$ in the composition of $f_{A}\left(t,\varphi\right)$.
Real physical observable surely repeats itself when $\varphi$ is
varied over $2\pi$. This suggests one includes the most basic periodic
functions $\cos\left(\varphi\right)$ and $\sin\left(\varphi\right)$
for fittings to $E_{em}^{p}\left(t,\varphi\right)$. So two forms
of $f_{A}\left(t,\varphi\right)$ are compared, namely,

\begin{equation}
f_{A1}\left(t,\varphi\right)=\beta_{0}\left(t\right)\cos\varphi+\beta_{1}\left(t\right)\sin\varphi,\label{pfit-f1}
\end{equation}
for which only the $2\pi$-periodicity is considered and
\begin{equation}
f_{A2}\left(t,\varphi\right)=\beta_{0}^{\prime}\left(t\right)\cos\varphi+\beta_{1}^{\prime}\left(t\right)\sin\varphi+c_{0}\left(t\right)\cos\left(2\varphi\right)+c_{1}\left(t\right)\sin\left(2\varphi\right),\label{pfit-f2}
\end{equation}
for which the particular mirror-asymmetric characters from the theoretical
consideration in Sec. \ref{sec-examples-Rashba} of the main text
are included too. 

Here we clarify which fitting function is applied to fit which signal.
By re-labeling $\alpha_{0/1}\left(t\right)$ in Eq. (\ref{sfit-f1})
as $\alpha_{0/1}^{s}\left(t\right)$ and renotating $f_{S}\left(t,\varphi\right)$
by $f_{S}^{s}\left(t,\varphi\right)$ as specification to the S-wave,
the data $E_{em}^{s}\left(t,\varphi\right)$ is fitted by $f_{S}^{s}\left(t,\varphi\right)$.
For the purpose of comparison, $E_{em}^{p}\left(t,\varphi\right)$
is fitted by three different functions, $f_{S}^{p}\left(t,\varphi\right)$,
$f_{A1}^{p}\left(t,\varphi\right)$ and $f_{A2}^{p}\left(t,\varphi\right)$
whose forms are given respectively by Eqs. (\ref{sfit-f1}), (\ref{pfit-f1})
and (\ref{pfit-f2}) with the fitting parameters specified for the
P-wave data. Note that the plane of interests is the $x$-$y$ plane
but the P-wave contains also an out-of-plane $z$-component as mentioned
before (recall the c-axis signal from Fig. \ref{exp-phi-scan}(a)).
By the theory of electromagnetism in medium considering the $x$-$z$
plane to the be incident plane, we have \cite{Shan04book},

\begin{equation}
E_{em}^{p}\propto\dot{J}_{z}-\gamma\dot{J}_{x},\label{SinPout_phbck}
\end{equation}
where the current $\boldsymbol{J}$ has all three $x$-, $y$- and
$z$-components. Here $\dot{J}_{x}$ and $\dot{J}_{z}$ are time derivatives
of the $x$- and the $z$-components of the current. The coefficient
$\gamma$ is related to the dielectric constant of the material \cite{Shan04book}.
Because $\varphi$ is an attribute of vectors on the $x$-$y$ plane
orthogonal to the $z$-axis, we assume that $J_{z}$ is independent
of $\varphi$. So fittings to $E_{em}^{p}\left(t,\varphi\right)$
are attempted by $f_{Ai}^{p}\left(t,\varphi\right)=f_{Ai}\left(t,\varphi\right)+f_{i}\left(t\right)$
for $i=1,2$ where $f_{i}\left(t\right)$ having no dependence on
$\varphi$ accounts for $\dot{J}_{z}$. The fitting parameters are
found by minimising the difference between the values obtained by
the fitting functions and those of the data themselves. The difference
functions are thus defined as $L^{s}\left(t\right)=\sum_{n=1}^{N_{in}}\left[f_{S}^{s}\left(t,\varphi_{n}\right)-E_{em}^{s}\left(t,\varphi_{n}\right)\right]^{2}$
for the S-wave data and $L_{X}^{p}\left(t\right)=\sum_{n=1}^{N_{in}}\left[f_{X}^{p}\left(t,\varphi_{n}\right)-E_{em}^{p}\left(t,\varphi_{n}\right)\right]^{2}$
for the P-wave data where $N_{in}$ is the number of angle values
collected. These difference functions are non-negative. Here $X=\left\{ S,A1,A2\right\} $
enumerates over different settings of $\varphi$-dependence characters
for fitting the P-wave data. The fitting parameters $\left\{ \alpha_{0}^{s}\left(t\right),\alpha_{1}^{s}\left(t\right)\right\} $
in $f_{S}^{s}\left(t,\varphi\right)$ are obtained by solving the
algebraic equations $\partial L^{s}\left(t\right)/\partial\alpha_{0}^{s}\left(t\right)=0$
and $\partial L^{s}\left(t\right)/\partial\alpha_{1}^{s}\left(t\right)=0$
for given $t$. Similar procedure is applied to obtain the fitting
parameters in $f_{X}^{p}\left(t,\varphi\right)$'s. To calibrate the
error of fitting as dimensionless quantities, we define $\Delta E_{n}^{s/p}=\text{max}\left\{ E_{em}^{s/p}\left(t_{i},\varphi_{n}\right)\right\} _{i=1}^{N_{T}}-\text{min}\left\{ E_{em}^{s/p}\left(t_{i},\varphi_{n}\right)\right\} _{i=1}^{N_{T}}$
where $N_{T}$ is the number of points in the time sequence probed
for the experimental run with fixed polarisation angle $\varphi_{n}$.
The error at each tested angle $\varphi_{n}$ is then given by $\delta_{n}^{s}=\left[\sqrt{\sum_{i=1}^{N_{T}}\left[f_{S}^{s}\left(t_{i},\varphi_{n}\right)-E_{em}^{s}\left(t_{i},\varphi_{n}\right)\right]^{2}/N_{T}}\right]/\Delta E_{n}^{s}$
for fitting to the S-wave and $\delta_{n}^{pX}=\left[\sqrt{\sum_{i=1}^{N_{T}}\left[f_{X}^{p}\left(t_{i},\varphi_{n}\right)-E_{em}^{p}\left(t_{i},\varphi_{n}\right)\right]^{2}/N_{T}}\right]/\Delta E_{n}^{p}$
for fitting to the P-wave. The overall error over all angles $\left\{ \varphi_{n}\right\} _{n=1}^{N_{in}}$
is given by the averages over these non-negative values, namely, $e_{s}=\sum_{n=1}^{N_{in}}\delta_{n}^{s}/N_{in}$
and $e_{pX}=\sum_{n=1}^{N_{in}}\delta_{n}^{pX}/N_{in}$. To see if
the error is particularly large for some values of $\varphi_{n}$,
standard deviations in the error distributions $\left\{ \delta_{n}^{s}\right\} _{n=1}^{N_{in}}$
and $\left\{ \delta_{n}^{pX}\right\} _{n=1}^{N_{in}}$ are extracted
and denoted by $\delta_{s}=\sqrt{\sum_{n=1}^{N_{in}}\left(\delta_{n}^{s}-e_{s}\right)^{2}/N_{in}}$
and $\delta_{pX}=\sqrt{\sum_{n=1}^{N_{in}}\left(\delta_{n}^{pX}-e_{pX}\right)^{2}/N_{in}}$
respectively. In the experiment, we have $N_{in}=36$ for which we
vary $\varphi_{in}$ from $0^{\circ}$ to $350^{\circ}$ by every
$10^{\circ}$ and the probed time window of 7 picoseconds is divided
into $N_{T}=106$ points for data acquisition. From the experimental
data and the above procedure, we get $e_{s}=4.2\%$, $e_{pS}=7.8\%$,
$e_{pA1}=5.9\%$, and $e_{pA2}=3.2\%$ and $\delta_{s}=2.3\%$, $\delta_{pS}=3.3\%$,
$\delta_{pA1}=2.6\%$ and $\delta_{pA2}=1.6\%$. The fitting to the
S-wave data by mirror-symmetric function $E_{em}^{\parallel}\left(t,\varphi\right)=f_{S}^{s}\left(t,\varphi\right)$
is thus plausible while the best fitting here to the P-wave data is
the mirror-asymmetric function given by $f_{A}^{p}\left(t,\varphi\right)=f_{A2}^{p}\left(t,\varphi\right)$.
The electric field component $E_{em}^{\perp}\left(t,\varphi\right)$
is then extracted accordingly by $E_{em}^{\perp}\left(t,\varphi\right)=f_{A2}^{p}\left(t,\varphi\right)-f_{A2}\left(t\right)$.
Hereby we have supplemented details of the fitting functions $E_{em}^{\parallel}\left(t,\varphi\right)$
and $E_{em}^{\perp}\left(t,\varphi\right)$ mentioned in Sec. \ref{SI-EXP-planeforSOC}
above.

\bibliographystyle{myunsrt} 
\bibliography{refs_bibexd-1} 

\begin{thebibliography}{10}
\bibitem{FootnoteRSOCkc}
In the literature we have surveyed, we can find values of $\lambda_{R}$ (Rashba
  parameter), $\lambda_{D}$ (Dresselhaus parameter) and the spin-split gap
  $\Delta$ so that the archetype Rashba-Dresselhaus SOC Hamiltonian described
  by
  $\mathcal{H}_{so}\left(\boldsymbol{k}\right)=\lambda_{R}\left(k_{x}\sigma_{y}-k_{y}\sigma_{x}\right)+\lambda_{D}\left(k_{x}\sigma_{y}+k_{y}\sigma_{x}\right)$
  can be used to perform an estimation of the inherent parameter
  $k_{c}^{-1}/\Delta$. Because it is not always immediately clear at which
  $\boldsymbol{k}$ the spin-split $\Delta$ is given, for the sole purpose of a
  rough estimation, we take the following steps to find an "averaged" value of
  $k_{c}^{-1}$. The difference between the two eigenenergies of
  $\mathcal{H}_{so}\left(\boldsymbol{k}\right)$ at
  $\left(k_{x},k_{y}\right)=\left(k_{ax},0\right)$ reads
  $\Delta_{x}=\left\vert\lambda_{R}+\lambda_{D}\right\vert k_{ax}$ and that at
  $\left(k_{x},k_{y}\right)=\left(0,k_{ay}\right)$ is
  $\Delta_{y}=\left\vert\lambda_{R}-\lambda_{D}\right\vert k_{ay}$. Matching
  $\Delta_{x/y}$ to $\Delta$ then gives $k_{ax/y}$ in terms of $\lambda_{R}$,
  $\lambda_{D}$ and $\Delta$. By diagonalising
  $\mathcal{H}_{so}\left(\boldsymbol{k}\right)$ at
  $\left(k_{x},k_{y}\right)=\left(k_{ax},0\right)$ to obtain the two band
  wavefunctions used in Eq. (\ref{BerryCon-def-Mk}) with $m=+$ and $n=-$ for
  the two bands of opposite spins, we find one estimation of $k_{c}^{-1}$ noted
  as $k_{cx}^{-1}$ also in terms of $\lambda_{R}$, $\lambda_{D}$ and $\Delta$.
  Similarly at $\left(k_{x},k_{y}\right)=\left(0,k_{ay}\right)$ one obtains
  another absolute value of Eq. (\ref{BerryCon-def-Mk}) denoted by
  $k_{cy}^{-1}$. We then let $k_{c}^{-1}=\sqrt{k_{cx}^{-1}k_{cy}^{-1}}$. Note
  that straightforwardly with Eq. (\ref{AbelianBrcv-defMk}), we see the Berry
  curvature is always zero for this type of Hamiltonian so the evaluation of
  the Berry connections becomes essential for assessing the geometric effects.
\bibitem{footnoteZeroBCat1D}
The Berry curvature Eq. (\ref{AbelianBrcv-defMk}) is mathematically only
  well-defined when the dimensionality of the BZ is $D\ge2$. Naively setting
  $D=1$ in Eq. (\ref{AbelianBrcv-defMk}) always leads to a zero Berry
  curvature.
\bibitem{FootnoteSSHNumSet}
The pulse-induced wave vector displacement is set to be
  $\mathcal{K}_{t}=\mathcal{K}^{*}\sin\left[\omega\left(t-t_{0}\right)\right]\exp\left\{
  -\left(t-t_{c}\right)^{2}/\tau_{P}^{2}\right\}$ peaked at $t=t_{c}$ chosen as
  the origin of the time axis $t_{c}=0.0$. To start the non-equilibrium
  transient processes, we assume the system stays at equilibrium till
  $t=t_{0}=t_{c}-2\tau_{P}$. The Gaussian decay of the waveform gives
  $t_{\text{off}}\simeq t_{c}+\tau_{P}$. Here the bonds $b_{1}$ and $b_{2}$ are
  in arbitrary unit $b$ which measures time by the unit of $t^{*}=b^{-1}$ and
  $\omega$ by $\omega^{*}=2\pi/t^{*}$. We measure $\mathcal{K}^{*}$ in unit of
  $\pi/a$ so $k_{t}a$ remains dimensionless. We use $\tau_{P}=t^{*}$ in all
  plots for SSH system.
\bibitem{FootnoteSSHsimDirac}
By its definition of Eq. (\ref{jdt-qcc}), the quantum acceleration operator
  $\vec{\mathbb{A}}\left(\boldsymbol{k}_{t}\right)$ (and therefore the
  particular part of the current-rate
  $\dot{\boldsymbol{j}}^{geo}\left(t\right)$) can reveal component of the
  driven spin texture not present in the native spin texture. The well-known 2D
  massless Dirac fermion given by $\mathcal{H}\left(\boldsymbol{k}\right)=\hbar
  v_{F}\boldsymbol{k}\cdot\boldsymbol{\sigma}$ is another example where the
  native spin texture is in-plane while
  $\vec{\mathbb{A}}\left(\boldsymbol{k}_{t}\right)\propto\sigma_{z}$ shows an
  out-of-plane component and is therefore anticipated to also show geometric
  effects in the current rate
  $\dot{\boldsymbol{j}}\left(t>t_{\text{off}}\right)=\dot{\boldsymbol{j}}^{geo}\left(t\right)$
  after turning off the exciting pulse.
\bibitem{footnoteNonScattProb}
In general, the longer $t$ is separated from $t^{\prime}$, the less the value
  $P\left(\boldsymbol{k}_{t},\boldsymbol{k}_{t^{\prime}}\right)$ takes.
  Therefore, with $t-t^{\prime}>t-\left(t^{\prime}+\text{d}t^{\prime} \right)$
  where $\text{d}t^{\prime}$ is an infinitesimal time interval around
  $t^{\prime}$, we have
  $P\left(\boldsymbol{k}_{t},\boldsymbol{k}_{t^{\prime}}\right)<P\left(\boldsymbol{k}_{t},\boldsymbol{k}_{t^{\prime}+\text{d}t^{\prime}}\right)$.
  By proportionality, one then has
  $P\left(\boldsymbol{k}_{t},\boldsymbol{k}_{t^{\prime}}\right)=P\left(\boldsymbol{k}_{t},\boldsymbol{k}_{t^{\prime}+\text{d}t^{\prime}}\right)\left(1-\tau_{\boldsymbol{k}_{t^{\prime}}}^{-1}\text{d}t^{\prime}\right)$
  which gives $\partial
  P\left(\boldsymbol{k}_{t},\boldsymbol{k}_{t^{\prime}}\right)/\partial
  t^{\prime}=-\tau_{\boldsymbol{k}_{t^{\prime}}}^{-1}P\left(\boldsymbol{k}_{t},\boldsymbol{k}_{t^{\prime}}\right)$.
  Consequently,
  $P\left(\boldsymbol{k}_{t},\boldsymbol{k}_{t^{\prime}}\right)=\exp\left[-\int_{t^{\prime}}^{t}\text{d}t^{\prime\prime}\tau_{\boldsymbol{k}_{t^{\prime\prime}}}^{-1}\right]$.
\bibitem{footnote-shortRlxT}
By assuming the smallness of $\tau_{ \boldsymbol{k}_{t^{\prime}} }$, the value
  of $P\left(t,t^{\prime}\right)f\left(t^{\prime}\right)$ becomes
  non-negligible only for $t^{\prime}$ near $t$ such that $\tau_{
  \boldsymbol{k}_{t^{\prime}} }$ is approximated by $\tau_{ \boldsymbol{k}_{t}
  }$.
\bibitem{footnote-partshortRlxT}
Conventionally, the second-order current is found by solving the Boltzmann
  equation for one band $n$, namely, $\partial
  f_{n}\left(\boldsymbol{k}_{t},t\right)/\partial
  t+\left(-e/\hbar\right)\boldsymbol{E}\left(t\right)\cdot \left(\partial
  f_{n}\left(\boldsymbol{k}_{t},t\right)/\partial \boldsymbol{k}_{t}\right)=
  -\left(f_{n}\left(\boldsymbol{k}_{t},t\right)-f_{n}^{0}\left(\boldsymbol{k}_{t}\right)\right)/\tau$
  with $f_{n}\left(\boldsymbol{k}_{t},t\right)$ considered up to the second
  order in the electric field and one computes the current by
  $\boldsymbol{j}\left(t\right)=-e\int\text{d}^{D}\boldsymbol{k}_{t}
  f_{n}\left(\boldsymbol{k}_{t},t\right)\boldsymbol{v}_{n}\left(\boldsymbol{k}_{t}\right)$
  where $\boldsymbol{v}_{n}\left(\boldsymbol{k}_{t}\right)$ is given by the
  right-hand side of Eq.~(\ref{adia-velo}). Following this approach with the
  partial application of the short-relaxation time approximation
  Eq.~(\ref{short-RT}) that leaves the form of
  $\boldsymbol{E}\left(t^{\prime}\right)$ intact in the relevant time integral,
  one then obtains the result of Eq.~(\ref{j2qnh}).
\bibitem{FootnoteIsoSpnTxtSdys}
These results are obtained by solving $\left(\hbar/2\right)\text{d}\left\langle
  \boldsymbol{\sigma}\right\rangle
  _{n_{-}}\left(\boldsymbol{k}_{t},t\right)/\text{d}t=\boldsymbol{\Lambda}_{so}\left(\boldsymbol{k}_{t}\right)\times\left\langle
  \boldsymbol{\sigma}\right\rangle _{n_{-}}\left(\boldsymbol{k}_{t},t\right)$,
  as a result of $i\hbar\left\vert
  \dot{\phi}_{n_{-},\boldsymbol{k}}\left(t,t_{0}\right)\right\rangle
  =\mathcal{H}\left(\boldsymbol{k}_{t}\right)\left\vert
  \phi_{n_{-},\boldsymbol{k}}\left(t,t_{0}\right)\right\rangle$, from the
  initial texture $\left.\left\langle \boldsymbol{\sigma}\right\rangle
  _{n_{-}}\left(\boldsymbol{k}_{t},t\right)\right\vert _{t=t_{0}}=\left\langle
  \boldsymbol{\sigma}\right\rangle _{n_{-}}^{0}\left(\boldsymbol{k}\right)$ to
  a later time $t=t_{1}=t_{0}+\delta t$ using the fourth order Runge-Kutta
  method with sufficiently small $\delta t$. Here in Eq.
  (\ref{t1nadspntxt-aniso}), we have $\overline{\delta
  s_{//}}=-\left(2/3\right)\left(\alpha_{R}^{2}\mathcal{K}_{t_{1/2}}k\right)\left(\delta
  t/\hbar\right)^{2}$ and $\overline{\delta
  s_{z}}=-\left(1/3\right)\left[\alpha_{R}^{3}\mathcal{K}_{t_{1/2}}\left(\mathcal{K}_{t_{1}}+\mathcal{K}_{t_{1/2}}\right)k\right]\left(\delta
  t/\hbar\right)^{3}$. They are dimemsionless amplitudes independent of
  $\theta$ and $\varphi_{e}$ in which $\mathcal{K}_{t}=\left\vert
  \boldsymbol{\mathcal{K}}_{t}\right\vert$ and $t_{1/2}=t_{0}+\delta t/2$.
\bibitem{FootnoteAniSPzp}
This result is seen from Eq. (\ref{t1nadspntxt-aniso}) where one finds
  $\int_{0}^{\pi}\text{d}\theta\overline{\left\langle
  \sigma_{\parallel/z}\right\rangle
  }\left(\boldsymbol{k},t_{1}\right)\propto\int_{0}^{\pi}\text{d}\theta\mathcal{G}_{1}\left(x_{\parallel/z},\theta,\varphi_{e}\right)=\int_{0}^{\pi}\text{d}\theta\left[\left(x_{\parallel/z}+x_{\parallel/z}^{-1}\right)\cos\left(2\theta\right)+\left(x_{\parallel/z}-x_{\parallel/z}^{-1}\right)\right]\sin\left(2\varphi_{e}\right)/\sqrt{r_{X}^{2}\cos^{2}\theta+r_{Y}^{2}\sin^{2}\theta}$
  where $x_{\parallel}=\sqrt{r_{A}}$ and $x_{z}=r_{A}$ by using that the
  function
  $\sin\left(2\theta\right)/\sqrt{r_{X}^{2}\cos^{2}\theta+r_{Y}^{2}\sin^{2}\theta}$
  is odd in the range of integration.
\end{thebibliography}
\end{document}